\documentclass[aps,prd,preprint,superscriptaddress,nofootinbib]{revtex4-1}
\usepackage{braket}
\usepackage{xargs}
\usepackage{mathtools}
\usepackage[english]{babel}
\usepackage[utf8]{inputenc}
\usepackage{amsmath,amssymb,braket,slashed,mathrsfs,amsthm}
\usepackage{pifont}
\usepackage{graphicx}
\usepackage{color,hyperref}
\usepackage{multirow,array}
\usepackage{bm}

\newcommand{\ba}{\begin{eqnarray}}
\newcommand{\ea}{\end{eqnarray}}
\newcommand{\vecp}[1]{\vec{#1}\hspace{1.2pt}}

\makeatletter
\newcommand{\sumint}{\mathop{\mathpalette\sum@int\relax}\slimits@}
\newcommand{\sum@int}[2]{%
	\ooalign{$\m@th#1\sum$\cr\hidewidth$\m@th#1\int$\hidewidth\cr}%
}
\newcommand{\sumintinline}{\mathop{\mathpalette\sum@intinline\relax}\slimits@}
\newcommand{\sum@intinline}[2]{%
	\ooalign{$\m@th#1\scalebox{0.8}{$\sum$}$\cr\hidewidth$\m@th#1\scalebox{1.1}{$\int$}$\hidewidth\cr}%
}
\makeatother

\newcommand\bnl{Physics Department, Brookhaven National Laboratory, Upton, NY 11973, USA}
\newcommand\cu{Physics Department, Columbia University, New York, NY 10027, USA}
\newcommand\riken{RIKEN-BNL Research Center, Brookhaven National Laboratory, Upton, NY 11973, USA}
\newcommand\edinb{School of Physics and Astronomy, The University of Edinburgh, Edinburgh EH9 3FD, UK}
\newcommand\uconn{Physics Department, University of Connecticut, Storrs, CT 06269-3046, USA}
\newcommand\soton{School of Physics and Astronomy, University of Southampton,  Southampton SO17 1BJ, UK}
\newcommand{\innovation}{Collaborative Innovation Center of Quantum Matter, Beijing 100871, China}
\newcommand{\chep}{Center for High Energy Physics, Peking University, Beijing 100871, China}
\newcommand{\pkuphy}{School of Physics, Peking University, Beijing 100871,
China}

\begin{document}
\title{Lattice Calculation of Light Meson Radiative Leptonic Decays}
\author{Peter Boyle}\affiliation{\bnl}\affiliation{\edinb}

\author{Norman H. Christ}\affiliation{\cu}

\author{Xu Feng}\affiliation{\pkuphy}\affiliation{\innovation}\affiliation{\chep}

\author{Taku Izubuchi}\affiliation{\bnl}\affiliation{\riken}

\author{Luchang Jin}\affiliation{\uconn}

\author{Christopher T. Sachrajda}\affiliation{\soton}

\author{Xin-Yu Tuo}\email{ttxxyy.tuo@gmail.com}\affiliation{\bnl}

\date{\today}

\begin{abstract}
In this work, we perform a lattice QCD calculation of the branching ratios and the form factors of radiative leptonic decays $P \to \ell \nu_\ell \gamma$ ($P = \pi, K$) using $N_f=2+1$ domain wall fermion ensembles generated by the RBC and UKQCD collaborations at the physical pion mass. We adopt the infinite-volume reconstruction (IVR) method, which extends lattice data to infinite volume and effectively controls the finite-volume effects.
This study represents a first step toward a complete calculation of radiative corrections to leptonic decays using the IVR method, including both real photon emissions and virtual photon loops.  
For decays involving a final-state electron, collinear radiative corrections, enhanced by the large logarithmic factors such as \(\ln(m_\pi^2/m_e^2)\) and \(\ln(m_K^2/m_e^2)\), can reach the level of \(O(10\%)\) and are essential at the current level of theoretical and experimental precision.
After including these corrections, our result for \(\pi \to e\nu_e\gamma\) agrees with the PIBETA measurement; for \(K \to e\nu_e\gamma\), our results are consistent with the KLOE data and exhibit a $1.7\sigma$ tension with E36; and for \(K \to \mu\nu_\mu\gamma\), where radiative corrections are negligible, our results confirm the previously observed discrepancies between lattice results and the ISTRA/OKA measurements at large photon energies, and with the E787 results at large muon–photon angles.
\end{abstract}

\maketitle

\section{Introduction\label{sec:Intro}}
Radiative leptonic decays of light pseudoscalar mesons \(P \to \ell\nu_\ell\gamma\) (\(P = \pi, K\)) involve all three interactions of the Standard Model, strong, weak, and electromagnetic, making them ideal low-energy processes to probe hadronic structure and test Standard Model predictions. 
Compared to the leptonic decays \(P \to \ell\nu_\ell\), the emission of an additional real photon reveals additional information about the internal structure of the meson. The decay amplitude receives contributions from the inner-bremsstrahlung (IB) term, proportional to the meson decay constant, and the structure-dependent (SD) term, characterized by the vector and axial-vector form factors \(F_V\) and \(F_A\)~\cite{Bijnens:1992en}. By measuring partial branching ratios from different regions of phase-space and comparing them with Standard Model predictions, one can search for signs of new physics. In this way, for example, possible tensor interactions~\cite{Chizhov:2004tu} can be constrained by experimental data~\cite{Bychkov:2008ws}.

In addition, this process provides important input for the determination of the CKM matrix elements \(V_{ud}\) and \(V_{us}\), and thereby for testing the unitarity of the matrix’s first row. 
The most precise deteminations, \(V_{ud}\) from superallowed nuclear \(\beta\) decays~\cite{Hardy:2020qwl} and \(V_{us}\) from leptonic and semileptonic decays, yield a first-row sum \(|V_{ud}|^2 + |V_{us}|^2 + |V_{ub}|^2=0.9983(6)(4)\), where the two uncertainties arise from \(|V_{ud}|^2\) and \(|V_{us}|^2\), respectively~\cite{ParticleDataGroup:2024cfk}. This result deviates from unitarity by \(2.4\sigma\), highlighting the need to better understand the origin of this tension. 
In order to use light meson leptonic decays to improve the determination of \(V_{us}\) and complement the extraction of \(V_{ud}\) from superallowed nuclear \(\beta\) decays, it is necessary to reduce the theoretical uncertainties associated with their radiative corrections.
According to the Bloch-Nordsieck theorem~\cite{Bloch:1937pw}, an infrared-safe prediction of radiative corrections includes both virtual-photon loop corrections and real-photon emissions. The real-photon process \(P \to \ell\nu_\ell\gamma\), studied in this work, is therefore an important component of the complete radiative corrections.
A full calculation of the radiative corrections at $O(\alpha)$ to leptonic decays $P\to\ell\nu_\ell(\gamma)$ using the IVR method is currently underway and the results will be presented in future publications\,~\footnote{The parentheses in $P\to\ell\bar{\nu}_\ell(\gamma)$ signify that the rates with and without a real photon in the final state are summed.}.

In theoretical studies, the hadronic structure of the meson is reflected in the momentum dependence of the form factors \(F_V\) and \(F_A\) defined in Eq.~(\ref{HM}) below. In the vector meson dominance (VMD) approach, this dependence is modeled by a pole-like form arising from low-lying resonances~\cite{Poblaguev:2002ug}. In chiral perturbation theory (ChPT), both form factors are constants at \(\mathcal{O}(p^4)\)~\cite{Bijnens:1992en}, with nontrivial momentum dependence entering only at \(\mathcal{O}(p^6)\)~\cite{Geng:2003mt,Ametller:1993hg}. Although ChPT does not incorporate the pole-like structure of VMD, it provides predictions at small momentum transfers typical of light meson radiative decays, where the form factors are often approximated as linear functions of the squared momentum transfer. To test the validity of this approximation and further reduce hadronic uncertainties, first-principles lattice QCD calculations are essential.

A major challenge in lattice studies of radiative decays is the limited number of discrete momenta available in finite-volume simulations, which makes it difficult to cover the full kinematic region. One strategy to overcome this limitation is to use twisted boundary conditions~\cite{Sachrajda:2004mi}, a method adopted by the Rome–Southampton collaboration to compute the radiative decays of both light mesons (\(\pi\), \(K\)) and charmed mesons (\(D\), \(D_s\))~\cite{Desiderio:2020oej,Frezzotti:2020bfa,Gagliardi:2022szw,DiPalma:2025iud,Frezzotti:2023ygt}.
In these works, the momentum dependence of the form factors was parameterized using linear or pole-like  Ansätze, fitted to results from multiple twisted momenta. These fits were then used to compute branching ratios in different regions of phase space for direct comparison with experimental data.
For \(\pi \to e \nu_e \gamma\) decays, the lattice predictions yield branching ratios that are larger than the PIBETA experimental measurements in certain regions of phase space, including the region O with the photon energy cut $E_\gamma> 10~\text{MeV}$~\cite{Frezzotti:2020bfa,Bychkov:2008ws}. 
For \(K \to e \nu_e \gamma\) decays, there is a discrepancy of up to $4\sigma$ between the measurements of the branching ratios obtained by the KLOE and E36 experiments with cuts on the electron momentum and photon energy of \(p_e > 200~\text{MeV}\) and \(E_\gamma > 10~\text{MeV}\) respectively~\cite{KLOE:2009urs,J-PARCE36:2022wfk}. The lattice results show better agreement with the E36 data~\cite{DiPalma:2025iud}.
For \(K \to \mu \nu_\mu \gamma\), the lattice predictions deviate from the ISTRA and OKA results at large photon energies, and from the E787 results at large muon–photon angles~\cite{DiPalma:2025iud,ISTRA:2010smy,OKA:2019gav,E787:2000ehx}.
Given these tensions, additional independent lattice QCD calculations are needed to cross-check these findings.

An alternative strategy is to work directly in the infinite volume, and then correct for the exponentially suppressed finite-volume effects introduced by the lattice calculation.
For instance, in lattice calculations of \(D_s \to \ell \nu \gamma^{(*)}\), the authors of 
Refs.~\cite{Giusti:2023pot,Giusti:2025ibe} propose the ``3d method'' to extrapolate time integrals to infinity using an exponential ansatz, 
and adopt the infinite-volume limit to access a wider range of photon momenta.
The infinite-volume reconstruction (IVR) method~\cite{Feng:2018qpx} was applied to the study of \(K \to \ell \nu \ell' \ell'\) decays in Ref.~\cite{Tuo:2021ewr}.  
This method reconstructs infinite-volume hadronic matrix elements from lattice data using ground-state dominance, thereby correcting for both temporal truncation and finite-volume effects.
In Ref.~\cite{Christ:2023lcc}  
the IVR framework was further developed for the complete computation of the radiative corrections to leptonic decays at $O(\alpha)$, incorporating both virtual-photon loops and real-photon emission within a unified formalism. 
Using this method, infrared divergences are subtracted directly in the weight functions, and the power-law finite-volume effects from the photon propagator are reduced to exponentially suppressed ones.

In this work, we initiate the application of the IVR method to the radiative decay \(P \to \ell \nu_\ell \gamma\) (\(P = \pi, K\)) which, in addition to the phenomenological significance of the results themselves, also provides a first step towards a complete computation of radiative corrections to leptonic decays at $O(\alpha)$ using the IVR method. Using two domain-wall fermion ensembles generated by the RBC and UKQCD collaborations at the physical pion mass, we determine the momentum dependence of the form factors and compute branching ratios with precision comparable to the Rome–Southampton results~\cite{Desiderio:2020oej,Frezzotti:2020bfa,DiPalma:2025iud}. Despite differences in lattice actions and computational methods, our results for the form factors are statistically consistent with those obtained by the Rome–Southampton collaboration (see Fig.~\ref{fig:fAVpi} and Fig.~\ref{fig:fAVK} in Sec.~\ref{sec4}). 

For decays involving a final-state electron, \(P \to e\nu_{e}\gamma\), we found that the $O(\alpha^2)$ collinear radiative corrections~\cite{Kuraev:2003gq} are significantly enhanced by two large logarithmic factors: (i) the collinear logarithm \(\ln(m_P^2/m_e^2)\), and (ii) the logarithm \(\ln(2E^*_{\gamma_2,\text{max}}/m_P)\) associated with the maximum energy (defined in the rest frame of $P$) of additional inner-bremsstrahlung photons emitted from the final-state electron, $E^*_{\gamma_2,\text{max}}$. The first term yields an enhancement of $\mathcal{O}(10)$ for both \(P = \pi\) and \(P = K\). The second logarithm becomes large in certain phase-space regions, or when a small experimental cut is imposed on the energy of the bremsstrahlung photons. Due to these two enhancements, the $O(\alpha^2)$ collinear radiative corrections can reach $\mathcal{O}(10\%)$ in the phase-space regions used by experiments, and are therefore non-negligible at the current level of experimental precision.

Our lattice results before applying such radiative corrections are consistent with those of the Rome–Southampton collaboration in Refs.~\cite{Frezzotti:2020bfa,DiPalma:2025iud}. 
Including these effects resolves the discrepancy between the lattice results and the PIBETA measurements for \(\pi \to e\nu_{e}\gamma\) (see Fig.~\ref{fig:Rpi_compare} in Sec.~\ref{sec4}). In the $K \to e\nu_e\gamma$ channel, the KLOE and E36 experiments differ in their treatment of additional inner-bremsstrahlung photons, leading to distinct radiative corrections that may, at least partially, explain the observed $4\sigma$ discrepancy between their results. Radiative corrections of order $O(10\%)$ are therefore essential for a meaningful comparison. Our lattice predictions, including collinear radiative corrections, agree well with the KLOE results (assuming that the energy cut on the second photon is independent of the angle of emission) and show a $1.7\sigma$ tension with E36 results (see Fig.~\ref{fig:RK_compare} in Sec.~\ref{sec4}). 
For \(K \to \mu\nu_{\mu}\gamma\), where collinear radiation from the final-state muon is negligible, our lattice results in the phase-space regions of E787, ISTRA, and OKA experiments are consistent with lattice calculation of the Rome–Southampton collaboration in Ref.~\cite{DiPalma:2025iud} (see Fig.~\ref{fig:cont_lim_Kmunug_E787} and Fig.~\ref{fig:cont_lim_Kmunug} in Sec.~\ref{sec4}). We confirm the previously observed tension between lattice predictions and the ISTRA and OKA measurements in the region of large photon energies, as well as the deviation from the E787 results at large angles between the muon and the photon.

Our numerical results are presented in detail in Sec.~\ref{sec4}. Here, we highlight results for the branching ratio of the decay $K \to e\nu_e\gamma$, evaluated with the kinematic cuts $E_\gamma > 10\,\mathrm{MeV}$ and $p_e > 200\,\mathrm{MeV}$. Our lattice prediction, inclusive with respect to additional inner-bremsstrahlung photons, is
\begin{equation}\label{eq:resultK}
  \frac{1}{B(K\to\mu\bar{\nu}_\mu(\gamma))}~B(K\to e\bar{\nu}_e\gamma(\gamma))\big|_{E_\gamma>10\,\mathrm{MeV},\,p_e>200\,\mathrm{MeV}} = 16.9\,(1.3)\times 10^{-6},
\end{equation}
which shows a $1.7\sigma$ tension with the E36 measurement of $19.8(1.1)\times 10^{-6}$~\cite{J-PARCE36:2022wfk}.
The KLOE experiment, by contrast, selects events with exactly one detected photon, imposing a laboratory-frame energy threshold $E^{\text{lab}}_{\gamma_2} < 20$~MeV~\cite{KLOE:2009urs} on the second photon. Under the simplified assumption of an angle-independent energy cut in the laboratory frame (the momentum of the kaon in the lab frame is chosen to be $\vec{p}^\text{~lab} = 100~\mathrm{MeV}$), our corresponding lattice result is
\begin{equation}\label{eq:resultK2}
  \frac{1}{B(K\to\mu\bar{\nu}_\mu(\gamma))}~B(K\to e\bar{\nu}_e\gamma(\gamma))\big|_{E_\gamma>10\,\mathrm{MeV},\,p_e>200\,\mathrm{MeV},\,E_{\gamma_2}^{\text{lab}} < 20\,\text{MeV}} = 16.1\,(1.3)\times 10^{-6},
\end{equation}
which is consistent with the KLOE measurement $14.83(67)\times 10^{-6}$ within $1\sigma$~\cite{KLOE:2009urs}. We also note that at the 2025 International Conference on Kaon Physics the NA62 experiment presented the preliminary result of $(15.9\pm0.2)\times10^{-6}$ for this quantity. However, the radiative corrections, including their effect
on the selection efficiency, are still to be fully evaluated~\cite{NA62Kaon2025}.
We stress again the importance of the $O(\alpha^2)$ collinear radiative corrections. Neglecting these corrections would shift the results in Eqs.~\eqref{eq:resultK} and~\eqref{eq:resultK2} to $18.6\,(1.4)\times 10^{-6}$, in agreement with the lattice prediction of Ref.~\cite{DiPalma:2025iud} which did not include these corrections.

This paper is organized as follows. In Sec.~\ref{sec2} we start by reviewing the calculation of branching ratios of meson radiative leptonic decays in Minkowski space. In Sec.~\ref{sec:RC}, we emphasize the necessity of including $O(\alpha^2)$ radiative corrections in electron channels. 
The application of the IVR method to computations of decay rates and form factors is explained in detail in Sec.~\ref{sec3}.
In Sec.~\ref{sec4}, we present our numerical results and compare them with lattice calculations of the Rome-Southampton collaboration and with the relevant experimental measurements.
We present our conclusions and prospects for future work in Sec.~\ref{sec5}. There are four appendices. 
In Appendix~\ref{sec:phasespace} we explain the cuts on the lepton and photon energies and momenta introduced in the experimental measurements. The formulae relevant for including collinear radiative corrections are summarized in Appendix~\ref{Appendix:RC}. Appendix~\ref{Appendix:scalar} contains the derivation of the hadronic functions using the scalar-function method which is an important element of our procedure (see Sec.~\ref{sec3}).   
Details of the finite-volume corrections used in our calculation are given in Appendix~\ref{Appendix:FV}.

\section{The Decay Amplitudes and Differential Branching Ratios\label{sec2}}
In this section, we present the decay amplitudes and differential branching ratios 
for the radiative process 
\[
P^+(p)\to \ell^+(p_\ell)\,\nu_\ell(p_{\nu_\ell})\,\gamma(k),
\]
where $P$ denotes a pion or a kaon, and $\ell$ an electron or a muon. 
The discussion in this section is formulated in Minkowski space.
The extraction of the same physical quantities from Euclidean correlators, 
as carried out in the lattice calculations, will be discussed in Sec.~\ref{sec3}. Since the following discussion applies to both pion and kaon decays, $P=\pi$ and $P=K$ respectively, for notational simplicity we omit the explicit label $P$ on the physical quantities.

\subsection{The Decay Amplitude}
\begin{figure} 
	\centering
	\includegraphics[width=0.65\textwidth]{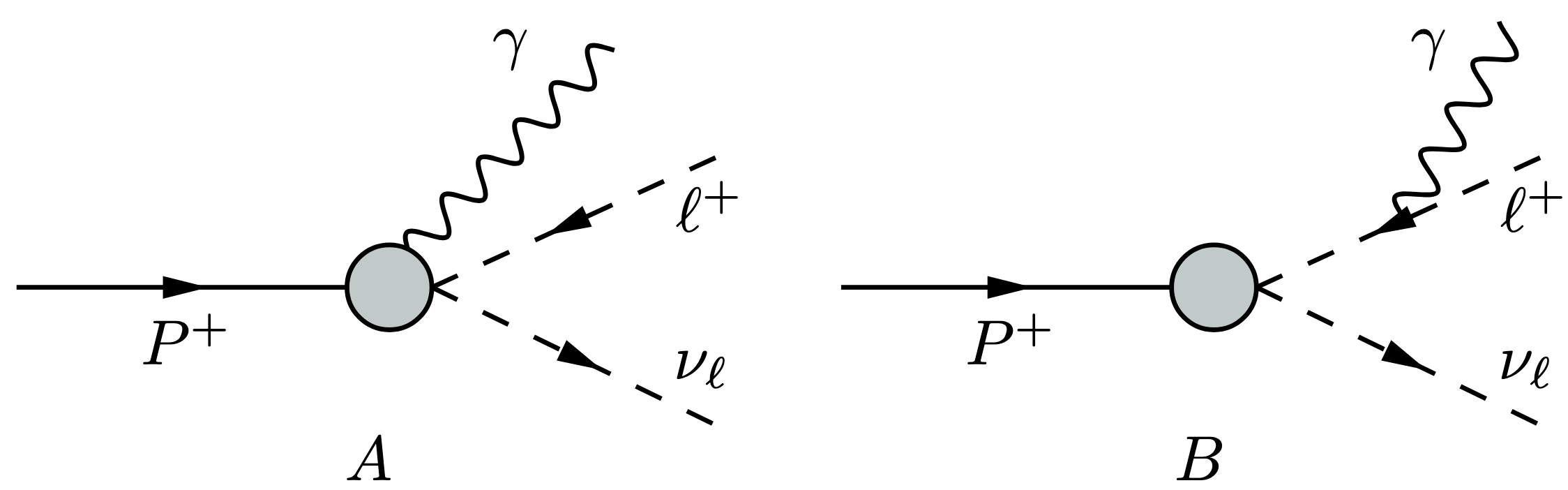}
	\caption{For the radiative decay $P^+\to \ell^+\nu_\ell\gamma$, the photon is emitted either from the initial-state meson (Diagram A) or from the final-state lepton (Diagram B). The diagrams correspond to the two terms in $\mathcal{M}^\mu$ in Eq.~\eqref{amp}.
		\label{fig:diagrams}
	}
\end{figure}

As shown in Fig.~\ref{fig:diagrams}, the decay amplitude consists of contributions from both initial-state meson radiation (diagram A) and final-state lepton radiation (diagram B). The total decay amplitude is
\begin{equation}\label{amp}
    \begin{aligned}
	   i\mathcal{M}[P\to \ell\nu_\ell\gamma]&=-\frac{G_F e V_\text{CKM}}{\sqrt{2}}\epsilon_\mu(k,\lambda)\mathcal{M}^\mu(k,p_\ell,p_{\nu_\ell}),\\
		\mathcal{M}^\mu(k,p_\ell,p_{\nu_\ell})&=f_P L^\mu(k,p_\ell,p_{\nu_\ell})-H^{\mu\nu}_M(k,p)\,l_\nu(p_\ell,p_{\nu_\ell}).
	\end{aligned}
\end{equation}
Here, $\epsilon_\mu(k,\lambda)$ denotes the photon's polarization vector, where $\lambda$ specifies its polarization state. The constant $G_F$ represents the Fermi coupling constant of the weak interaction, and $e$ is the electric charge. $V_\text{CKM}$ refers to the corresponding Cabibbo–Kobayashi–Maskawa (CKM) matrix element, which is $V_{ud}^*$ for the pion and $V_{us}^*$ for the kaon. The two terms in $\mathcal{M}^\mu(k,p_\ell,p_{\nu_\ell})$ represent the contributions from diagram B and diagram A in Fig.~\ref{fig:diagrams}, respectively. 

The contribution from diagram B is proportional to the meson decay constant $f_P$. Its leptonic factor is defined by
\begin{equation}\label{Lpart}
	\begin{aligned}
		L^\mu(k,p_\ell,p_{\nu_\ell}) &= l^\mu(p_\ell,p_{\nu_\ell}) + L^{\prime\mu}(k,p_\ell,p_{\nu_\ell}),\\[1mm]
		l^\mu(p_\ell,p_{\nu_\ell}) &= \bar{u}(p_{\nu_\ell})\gamma^\mu(1-\gamma_5)v(p_\ell),\\[1mm]
		L^{\prime\mu}(k,p_\ell,p_{\nu_\ell}) &= m_\ell\,\bar{u}(p_{\nu_\ell})(1+\gamma_5)\frac{2p_\ell^\mu+\slashed{k}\gamma^\mu}{m_\ell^2-(k+p_\ell)^2} v(p_\ell).
	\end{aligned}
\end{equation}

To describe the contribution from Diagram A, we define the hadronic matrix element in Minkowski space and in the rest frame of $P$ by
\begin{equation}
    H_M^{\mu \nu}(k, p)=\int d^4 x\, e^{i k\cdot x}\langle 0|T\{J_{\text{em},M}^\mu(x) J_{W,M}^\nu(0)\}| P(p)\rangle,
\end{equation}
where $k=(E,\vecp{k})$ and $p=(m_P,\vecp{0})$ denote the four-momenta of the photon and the initial-state meson respectively and $m_P$ is the the mass of $P$. The electromagnetic current in Minkowski space is given by
$J_{\text{em},M}^\mu=\frac 23 \bar{u}\gamma^\mu u-\frac 13 \bar{d}\gamma^\mu d-\frac 13 \bar{s}\gamma^\mu s$. 
For the pion, the weak current in Minkowski space is $J_{W,M}^\nu=\bar{d}\gamma^\nu(1-\gamma_5)u$, while for kaon it is $J_{W,M}^\nu=\bar{s}\gamma^\nu(1-\gamma_5)u$. 

The Ward identity $k_{\mu}H_M^{\mu\nu}(k,p)=f_P \,p^\nu$ implies that the hadronic matrix element $H_M^{\mu \nu}(k, p)$ can be expressed in terms of four structure-dependent form factors and the decay constant as follows~\cite{Bijnens:1992en}
\begin{equation}\label{HM}
	\begin{aligned}
		H_M^{\mu \nu}(k, p)= & \frac{R_1}{m_P}\left[k^2 g^{\mu \nu}-k^\mu k^\nu\right]+\frac{R_2}{m_P^3}\left[\left(k \cdot p-k^2\right) k^\mu-k^2(p-k)^\mu\right](p-k)^\nu \\
		& -i \frac{F_V}{m_P} \varepsilon^{\mu \nu \alpha \beta} k_\alpha p_\beta+\frac{F_A}{m_P}\left[\left(p \cdot k-k^2\right) g^{\mu \nu}-(p-k)^\mu k^\nu\right] \\
		& +f_P\left[g^{\mu \nu}+\frac{(2 p-k)^\mu(p-k)^\nu}{2 p \cdot k-k^2}\right],
	\end{aligned}
\end{equation}
where the convention of the Levi-Civita symbol in Minkowski space is chosen to be $\varepsilon^{0123}=1$ and $\varepsilon_{0123}=-1$; note that different conventions of the Levi-Civita symbol can lead to the opposite sign in front of $F_V$ in the literature. $R_1, R_2, F_V, F_A$ denote four dimensionless form factors that depends on $k^2$ and $p\cdot k$.
For the decay $P \to \ell\nu_\ell\gamma$, the contributions from $R_1$ and $R_2$ vanish because the final-state photon is on-shell and only the form factors $F_V$ and $F_A$, together with the decay constant $f_P$, appear in the $P \to \ell\nu_\ell\gamma$ decay amplitude. 
The amplitude \(\mathcal{M}^\mu(k,p_\ell,p_{\nu_\ell})\) satisfies the Ward identity $k_{\mu}\mathcal{M}^\mu(k,p_\ell,p_{\nu_\ell})=0$.

\subsection{Differential Branching Ratios}
The three-body phase space can be written as~\cite{Bijnens:1992en}
\begin{equation}
\begin{aligned}
	 d\Phi_3=\frac{m_P^2}{128 \pi^3} \,d x_\gamma \,d y_\ell,
\end{aligned}
\end{equation}
with the kinematic variables $(x_\gamma,y_\ell)$ defined as
\begin{equation}\label{xy_def}
	x_\gamma=\frac{2p\cdot k}{m_P^2},\quad y_\ell=\frac{2p\cdot p_\ell}{m_P^2}.
\end{equation}
They satisfy
\begin{equation}\label{phasespace}
		0\leq x_\gamma\leq 1-r_\ell,\quad
		1-x_\gamma+\frac{r_\ell}{1-x_\gamma}\leq y_\ell\leq 1+r_\ell,
\end{equation}
where $r_\ell=m_\ell^2/m_P^2$. In the lattice calculation we evaluate the amplitude in the same regions of phase space as those used in the experiments~\cite{Bychkov:2008ws,KLOE:2009urs,J-PARCE36:2022wfk,ISTRA:2010smy,OKA:2019gav,E787:2000ehx}; the details are provided in Appendix~\ref{sec:phasespace}. 

The differential branching ratio for the decay $P\to \ell\nu_\ell\gamma$ at $O(\alpha)$ is given by
\begin{equation}\label{B}
\begin{aligned}
	\frac{d^2B[P\to \ell\nu_\ell\gamma]}{dx_\gamma\,dy_\ell} &= \frac{\alpha}{2\pi(1-r_\ell)^2}\,B^{(0)}[P\to \ell\nu_\ell]\, A(x_\gamma,y_\ell),\\[1mm]
	A(x_\gamma,y_\ell) &= \frac{1}{4m_\ell^2 f_P^2}\sum_{\lambda,\text{spin}}\left(\epsilon_\mu(k,\lambda)\mathcal{M}^\mu\right)\left(\epsilon_{\rho}(k,\lambda)\mathcal{M}^{\rho}\right)^*.
\end{aligned}
\end{equation}
Here, $B^{(0)}[P\to \ell\nu_\ell] = G_F^2 |V_\text{CKM}|^2 f_P^2 m_P^3 r_\ell (1-r_\ell)^2 / (8\pi\Gamma_P)$ denotes the branching ratio of the leptonic decay $P\to \ell\nu_\ell$ in the absence of electromagnetic corrections. $\Gamma_P$ is the total decay width of the meson $P$. The reduced squared amplitude $A(x_\gamma,y_\ell)$ is a dimensionless quantity that can be expressed in terms of the form factors $(F_V, F_A)$ in the commonly used form~\cite{Bijnens:1992en,ParticleDataGroup:2024cfk}:
\begin{equation}
	\begin{aligned}
	A(x_\gamma,y_\ell)=&\,f_{\mathrm{IB}}(x_\gamma,y_\ell)+\frac{1}{r_\ell}\left(\frac{m_P}{2f_P}\right)^2\left[\left(F_V+F_A\right)^2 f_{\mathrm{SD}^{+}}(x_\gamma,y_\ell)+\left(F_V-F_A\right)^2 f_{\mathrm{SD}^{-}}(x_\gamma,y_\ell)\right]\\
	-&\left(\frac{m_P}{f_P}\right)\left[\left(F_V+F_A\right) f_{\mathrm{INT}^{+}}(x_\gamma,y_\ell)+\left(F_V-F_A\right) f_{\mathrm{INT}^{-}}(x_\gamma,y_\ell)\right].\\
	\end{aligned}
\end{equation}
The functions appearing in the above expression are 
\begin{equation}\label{f0}
	\begin{aligned}
		 f_{\mathrm{IB}}(x_\gamma,y_\ell)&=\left[\frac{1-y_\ell+r_\ell}{x_\gamma^2\left(x_\gamma+y_\ell-1-r_\ell\right)}\right]
		\biggl[x_\gamma^2+2(1-x_\gamma)\left(1-r_\ell\right)
		-\frac{2 x_\gamma r_\ell\left(1-r_\ell\right)}{x_\gamma+y_\ell-1-r_\ell}\biggr], \\[6pt]
		 f_{\mathrm{SD}^{+}}(x_\gamma,y_\ell)&=\bigl[x_\gamma+y_\ell-1-r_\ell\bigr]
		\bigl[(x_\gamma+y_\ell-1)(1-x_\gamma)-r_\ell\bigr], \\[6pt]
		 f_{\mathrm{SD}^{-}}(x_\gamma,y_\ell)&=\bigl[1-y_\ell+r_\ell\bigr]
		\bigl[(1-x_\gamma)(1-y_\ell)+r_\ell\bigr], \\[6pt]
		 f_{\mathrm{INT}^{+}}(x_\gamma,y_\ell)&=\left[\frac{1-y_\ell+r_\ell}{x_\gamma\left(x_\gamma+y_\ell-1-r_\ell\right)}\right]
		\bigl[(1-x_\gamma)(1-x_\gamma-y_\ell)+r_\ell\bigr], \\[6pt]
		 f_{\mathrm{INT}^{-}}(x_\gamma,y_\ell)&=\left[\frac{1-y_\ell+r_\ell}{x_\gamma\left(x_\gamma+y_\ell-1-r_\ell\right)}\right]
		\bigl[x_\gamma^2-(1-x_\gamma)(1-x_\gamma-y_\ell)-r_\ell\bigr].
	\end{aligned}
\end{equation}

In experimental measurements, the decay rates are usually normalized using the leptonic decay branching ratio \(B[P\to \mu\nu_\mu(\gamma)]\).
For ease of comparison, following Refs.~\cite{Frezzotti:2020bfa,DiPalma:2025iud}, we define the normalized differential branching ratio at \(O(\alpha)\) (denoted as “wo. RC” to distinguish it from the result including radiative corrections in the next section) as
\begin{equation}\label{R0}
\begin{aligned}
    \frac{d^2R_\gamma(x_\gamma,y_\ell)}{dx_\gamma\,dy_\ell} &=\frac{1}{B^{(0)}[P\to\mu\nu_\mu]}\frac{d^2 B[P\to\ell\nu_\ell\gamma]}{dx_\gamma\,dy_\ell}\\&= \frac{\alpha}{2\pi(1-r_\ell)^2}\,R^{(0)}_{\ell/\mu}\, A(x_\gamma,y_\ell),
\end{aligned}
\end{equation}
where for $\ell=\mu$, $R^{(0)}_{\ell/\mu}=1$; for $\ell=e$,
\(R^{(0)}_{e/\mu}=\Gamma^{(0)}[P\to e \nu_e]/\Gamma^{(0)}[P\to \mu \nu_\mu]\) is the ratio of the $P_{e2}$ and $P_{\mu 2}$ decay widths in the absence of electromagnetic corrections.

\section{The Importance of Radiative Corrections\label{sec:RC}}
According to the Bloch-Nordsieck theorem~\cite{Bloch:1937pw}, the infrared-safe inclusive rate for $P\to \ell\nu_\ell\gamma(\gamma)$ receives radiative corrections at $O(\alpha^2)$ from two sources: (i) diagrams involving both a virtual-photon loop and the emission of a real photon and (ii) the emission of two real photons, $P\to \ell\nu_\ell\gamma\gamma$. The infrared divergences in these two contributions cancel in the sum. 
When the second photon, either the photon in the virtual-photon loop or the second real photon, is emitted collinearly from the charged lepton, both (i) and (ii) contributions develop collinear divergences in the $m_\ell \to 0$ limit. These singularities are regulated by the finite lepton mass, resulting in radiative corrections proportional to $\ln(m_P^2/m_\ell^2)$~\cite{Kuraev:2003gq}.
For electron final states, the logarithmic factors $\ln(m_\pi^2/m_e^2) \approx 11$ and $\ln(m_K^2/m_e^2) \approx 14$ lead to a significant enhancement of the collinear radiative corrections.
In contrast, for muon final states, such collinear radiative corrections can safely be neglected.

Another source of enhancement arises from the large logarithm \(\ln(2E^*_{\gamma_2,\text{max}}/m_P)\), where \(E^*_{\gamma_2,\text{max}}\) denotes the maximum energy of inner-bremsstrahlung photons emitted from the final-state electron, defined in the rest frame of the meson \(P\). Such logarithmic terms are typical in radiative corrections. This logarithm becomes large when the electron energy approaches its kinematic endpoint, reducing the available phase space for bremsstrahlung emission (\(E^*_{\gamma_2,\text{max}} \to 0\)), or when a tight experimental cut is imposed on the bremsstrahlung photon energy.

Following Ref.~\cite{Kuraev:2003gq}, we provide quantitative estimates of the radiative corrections in $\pi\to e\nu_e\gamma$ and $K\to e\nu_e\gamma$, showing that these effects are as large as $O(10\%)$ and are therefore non-negligible at the current level of experimental precision. In particular, for $K\to e\nu_e\gamma$, the treatment of the additional inner-bremsstrahlung photons differs between the KLOE~\cite{KLOE:2009urs} and E36~\cite{J-PARCE36:2022wfk} analyses; consequently, the associated radiative corrections also differ, which may partially account for the discrepancies between their measurements.

\subsection{Collinear Radiative Corrections}
The $O(\alpha^2)$ radiative corrections to the decay $P\to e\nu_e\gamma(\gamma)$ which is inclusive with respect to the second photon, have been studied in detail in Ref.~\cite{Kuraev:2003gq}, and we review them in Appendix~\ref{Appendix:RC}. For this inclusive decay, the normalized differential branching ratio that includes the $O(\alpha^2)$ collinear radiative corrections is
\begin{equation}\label{R_RC}
	\frac{d^2R_\gamma^{\mathrm{RC}}(x_\gamma,y_e)}{dx_\gamma\,dy_e} = \frac{\alpha}{2\pi(1-r_e)^2}\,R_{e/\mu} \left(A(x_\gamma,y_e)+\frac{\alpha}{2\pi}(L_e-1)A^{\mathrm{RC}}(x_\gamma,y_e)\right)+O(\alpha^2 L_e^0).
\end{equation}
Here, $R_{e/\mu} = \Gamma[P \to e\nu_e(\gamma)]/\Gamma[P \to \mu\nu_\mu(\gamma)]$ is defined as the ratio of the leptonic decay widths $P\to e\nu_e(\gamma)$ and $P\to \mu\nu_\mu(\gamma)$. The denominator $\Gamma[P \to \mu\nu_\mu(\gamma)]$ arises from the normalization used in $R_\gamma$.  The quantity $L_e=\ln (y_e^2 m_P^2/m_e^2)$ is the large logarithm related to collinear radiative corrections. The term $O(\alpha^2 L_e^0)$ denotes subleading $O(\alpha^2)$ contributions that are not enhanced by $L_e$; these effects are at the sub-percent level and are not the focus here.

The function \(A^{\mathrm{RC}}(x_\gamma, y_e)\) was derived in Ref.~\cite{Kuraev:2003gq} by integrating over the allowed kinematics of the second photon. Its explicit form is reviewed in Appendix~\ref{Appendix:RC}. As is typical for radiative corrections, $A^{\mathrm{RC}}(x_\gamma,y_e)$ contains terms proportional to $\ln(2E_{\gamma_2,\text{max}}^*/m_P)$, where $E_{\gamma_2,\text{max}}^*$ is the maximal energy of the second emitted photon in the rest frame of the meson $P$. We explain the origin of this logarithm in Appendix~\ref{Appendix:RC}. 
In the function \(A^{\mathrm{RC}}(x_\gamma, y_e)\), we focus explicitly on the collinear radiative correction, where the photon is emitted from the final-state electron in a direction nearly parallel to its momentum, leading to the kinematic constraint $E_{\gamma_2,\text{max}}^* = m_P(1 - y_e)/2$ in the $m_e\to0$ limit~\footnote{Considering the two-step process $P \to e(t)\nu\gamma(x_\gamma)$ followed by the emission of a collinear photon $e(t)\to e(y_e)\gamma(x_{\gamma_2})$ with $x_{\gamma_2}=2E_{\gamma_2}^*/m_P$, the intermediate electron energy fraction satisfies $y_e+x_{\gamma_2}<t<1$ in the $m_e\to0$ limit, which gives $x_{\gamma_2,\max}=1-y_e$.}.
Therefore, compared to the $O(\alpha)$ decay rate proportional to $A(x_\gamma,y_e)$, if $y_e$ is close to 1, the collinear radiative correction is strongly enhanced by
\begin{equation}
\begin{aligned}
    &\left(\frac{\alpha}{2\pi}\right)(L_e-1)A^{RC}(x_\gamma,y_e)/A(x_\gamma,y_e)\\\sim&\left(\frac{\alpha}{2\pi}\right) L_e \ln\frac{2E_{\gamma_2,\text{max}}^*}{m_P}\\
    = &\left(\frac{\alpha}{2\pi}\right) \ln \frac{y_e^2 m_P^2}{m_e^2} \ln(1-y_e).
\end{aligned}
\end{equation}
Although the collinear radiative corrections are suppressed by a small prefactor of $\alpha/(2\pi) \sim 0.12\%$ relative to the $O(\alpha)$ result, they are significantly enhanced for several different reasons:
\begin{itemize}
    \item The difference between $R^{(0)}_{e/\mu}$ used in the formula without radiative corrections (Eq.~(\ref{R0})) and $R_{e/\mu}$ used in the formula with $O(\alpha^2)$ radiative corrections (Eq.~(\ref{R_RC}));
    \item The collinear logarithm \(L_e=\ln (y_e^2 m_P^2/m_e^2) \sim \mathcal{O}(10)\);
    \item The logarithm \(\ln(1 - y_e)\), which becomes large as \(y_e \to 1\). Physically, this enhancement arises because when the electron energy approaches its kinematic endpoint ($y_e\to 1$), the available phase space for the second emitted photon shrinks (\(E_{\gamma_2,\text{max}}^* \to 0\)), leading to an enhancement due to the logarithm \(\ln (2E_{\gamma_2,\text{max}}^*/m_P)\).
\end{itemize}
As shown numerically in Sec.~\ref{sec:RC,pi} and Sec.~\ref{sec:RC,K}, such collinear radiative corrections can reach the level of \(\mathcal{O}(10\%)\) in the phase-space regions relevant to experiments.

In experimental measurements, aside from the fully inclusive radiative process $P \to e \nu_e \gamma(\gamma)$, it is also common to impose an energy cut on the second emitted photon, thereby vetoing events in which this photon is sufficiently energetic. For example, in the KLOE measurement of $K \to e \nu \gamma$, events are selected by requiring that exactly one photon is detected, with a detection condition of $E_\gamma^{\text{lab}} > 20~\text{MeV}$ in the laboratory frame~\cite{KLOE:2009urs}. In other words, events containing a second photon with laboratory-frame energy above $20~\text{MeV}$ are excluded.
Consequently, KLOE effectively removes part of the collinear radiative corrections arising from hard bremsstrahlung photon emission. 

In realistic measurements, the laboratory-frame energy cut for the second photon can have nontrivial angular dependence due to detector geometry and response. For theoretical estimates of the cut’s impact, we neglect these detector-specific complexities. As a simplifying approximation, we impose a angle-independent, laboratory-frame energy cutoff on the second emitted photon and use it to estimate the collinear radiative corrections.

We define $\vec{p}^{~\text{lab}}$ as the momentum of the meson $P$ in the laboratory frame, and impose the laboratory-frame energy restriction on the second photon $E_{\gamma_2}^{\text{lab}}<E^{\text{lab}}_{\gamma_2,\text{cut}}$. Then the normalized differential branching ratio including the $O(\alpha^2)$ collinear radiative corrections with this cut, is
\begin{equation}\label{R_RC_cut}
\begin{aligned}
	&\frac{d^2 R_\gamma^{\mathrm{RC,cut}}\left(x_\gamma, y_e;\vec{p}^{~\text{lab}},E^{\text{lab}}_{\gamma_2,\text{cut}}\right)}{d x_\gamma d y_e} \\=& \frac{\alpha}{2\pi(1-r_e)^2}\,R_{e/\mu}  \left(A(x_\gamma,y_e)+\frac{\alpha}{2\pi}(L_e-1)A^{\mathrm{RC}}_{\text{cut}}(x_\gamma,y_e;\vec{p}^{~\text{lab}},E^{\text{lab}}_{\gamma_2,\text{cut}})\right)+O(\alpha^2 L_e^0).
\end{aligned}
\end{equation}
Here, $A^{\mathrm{RC}}_{\text{cut}}(x_\gamma,y_e;\vec{p}^{~\text{lab}},E^{\text{lab}}_{\gamma_2,\text{cut}})$ denotes the collinear radiative correction with the imposed energy cutoff on the second photon; It depends on the meson momentum $\vec{p}^{~\text{lab}}$ and the laboratory-frame cutoff $E^{\text{lab}}_{\gamma_2,\text{cut}}$. The expression for $A^{\mathrm{RC}}_{\text{cut}}(x_\gamma,y_e;\vec{p}^{~\text{lab}},E^{\text{lab}}_{\gamma_2,\text{cut}})$ is derived in Appendix~\ref{Appendix:RC}. As before, the $O(\alpha^2 L_e^0)$ term denotes subleading radiative corrections that are not enhanced by the large logarithm $L_e$ and is neglected here. 

Similar to the case without an energy cut, the collinear radiative corrections are strongly enhanced by
\begin{equation}\label{LLcut}
\begin{aligned}
    &\left(\frac{\alpha}{2\pi}\right)(L_e-1)A_{\text{cut}}^{RC}(x_\gamma,y_e;\vec{p}^{~\text{lab}},E^{\text{lab}}_{\gamma_2,\text{cut}})/A(x_\gamma,y_e)\\\sim&\left(\frac{\alpha}{2\pi}\right) \ln\frac{y_e^2 m_P^2}{m_e^2} \ln\frac{2E_{\gamma_2,\text{max}}^*}{m_P},
\end{aligned}
\end{equation}
where \(E_{\gamma_2,\text{max}}^* = \min\{m_P(1 - y_e)/2,\, E_{\gamma_2,\text{cut}}^*\}\) is determined by both the kinematic constraint $m_P(1 - y_e)/2$, and the energy cutoff $E^*_{\gamma_2,\text{cut}}$ for the second photon, defined by Lorentz-boosting the laboratory-frame cutoff $E^{\text{lab}}_{\gamma_2,\text{cut}}$ to the rest frame of the meson. The explicit form of \(E_{\gamma_2,\text{cut}}^*\) is provided in Appendix~\ref{Appendix:RC}. This expression implies that imposing a more stringent energy cut (i.e., a smaller \(E_{\gamma_2,\text{cut}}^*\) or \(E_{\gamma_2,\text{cut}}^{\text{lab}}\)) leads to enhanced collinear radiative corrections, as numerically confirmed in Sec.~\ref{sec:RC,K}.

By comparing Eq.~\eqref{R_RC}, Eq.~\eqref{R_RC_cut}, and \eqref{R0}, we can obtain the $O(\alpha^2 L_e)$ radiative correction to the normalized differential branching ratio of $P\to e\nu_e \gamma(\gamma)$. For example, when the energy restriction is imposed on the second photon, the resulting radiative correction is
\begin{equation}\label{R_RC_diff}
	\frac{d^2 R_\gamma^{\mathrm{RC,cut}}\left(x_\gamma, y_e;\vec{p}^{~\text{lab}},E^{\text{lab}}_{\gamma_2,\text{cut}}\right)}{d x_\gamma d y_e}-\frac{d^2 R_\gamma\left(x_\gamma, y_e\right)}{d x_\gamma d y_e}.
\end{equation}
Here, $\frac{d^2 R_\gamma\left(x_\gamma, y_e\right)}{d x_\gamma d y_e}$ denotes the $O(\alpha)$ normalized differential branching ratio in the absence of radiative corrections, as defined in Eq.~\eqref{R0}. In the following subsections, we provide explicit numerical estimates of the collinear corrections for both $\pi\to e\nu_e\gamma$ and $K\to e\nu_e\gamma$ decays.

\subsection{Radiative Corrections in $\pi\to e\nu_e\gamma$\label{sec:RC,pi}}
We examine the collinear radiative corrections relevant to the PIBETA experiment’s determination of the branching ratio for $\pi\to e\nu_e\gamma$. In this experiment, candidate events require simultaneous detection of a neutral shower (photon cluster) and a positron ($e^+$) track. Events with multiple neutral showers are retained; the $(e,\gamma)$ pair with the smallest time difference is recorded~\cite{Bychkov:2008ws}. Consequently, the measurement is inclusive of events with a second real photon. Accordingly, the PIBETA collaboration adopted the inclusive correction formula for $\pi\to e\nu_e\gamma(\gamma)$ from Ref.~\cite{Kuraev:2003gq} (identical to Eq.~\eqref{R_RC} here) to compute theoretical predictions (see Ref.~\cite{Bychkov:2008ws}, Table III, $B_{\text{the}}$), which agree well with the experimental measurements. To obtain these theoretical predictions, the authors of Ref.~\cite{Bychkov:2008ws} first performed a fit of the $F_A$ value to the experimental data, using the fixed value and slope of $F_V$ taken from Refs.~\cite{ParticleDataGroup:2008zun,Mateu:2007tr}, and then used these inputs to compute the theoretical results.

In Table~\ref{table:piRC}, we reproduce the calculation of radiative corrections and report numerical values for the phase space regions used by the PIBETA experiment. Definitions of these regions are given in Table~\ref{table:PSpi} of Appendix\,\ref{sec:phasespace}. In the computation, we use the form factors $(F_V, F_A)$ determined from our lattice QCD calculation, with the method and results discussed in Sections~\ref{sec3} and~\ref{sec4}. 
Table~\ref{table:piRC} also lists the PIBETA measurements and the radiative corrections expressed as percentages of those measurements. The corrections shift the branching ratios by approximately 6\% to 10\%, which are non-negligible at the current level of experimental precision. In Sec.~\ref{sec4}, we further point out that including these corrections explains the discrepancy between lattice QCD predictions without radiative corrections and the PIBETA measurements.

\begin{table}
	\centering
	\begin{tabular}{cccc}
		\hline \hline  Regions& PIBETA & $O(\alpha^2 L_e)$ RC & RC/PIBETA\\
		\hline 
		A & $2.614(21)$&  $-0.17$& $-6.5\%$ \\
		B & $14.46(22)$ &  $-0.83$ & $-5.7\%$ \\
		C& $37.69(46)$ &  $-3.9$ & $-10.3\%$ \\
		O & $73.86(54)$ &  $-6.4$ & $-8.7\%$ \\
		\hline \hline
	\end{tabular}
	\caption{\label{table:piRC} $O(\alpha^2 L_e)$ radiative corrections in the $\pi\to e\nu_e\gamma$ process for phase-space regions (A, B, C, O) as defined in Table~\ref{table:PSpi}. The column ``$O(\alpha^2 L_e)$ RC'' shows the collinear radiative corrections. We also list the PIBETA measurements and the ratio RC/PIBETA (in percent), showing corrections at the $6$--$10\%$ level across the regions.}
\end{table}

\subsection{Radiative Corrections in $K\to e\nu_e\gamma$\label{sec:RC,K}}
We note that the KLOE~\cite{KLOE:2009urs} and E36~\cite{J-PARCE36:2022wfk} experiments use different treatments of a second emitted photon in their event selection. As noted above, KLOE requires one and only one detected photon, with the energy cut in laboratory frame $E_\gamma^{\text{lab}}>20~\text{MeV}$ as the detection condition~\cite{KLOE:2009urs}. In contrast, E36 reports a measurement that is inclusive of inner-bremsstrahlung photons~\cite{J-PARCE36:2022wfk}. Consequently, the two experiments inevitably incorporate different radiative corrections. Because these corrections can be as large as $O(10\%)$, accounting for the differing treatments of the second photon is essential for a meaningful comparison between the two experiments.

\begin{figure} 
	\centering
	\includegraphics[width=0.9\textwidth]{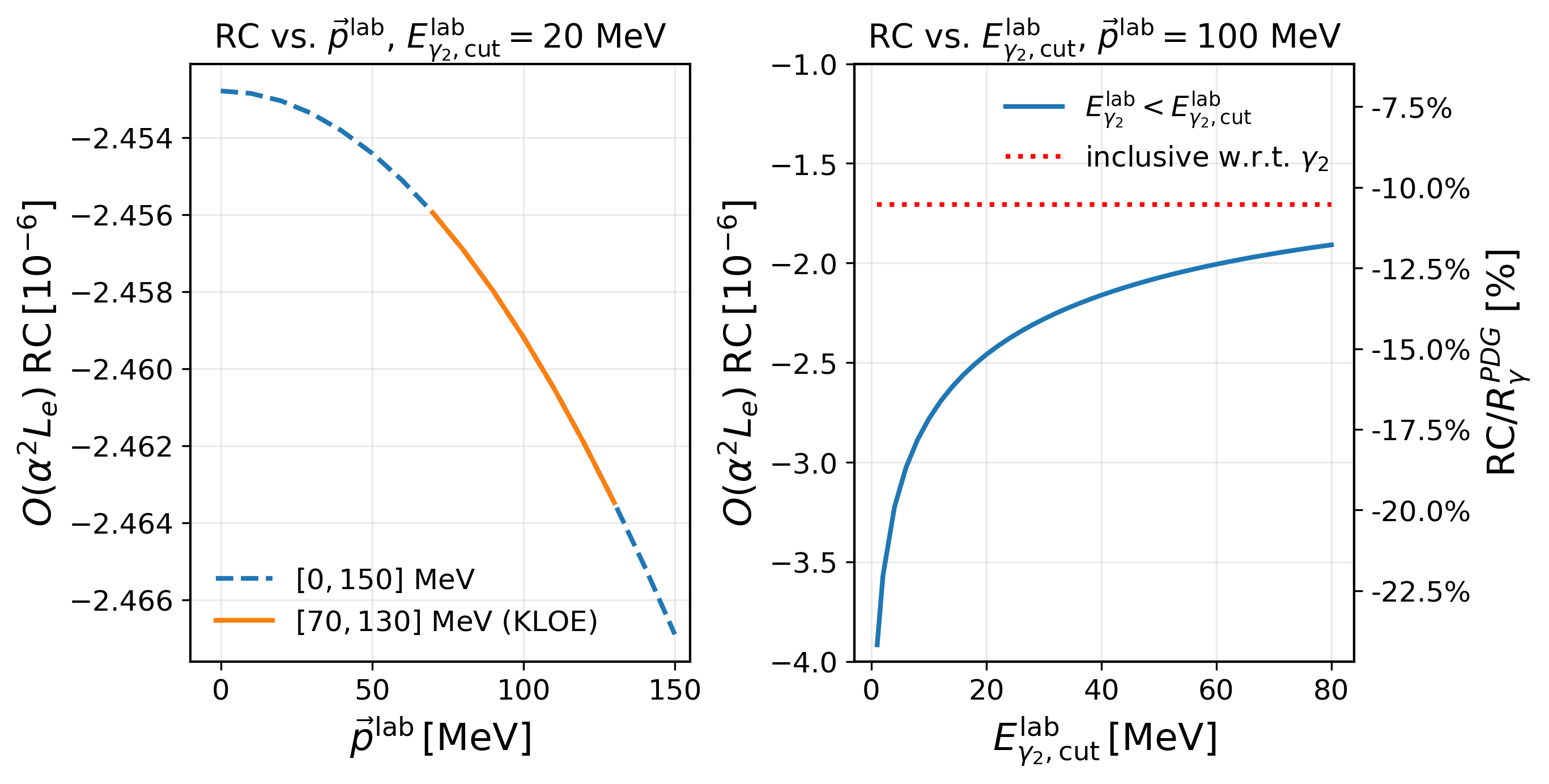}
	\caption{Radiative corrections evaluated for (i) an inclusive treatment of the second photon (denoted as ``inclusive w.r.t $\gamma_2$'') and (ii) an angle-independent, laboratory-frame photon-energy cutoff $E_{\gamma_2}^{\text{lab}} < E_{\gamma_2,\text{cut}}^{\text{lab}}$.  
    Left: Dependence of the correction for case (ii) on the kaon momentum $\vec{p}^{~\text{lab}}$ at fixed $E_{\gamma_2,\text{cut}}^{\text{lab}} = 20$~MeV; the KLOE momentum range is indicated in orange.  
    Right: The blue solid line shows the correction for case (ii) as a function of $E_{\gamma_2,\text{cut}}^{\text{lab}}$ at fixed ${p}^{\text{lab}} = 100$~MeV, while the red dotted line shows the correction for case (i). 
    The left and right vertical axes display, respectively, the value of the correction and its percentage relative to the PDG branching ratio $R_\gamma^{\text{PDG}} = 1.62(22)\times 10^{-6}$~\cite{ParticleDataGroup:2024cfk}.	\label{fig:RC_cut}
	}
\end{figure}

We provide theoretical estimates of the radiative corrections in two settings: (i) an inclusive treatment of the second photon and (ii) a selection with a laboratory-frame energy cut on the second photon. As noted above, we assume an angle-independent, laboratory-frame energy cutoff for the second photon, namely $E_{\gamma_2}^{\text{lab}} < E_{\gamma_2,\text{cut}}^{\text{lab}}$. In Fig.~\ref{fig:RC_cut}, we present the radiative corrections and illustrate their dependence on the kaon momentum $\vec{p}^{~\text{lab}}$ and the energy cutoff $E_{\gamma_2,\text{cut}}^{\text{lab}}$ in the laboratory frame. 
The phase space used here is the region 1–5 in Table~\ref{table:PSK}, which is defined by 
$20\,\mathrm{MeV}<E_\gamma<250\,\mathrm{MeV}$ and $|\vec{p}_e|>200\,\mathrm{MeV}$
in the rest frame of the kaon.

In the left panel of Fig.~\ref{fig:RC_cut}, we fix the cutoff to $E_{\gamma_2,\text{cut}}^{\text{lab}} = 20$~MeV (as in the KLOE experiment~\cite{KLOE:2009urs}) and plot the radiative correction given by Eq.~\eqref{R_RC_cut} as a function of the laboratory-frame kaon momentum $\vec{p}^{~\text{lab}}$. The orange solid line marks the KLOE momentum interval $\vec{p}^{~\text{lab}} \in [70,130]$~MeV~\cite{KLOE:2009urs}. The point at $\vec{p}^{~\text{lab}} = 0~$MeV corresponds to applying the energy cutoff in the kaon rest frame. For fixed $E_{\gamma_2,\text{cut}}^{\text{lab}}$, the collinear radiative correction depends only weakly on $\vec{p}^{~\text{lab}}$. Hence, it suffices to evaluate the correction at a representative momentum, e.g., $\vec{p}^{~\text{lab}}\sim 100$~MeV as in the KLOE experiment.

In the right panel of Fig.~\ref{fig:RC_cut}, the blue solid line shows the radiative correction from Eq.~\eqref{R_RC_cut} as a function of the photon-energy cutoff $E_{\gamma_2,\text{cut}}^{\text{lab}}$, evaluated at fixed kaon momentum $\vec{p}^{~\text{lab}} = 100$~MeV. The red dotted line shows the radiative correction inclusive of the second photon (denoted “inclusive w.r.t. $\gamma_2$”), as given by Eq.~\eqref{R_RC}. The left and right vertical axes display, respectively, the magnitude of the radiative correction and its value as a percentage of the PDG result $R_\gamma^{\text{PDG}} = 1.62(22)\times 10^{-6}$~\cite{ParticleDataGroup:2024cfk}. As $E_{\gamma_2,\text{cut}}^{\text{lab}} \to 0$~MeV, the correction develops an infrared divergence, as predicted by Eq.~\eqref{LLcut} by setting $E_{\gamma_2,\text{max}}^*=E_{\gamma_2,\text{cut}}^* \to 0$, with $E_{\gamma_2,\text{cut}}^*$ the corresponding energy cut in the rest frame (defined in Appendix~\ref{Appendix:RC}). From the figure, the $O(\alpha^2 L_e)$ radiative corrections exceed the 10\% level irrespective of whether a photon-energy cutoff is applied. Imposing a photon-energy cutoff increases the magnitude of the correction. This behavior is expected: the radiative correction from the sum of virtual-photon loop diagrams and the emission of a second real photon is negative; the latter contributes positively, so removing part of its contribution via an energy cutoff increases the magnitude of the net negative correction.

Comparing the inclusive correction (red dotted line in the right panel of Fig.~\ref{fig:RC_cut}) with the correction with a photon-energy cut (blue solid line in the same panel) at $E_{\gamma_2,\text{cut}}^{\text{lab}} = 20$~MeV, the difference can be as large as $4.6\%$ of the PDG value $R_\gamma^{\text{PDG}} = 1.62(22)\times 10^{-6}$. Our simplified estimate, assuming an angle-independent laboratory-frame cutoff, indicates that the different treatments of the second photon lead to different radiative corrections. This may play a non-negligible role in the observed discrepancy between KLOE and E36 measurements.

In this section we have emphasised the fact that the radiative corrections for decays with an electron in the final state are significant, typically of $O(10\%)$. In order to compare the theoretical predictions for the decay rates with the corresponding experimental results, a precise understanding of the experimental treatment of addition final-state photons is necessary. For the PIBETA and E36 experiments, in which no cuts are imposed on the additional photons, we were able to apply the leading radiative corrections to our lattice QCD results to derive the corresponding theoretical prediction. For experiments, such as NA62, in which vetoes of events with a second photon depend on selection efficiencies, it will be necessary for the experiments themselves to apply Monte Carlo simulations to derive physical observables that can be reliably calculated theoretically (e.g. to determine the form factors $F_V$ and $F_A$, which can then be compared directly with our results in Section~\ref{sec4})\,\footnote{We thank E.Goudzovski, T.Husak and A.Romano from the NA62 collaboration for discussions on this point.}.

\section{Computing radiative decay rates using infinite-volume reconstruction\label{sec3}}
Our goal in this this paper is to determine the form factors $(F_V,F_A)$ and the reduced squared amplitude $A(x_\gamma,y_\ell)$ from lattice QCD computations and subsequently to integrate Eqs.~(\ref{R0}), (\ref{R_RC}) or (\ref{R_RC_cut}) over the phase space to obtain the branching ratio for the decay $P\to \ell\nu_\ell\gamma$. Our method is based on the infinite-volume reconstruction (IVR) technique~\cite{Feng:2018qpx} and is described in detail in Ref.~\cite{Tuo:2021ewr} where it is also applied to the emission of a virtual photon, i.e. to the processes $K\to\ell\bar\nu_\ell\,(\ell^{\prime+}\ell^{\prime-})$, where $\ell$ and $\ell^{\prime\pm}$ are charged leptons. We now briefly summarize the method, focusing on the case of real photon emission.

\subsection{The Hadronic Matrix Element\label{sec:H}}
We present our lattice method for calculating the hadronic matrix element \(H_M^{\mu\nu}(k,p)\). We begin by formulating the method in the infinite-volume and continuum limit, where the Euclidean momentum-space hadronic function is defined as
\begin{equation}\label{Direct0}
\begin{aligned}
    H_E^{\mu \nu}(k_E, p_E)
    &=-i\int_{-\infty}^{\infty} dt \int d^3 \vec{x}\,
    e^{k^0 t-i \vec{k} \cdot \vec{x}}\, H_E^{\mu \nu}(x)\,,\\
    H_E^{\mu \nu}(x)
    &=\langle 0 \vert T\{J_{\text{em},E}^\mu(\vec{x},t)\, J_{W,E}^\nu(0)\}\vert P(p_E)\rangle\,.
\end{aligned}
\end{equation}
Here \(k_E=(ik^0,\vec{k})\) and \(p_E=(im_P,\vec{0})\) denote the Euclidean momenta of the photon and the initial-state meson, respectively, with \(k^0=|\vec{k}|=m_P x_\gamma/2\). The coordinate \(x=(\vec{x},t)\) is defined in Euclidean space. The Euclidean electromagnetic and weak currents, \(J_{\text{em},E}^\mu\) and \(J_{W,E}^\nu\), are defined by replacing the Minkowski gamma matrices with their Euclidean counterparts, \(\gamma_{E,0}=\gamma_{M,0}\) and \(\gamma_{E,i}=-i\gamma_{M,i}\) for \(i=1,2,3\).

The above definition is valid only if there are no intermediate hadronic states lighter than the initial-state meson. This condition is satisfied for \(P\to \ell\nu_\ell\gamma\). Otherwise, the temporal integral contains exponentially growing contributions, which must be subtracted~\cite{Tuo:2024bhm}. Under this condition, the Minkowski hadronic function is related to its Euclidean counterpart through
\begin{equation}\label{Direct2}
    H_M^{\mu \nu}(k, p)=c_{E\to M}^{\mu\nu} H_E^{\mu \nu}(k_E, p_E),
\end{equation}
where \(c_{E\to M}^{\mu\nu}\) accounts for the difference between the Euclidean and Minkowski gamma-matrix conventions, with \(c_{E\to M}^{00}=1\), \(c_{E\to M}^{ij}=-1\), and \(c_{E\to M}^{0i}=c_{E\to M}^{i0}=-i\).

In practical lattice calculations, Eq.~\eqref{Direct0} is evaluated using the finite-volume, finite-lattice-spacing hadronic function $H_E^{(L),\mu \nu}(x)$,
\begin{equation}\label{Direct}
\begin{aligned}
    H_E^{\mu \nu}(k_E, p_E)
    =-i\lim_{T,L\to\infty}\lim_{a\to 0}\sum_{t=-T/2}^{T/2}\sum_{\vec{x}}
    e^{k^0 t-i \vec{k} \cdot \vec{x}}\, H_E^{(L),\mu \nu}(x)\,,
\end{aligned}
\end{equation}
where the sums run over the discretized lattice points in the finite volume. The finite-volume effects associated with finite $(T,L)$ are corrected using the infinite-volume reconstruction described in Sec.~\ref{sec:IVR}, while lattice artifacts are accounted for through the continuum extrapolation discussed in Sec.~\ref{sec4}. The hadronic function \(H_E^{(L),\mu \nu}(x)\) is extracted from lattice three-point correlation functions as
\begin{equation}\label{HV}
\begin{aligned}
    H_E^{(L),\mu \nu}(x)
    &=\langle 0 \vert T\{J_{\text{em},E}^\mu(\vec{x},t)\, J_{W,E}^\nu(0)\}\vert P(p_E)\rangle^{(L)}\\
    &=\begin{cases}
        N_P^{-1} Z_V Z_W\, e^{m_P \Delta T}
        \langle J_{\text{em},E}^\mu(\vec{x}, t)\, J_{W,E}^\nu(\vec{0}, 0)\,\phi_P^\dagger(-\Delta T)\rangle^{(L)},
        & t \ge 0,\\[6pt]
        N_P^{-1} Z_V Z_W\, e^{m_P(\Delta T-t)}
        \langle J_{W,E}^\nu(\vec{0}, 0)\, J_{\text{em},E}^\mu(\vec{x}, t)\,\phi_P^\dagger(t-\Delta T)\rangle^{(L)},
        & t < 0.
    \end{cases}
\end{aligned}
\end{equation}
Here the initial-state meson with Euclidean four-momentum \(p_E=(im_P,\vec{0})\) is created by the wall-source interpolating operator \(\phi_\pi^\dagger(t)=i\bar{u}\gamma_5 d(t)\) for \(P=\pi\) and \(\phi_K^\dagger(t)=i\bar{u}\gamma_5 s(t)\) for \(P=K\). The source-sink separation \(\Delta T\) is chosen sufficiently large to ensure ground-state dominance, as discussed in Appendix~\ref{Appendix:deltaT}. 
The meson mass \(m_P\) and the overlap factor \(N_P=|\langle 0\vert \phi_P\vert P\rangle|/(2m_P)\) are determined from the two-point function \(\langle \phi_P(t)\phi_P^\dagger(0)\rangle\). The superscript \((L)\) on \(\langle\cdots\rangle^{(L)}\) indicates that the matrix elements are evaluated in a finite volume \(V=L^3\). 
The factor $Z_V$ is the renormalization coefficient of the local vector current and is taken from Ref.~\cite{RBC:2014ntl}. We denote the renormalization coefficient of the weak current by $Z_W$, which is $Z_W=Z_V$ for its vector component and $Z_W=Z_A$ for its axial-vector component. In the present work, we set $Z_A=Z_V$, following Appendix~B of Ref.~\cite{RBC:2010qam}, where it is shown that this choice avoids the $O(a m_{\mathrm{res}})$ effects. The uncertainty in $Z_V$ is about $0.03\%$, which is negligible compared with the statistical uncertainty in this work.

In the numerical calculation, the three-point function is constructed using Coulomb-gauge-fixed wall-source propagators and point-source propagators. Wall-source propagators are computed on all time slices, and for each gauge configuration we also compute about 2000 point-source propagators with randomly chosen source locations. For each point source, we use translational invariance to define it as the origin, where the weak current is inserted. We then loop over all lattice sites $x=(\vec{x},t)$ relative to that point and choose the wall-source time slice as
\begin{equation}
    t_{\mathrm{src}}=\min(t,0)-\Delta T.
\end{equation}
The corresponding contraction is then performed to obtain the three-point function.

To reduce data storage requirements and accelerate numerical calculations, the authors of Ref.~\cite{Tuo:2021ewr} 
proposed the ``scalar function method'' for evaluating Eq.~\eqref{Direct}. 
Here we present the formulae for decays with a real-photon in the final state. The detailed derivation is provided in Ref.~\cite{Tuo:2021ewr} and reviewed in Appendix~\ref{Appendix:scalar}. We again first consider the infinite-volume and continuum limit, where the hadronic function \(H_E^{\mu\nu}(x)\) can be decomposed into six coordinate-space scalar functions,
\begin{equation}\label{IE}
    \begin{aligned} 
     I_1\left(|\vec{x}|, t\right)&=\delta^{\mu \nu} H_{E}^{\mu\nu}(x), \\ 
     I_2\left(|\vec{x}|, t\right)&=-\frac{p_E^\mu p_E^\nu}{m_P^2} H_{E}^{\mu\nu}(x)=H_{E}^{00}(x), \\ 
     I_3\left(|\vec{x}|, t\right)&=\frac{x^\mu p_E^\nu}{i m_P} H_{E}^{\mu\nu}(x)-\frac{x \cdot p_E}{i m_P} I_2\left(|\vec{x}|, t\right)=x^i H_E^{ i 0}(x), \\ 
     I_4\left(|\vec{x}|, t\right)&=\frac{x^\nu p_E^\mu}{i m_P} H_{E}^{ \mu\nu}(x)-\frac{x \cdot p_E}{i m_P} I_2\left(|\vec{x}|, t\right)=x^i H_E^{ 0 i}(x), \\ 
     I_5\left(|\vec{x}|, t\right)&=x^\mu x^\nu H_E^{ \mu \nu}(x)-\frac{x \cdot p_E}{i m_P}\left(I_3\left(|\vec{x}|, t\right)+I_4\left(|\vec{x}|, t\right)\right)-\left(\frac{x \cdot p_E}{i m_P}\right)^2 I_2\left(|\vec{x}|, t\right)\\
    &=x^i x^j H_E^{ i j}(x), \\ 
     I_6\left(|\vec{x}|, t\right)&=\varepsilon_E^{\mu \nu \alpha \beta} \frac{x^\alpha p_E^\beta}{im_P} H_E^{ \mu \nu}(x)=\varepsilon_E^{\mu \nu \alpha 0} x^\alpha H_E^{ \mu \nu}(x).
    \end{aligned}
\end{equation}
Here, the indices $i,j$ take values in $\{1,2,3\}$. In these equations, $I_i\left(|\vec{x}|, t\right)$ for $i=1,\cdots,6$ are first defined as scalar functions, invariant under 4D Euclidean transformations in the infinite-volume limit, and then simplified in the rest frame of the meson. They depend only on the scalar variables $\left(|\vec{x}|,t\right) = \left(\sqrt{x^2-(x\cdot p_E)/(i m_P)},(x\cdot p_E)/(i m_P)\right)$. The Levi-Civita symbol in Euclidean space follows the convention $\varepsilon_E^{1230}=\varepsilon_{E,1230}=1$. In practical lattice calculations, we use the scalar functions $I_i^{(L)}(|\vec{x}|,t)$ obtained from Eq.~\eqref{IE}, with $H_E^{\mu\nu}(x)$ replaced by the lattice hadronic function $H_E^{(L)\,\mu\nu}(x)$ defined in Eq.~\eqref{Direct}.

Next, we also express $H_M^{\mu \nu}(k, p)$ in terms of momentum-space scalar functions:
\begin{equation}\label{IM}
	\begin{aligned}
		& \tilde{I}_1(x_\gamma)=g_{\mu\nu}\,p^2\,H_M^{\mu \nu}(k, p), 
		\quad \tilde{I}_2(x_\gamma)=p_\mu\,p_\nu\,H_M^{\mu \nu}(k, p), \\
		& \tilde{I}_3(x_\gamma)=k_\mu\,p_\nu\,H_M^{\mu \nu}(k, p), 
		\quad \tilde{I}_4(x_\gamma)=p_\mu\,k_\nu\,H_M^{\mu \nu}(k, p), \\
		& \tilde{I}_5(x_\gamma)=k_\mu\,k_\nu\,H_M^{\mu \nu}(k, p), 
		\quad \tilde{I}_6(x_\gamma)=-i\,\varepsilon_{\mu\nu \alpha \beta}\,k^\alpha p^\beta\,H_M^{\mu \nu}(k, p).
	\end{aligned}
\end{equation}
The convention for the Levi-Civita symbol in Minkowski space is chosen to be $\varepsilon^{0123}=1$ and $\varepsilon_{0123}=-1$. In general, i.e. allowing for the final-state photon to be virtual which is the case for $P\to\ell\bar{\nu}_\ell(\ell^{\prime+}\ell^{\prime-})$ decays, these scalar functions depend on the variables $\rho_1 = k^2/m_P^2$ and $\rho_2 = (p-k)^2/m_P^2$~\cite{Tuo:2021ewr}. For the decays with a real photon in the final state which are considered in this work, $\rho_1 = 0$ and $\rho_2 = 1 - x_\gamma$, so both these variables depend solely on $x_\gamma$. The $\tilde{I}_i(x_\gamma)$ can be obtained from the $I_{j}(|\vec{x}|^2,t)$ by using
\begin{equation}\label{scalar_calc}
\begin{aligned}
    \tilde{I}_i(x_\gamma)
	&= -im_P^2
	\Biggl( \sum_{j=1}^6 \int_{-\infty}^{\infty} dt\int d^3 x \,e^{k^0 t}\,
	\phi_{i j}\bigl(x_\gamma,|\vec{x}|\bigr)\,
	I_{j}\bigl(|\vec{x}|,t\bigr)\Biggr),\\
    &= -im_P^2
	\lim_{T,L\to\infty}\lim_{a\to 0}\Biggl( \sum_{j=1}^6 \sum_{t=-T/2}^{T/2}\sum_{\vec{x}} \,e^{k^0 t}\,
	\phi_{i j}\bigl(x_\gamma,|\vec{x}|\bigr)\,
	I^{(L)}_{j}\bigl(|\vec{x}|,t\bigr)\Biggr),
\end{aligned}
\end{equation}
which in turn allows for the reconstruction of the Minkowski-space matrix element:
\begin{equation}\label{HfromI}
    H_M^{\mu\nu}(k,p)
	=\sum_{i=1}^6 \omega_i^{\mu\nu}(k,p)\,\tilde{I}_i(x_\gamma).
\end{equation}
Here, $\omega_i^{\mu\nu}(k,p)$ denote kinematic factors. The derivation of $\phi_{ij}\left(x_\gamma, |\vec{x}|\right)$ and $\omega_i^{\mu\nu}(k,p)$ are discussed in detail in Ref.~\cite{Tuo:2021ewr}, and we review them in Appendix~\ref{Appendix:scalar}. The weight functions 
$\phi_{i j}\bigl(x_\gamma,|\vec{x}|\bigr)$ are given by
\begin{equation}\label{jphi}
	\phi_{ij}(x_\gamma,|\vec{x}|)
	=
	\begin{pmatrix}
		j_0(\varphi) & 0 & 0 & 0 & 0 & 0 \\
		0 & j_0(\varphi) & 0 & 0 & 0 & 0 \\
		0 & \tfrac{x_\gamma}{2}j_0(\varphi) & -\tfrac{m_P x_\gamma^2}{4}\frac{j_1(\varphi)}{\varphi} & 0 & 0 & 0 \\
		0 & \tfrac{x_\gamma}{2}j_0(\varphi) & 0 & -\tfrac{m_P x_\gamma^2}{4}\frac{j_1(\varphi)}{\varphi} & 0 & 0 \\
		-\tfrac{x_\gamma^2}{4}\frac{j_1(\varphi)}{\varphi} & \tfrac{x_\gamma^2}{4}\left(j_0(\varphi)+\frac{j_1(\varphi)}{\varphi}\right) & -\tfrac{m_P x_\gamma^3}{8}\frac{j_1(\varphi)}{\varphi} & -\tfrac{m_P x_\gamma^3}{8}\frac{j_1(\varphi)}{\varphi} & \tfrac{m_P^2 x_\gamma^4}{16}\frac{j_2(\varphi)}{\varphi^2} & 0 \\
		0 & 0 & 0 & 0 & 0 & \tfrac{m_P x^2_\gamma}{4}\frac{j_1(\varphi)}{\varphi}
	\end{pmatrix},
\end{equation}
where the functions $j_i(\varphi)$, with $\varphi=|\vec{k}||\vec{x}|=m_P x_\gamma |\vec{x}|/2$, are spherical Bessel functions:
\begin{equation}\label{eq:j123def}
    j_0(\varphi)=\frac{\sin\varphi}{\varphi}, 
\quad 
j_1(\varphi)=\frac{\sin\varphi-\varphi\cos\varphi}{\varphi^2},
\quad
j_2(\varphi)=\frac{\bigl(3-\varphi^2\bigr)\sin\varphi-3\varphi\cos\varphi}{\varphi^3}.
\end{equation}

\subsection{Infinite-Volume Reconstruction\label{sec:IVR}}
\begin{figure} 
	\centering
	\includegraphics[width=0.3\textwidth]{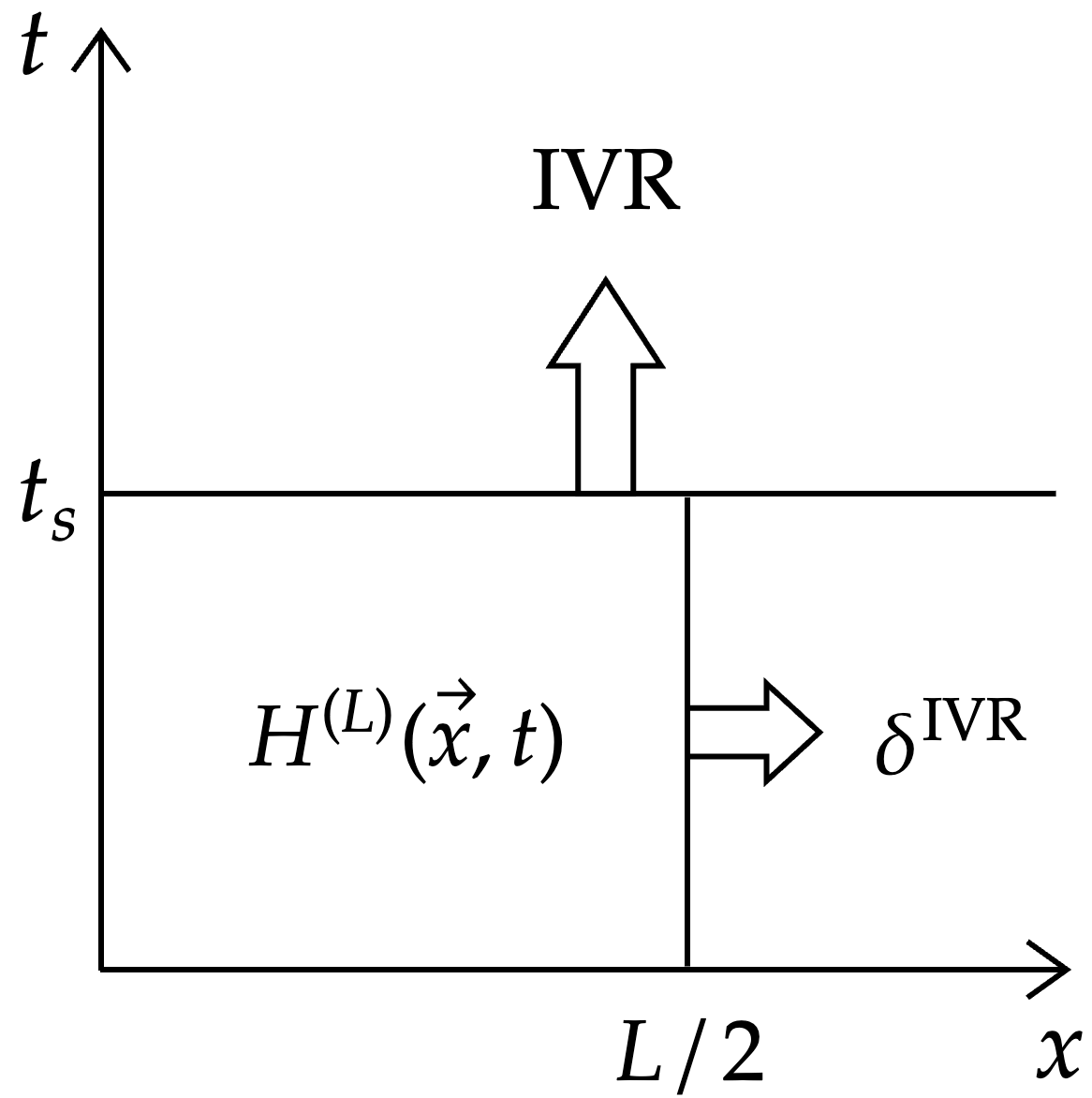}
	\caption{Idea of the IVR method: Reconstruct the infinite-volume hadronic function from finite-volume lattice data via (i) a temporal reconstruction (IVR) and (ii) a spatial reconstruction ($\delta^{\mathrm{IVR}}$). 		\label{fig:IVR}
	}
\end{figure}
In practical lattice calculations, where the lattice data is generated in a finite volume $(T,L)$, we use the infinite-volume reconstruction (IVR) method to compute Eq.~\eqref{scalar_calc}. As shown in Fig.~\ref{fig:IVR}, this method reconstructs the hadronic function in long-distance regions along both temporal and spatial directions (denoted as IVR and $\delta^{\mathrm{IVR}}$, respectively), thereby extending the finite-volume lattice data to obtain the infinite-volume hadronic function. Eq.~(\ref{scalar_calc}) is evaluated using the IVR method as follows~\cite{Tuo:2021ewr}
\begin{equation}\label{scalar_calc2}
	\begin{aligned}
		\tilde{I}_i(x_\gamma)&=\tilde{I}^{\text{IVR}}_i(x_\gamma;L)+\delta_i^{\text{IVR}}(L),\\
        &=\tilde{I}^{(s)}_i(x_\gamma;L)
		+\tilde{I}^{(l)}_i(x_\gamma;L)
		+\delta_i^{\text{IVR}}(L),\\
		\tilde{I}^{(s)}_i(x_\gamma;L)&=-im_P^2\sum_{j=1}^6 \sum_{t=-(t_s-1)}^{T/2} 
		\,\sum_{\vec{x}}\,e^{k^0 t}\,\phi_{ij}\bigl(x_\gamma, |\vec{x}|\bigr)\, I^{(L)}_{j}\bigl(|\vec{x}|,t\bigr),\\
		\tilde{I}^{(l)}_i(x_\gamma;L)&=-im_P^2\sum_{j=1}^6 \sum_{\vec{x}}\, 
		\frac{e^{-k^0 t_s}}{k^0+E_P(\vec{k})-m_P}\,
		\phi_{ij}\bigl(x_\gamma, |\vec{x}|\bigr)\,
		I^{(L)}_{j}\bigl(|\vec{x}|,-t_s\bigr).
	\end{aligned}
\end{equation}
Here, $k^0=m_P x_\gamma/2$ is the photon's energy and $E_P(\vec{k})=\sqrt{\vec{k}^2+m_P^2}$ is the energy of the meson with momentum $-\vec{k}$. The time cutoff $-t_s$ separates the short-range contribution with $t>-t_s$, $\tilde{I}^{(s)}_i(x_\gamma;L)$, from the long-range contribution with $t\leq-t_s$, $\tilde{I}^{(l)}_i(x_\gamma;L)$. In the long-range region, ground-state dominance allows us to reconstruct $\tilde{I}^{(l)}_i(x_\gamma;L)$ using data obtained at $t = -t_s$. 

The term $\delta_i^{\text{IVR}}(L)$ accounts for the residual exponentially suppressed finite-volume effects in the IVR method. We calculate $\delta_i^{\text{IVR}}(L)$ arising from the contributions shown in Fig.~\ref{fig:intstate}. 
As shown in Appendix~\ref{Appendix:FV}, when the electromagnetic form factor in these diagrams is set to $F^{(P)}(q^2) = 1$, these two diagrams correspond to the point-particle approximation, and the resulting finite-volume correction is denoted by \( \delta^{\mathrm{IVR}}_{i,\mathrm{pt}} \). Alternatively, structure-dependent information, such as the charge radius, can be included in the estimation of the finite-volume effect, yielding a correction denoted by \( \delta^{\mathrm{IVR}}_{i,\mathrm{SD}} \). 
Numerically we found that $\delta_i^{\text{IVR}}(L)$ is dominated by the point-particle approximation and the difference between \( \delta^{\mathrm{IVR}}_{i,\mathrm{pt}} \) and \( \delta^{\mathrm{IVR}}_{i,\mathrm{SD}} \) is much smaller than the statistical error. More details of finite-volume corrections can be found in Ref.~\cite{Tuo:2021ewr} and Appendix~\ref{Appendix:FV}. 
\begin{figure}
	\centering
	\includegraphics[width=0.7\textwidth]{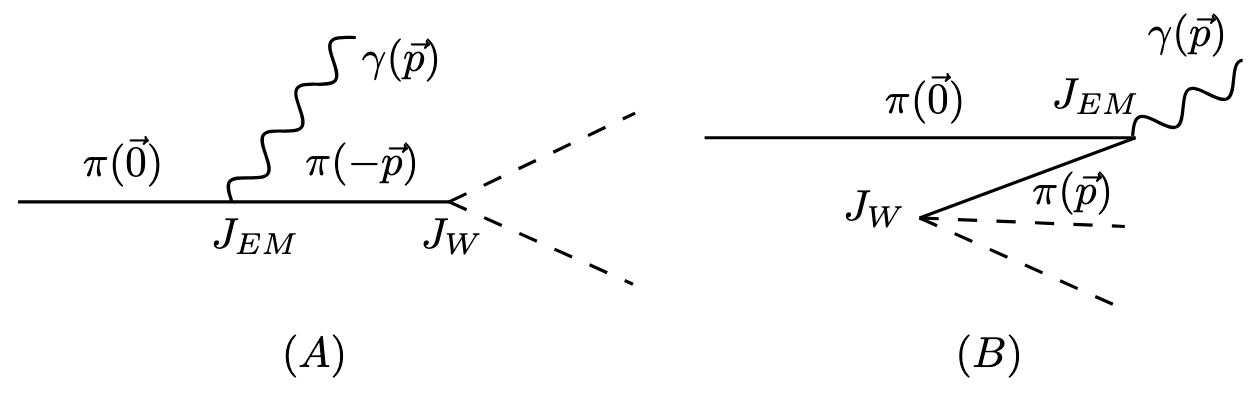}
	\caption{To estimate finite-volume effects, we consider two intermediate states in the decay \(P\to e\nu_e\gamma\). In the first case (diagram A), the process \(P\to P\gamma \to e\nu_e\gamma\) occurs with time ordering \(t < 0\). In the second case (diagram B), the process \(P\to 2Pe\nu_e\to e\nu_e\gamma\) occurs with time ordering \(t > 0\).
		\label{fig:intstate}
	}
\end{figure}

\subsection{The Decay Amplitude and Branching Ratio\label{sec2.3}}
Having obtained the hadronic matrix element as described above, we next 
construct the squared amplitude and branching ratio. In the decay amplitude $\mathcal{M}^\mu$ in Eq.~\eqref{amp}, the contribution from the term $f_P\,g^{\mu\nu}l_\nu(p_\ell,p_{\nu_\ell})$ in the contribution from the hadronic matrix element of Fig.~\ref{fig:diagrams}A cancels the contribution from $f_P\,l^\mu(p_\ell, p_{\nu_\ell})$ in Fig.~\ref{fig:diagrams}B. However, due to differences between the decay constant extracted from two and three-point correlation functions, $f_P^{2\text{pt}}$ and $f_P^{3\text{pt}}$ respectively, this cancellation is not exact in lattice computations~\cite{Desiderio:2020oej}. To preserve this cancellation, we adopt a modified expression for the amplitude:
\begin{equation}\label{amp_sub}
	\begin{aligned}
		\mathcal{M}^\mu(k,p_\ell,p_{\nu_\ell})
		&= f_P\,L^{\prime\mu}(k,p_\ell,p_{\nu_\ell})
		-\bar{H}^{\mu\nu}_M(k,p)\,l_\nu(p_\ell,p_{\nu_\ell}),\\
		\bar{H}^{\mu\nu}_M(k,p)
		&= H^{\mu\nu}_M(k,p) 
		- f_P^{\text{3pt}}\,g^{\mu\nu},
	\end{aligned}
\end{equation}
where $L^{\prime\mu}(k,p_\ell,p_{\nu_\ell})$ is defined in Eq.~(\ref{Lpart}).
Following the method of Ref.~\cite{Desiderio:2020oej}, $f_P^{\text{3pt}}$ can be extracted from the hadronic function in the $x_\gamma\to 0$ limit. In our method, this limit is equivalent to the determination of $f_P^{\text{3pt}}$ using
\begin{equation}\label{f3pt}
     f_P^{\text{3pt}}=-\frac{i}{3}\sum_{i=1}^3 \sum_{t=-(t_s-1)}^{T/2}\sum_{\vec{x}}~ H_E^{(L),ii}(x).
\end{equation}
As discussed in Ref.~\cite{Desiderio:2020oej}, subtracting $f_P^{\text{3pt}}$ cancels both statistical errors and lattice artifacts. In the numerical calculation, we therefore adopt the same temporal cutoff $t>-t_s$ as in the $x_\gamma\neq 0$ case to preserve this cancellation.

Using the amplitude, we compute the reduced squared amplitude as
\begin{equation}\label{Atrans}
\begin{aligned}
	A(x_\gamma,y_\ell)
	&= \frac{1}{4m_\ell^2\,f_P^2}
	\sum_{\lambda,\mathrm{spin}}
	\Bigl(\epsilon_\mu(k,\lambda)\,\mathcal{M}^\mu\Bigr)
	\Bigl(\epsilon_\rho(k,\lambda)\,\mathcal{M}^{\rho}\Bigr)^*\\
	&= \frac{1}{4m_\ell^2\,f_P^2}
	\sum_{\lambda=1}^2 
	\epsilon_\mu^{\perp}(k,\lambda)\,\epsilon_\rho^{\perp}(k,\lambda)
	\sum_{\mathrm{spin}}
	\bigl(\mathcal{M}^\mu \,\mathcal{M}^{\rho*}\bigr),
\end{aligned}
\end{equation}
where, for real-photon radiation, we sum over the two physical transverse polarizations $\epsilon_\mu^{\perp}(k,\lambda)$. In the infinite‐volume and continuum limit, restricting to transverse polarizations removes the ground‐state contribution from the intermediate \(P\) state in the combination $\epsilon_\mu^\perp(k,\lambda)\,H_M^{\mu\nu}(k,p)$, because this contribution is proportional to \((2p-k)^\mu (p-k)^\nu\) and \(\epsilon_\mu^\perp(k,\lambda)\) is orthogonal to \(p\) and \(k\). 
In the numerical calculation, we found that this contribution doesn't vanish exactly due to both finite-volume effects and lattice artifacts. 
Consequently, although the temporal reconstruction (i.e., \(\tilde{I}_i^{(l)}\)) vanishes for the combination $\epsilon_\mu^\perp(k,\lambda)\,H_M^{\mu\nu}(k,p)$, the finite-volume correction from the ground-state contribution remains non-negligible and is addressed through the spatial reconstruction (i.e., \(\delta^{\mathrm{IVR}}_i\)). Residual lattice artifacts are controlled by comparing ensembles with different lattice spacings but similar physical volumes, as detailed in Sec.~\ref{sec4}. 

It should be noted that, in calculations of the complete radiative corrections to leptonic decays including virtual-photon loops, the inclusion of all photon polarization states in real-photon emissions is required to ensure infrared cancellation with the virtual-photon loop contributions in the Feynman gauge~\cite{Christ:2023lcc}. Temporal reconstruction (i.e., \(\tilde{I}_i^{(l)}\)) is therefore necessary in that case.

We numerically implement the polarization vectors and spinor matrices, then evaluate the squared amplitude \(A(x_\gamma,y_\ell)\) through matrix products. Substituting \(A(x_\gamma,y_\ell)\) into Eq.~(\ref{R0}) and Eq.~(\ref{R_RC}) (or Eq.~(\ref{R_RC_cut})) yields the normalized differential branching ratio \(\tfrac{d^2R_\gamma}{dx_\gamma\,dy_\ell}\) for the cases without and with radiative corrections, respectively. By integrating over the phase space in the same experimental regions listed in Appendix~\ref{sec:phasespace}, we obtain the final results for \(R_\gamma\).

\subsection{Form Factors \(F_V\) and \(F_A\)}
To provide more information of the meson structure using lattice QCD, we present the expressions for determining the form factors $F_V$ and $F_A$. Specifically, the vector form factor \(F_V(x_\gamma)\) is directly related to \(\tilde{I}_6(x_\gamma)\) by
\begin{equation}\label{FV}
    F_V(x_\gamma) = -\frac{2\tilde{I}_6(x_\gamma)}{m_P^3\, x_\gamma^2}.
\end{equation}

To extract the axial form factor \(F_A(x_\gamma)\), we contract the axial-vector part of the hadronic matrix element \(\bar{H}_{M,A}^{\mu\nu}(k,p)\) defined in Eq.~\eqref{amp_sub} with the physical transverse polarization vector \(\epsilon_\mu^\perp(k,\lambda)\):
\begin{equation}
    \epsilon_\mu^\perp(k,\lambda)\,\bar{H}_{M,A}^{\mu\nu}(k,p) = \epsilon^{\perp,\nu}(k,\lambda)\,\frac{x_\gamma}{2}\,F_A(x_\gamma)\,m_P.
\end{equation}
Alternatively, by substituting the scalar function decomposition of \(\bar{H}_M^{\mu\nu}(k,p)\) in Eq.~\eqref{HfromI}, we obtain
\begin{equation}
\begin{aligned}
    &\epsilon_\mu^\perp(k,\lambda)\,\bar{H}_{M,A}^{\mu\nu}(k,p) = \epsilon^{\perp,\nu}(k,\lambda)\\ \times&\left(\frac{x_\gamma\,\tilde{I}_1(x_\gamma) - 2\,\tilde{I}_4(x_\gamma) - 2\,\tilde{I}_3(x_\gamma) + \frac{4\,\tilde{I}_5(x_\gamma)}{x_\gamma}}{2\,m_P^2\,x_\gamma} - f_P^{\text{3pt}}\right).
\end{aligned}    
\end{equation}
Comparing these two expressions yields the formula for \(F_A(x_\gamma)\) in terms of the scalar functions:
\begin{equation}\label{FA}
    F_A(x_\gamma) = \frac{2}{x_\gamma}\left(\frac{x_\gamma\,\tilde{I}_1(x_\gamma) - 2\,\tilde{I}_4(x_\gamma) - 2\,\tilde{I}_3(x_\gamma) + \frac{4\,\tilde{I}_5(x_\gamma)}{x_\gamma}}{2\,m_P^3\,x_\gamma} - \frac{f_P^{\text{3pt}}}{m_P}\right).
\end{equation}
$f_P^{\text{3pt}}$ is extracted using Eq.~(\ref{f3pt}). Using the scalar functions defined in Sec.~\ref{sec:H}, we can compute \(F_V(x_\gamma)\) and \(F_A(x_\gamma)\) for any value of \(x_\gamma\).

Another approach to accessing a wide range of photon momenta is though twisted boundary condition~\cite{Desiderio:2020oej,Frezzotti:2020bfa,Gagliardi:2022szw,DiPalma:2025iud,Frezzotti:2023ygt}. In cases where more form factors contribute to the amplitude—for instance, the decay \(K^+\to \ell\nu_\ell \ell'\ell'\) involves four form factors (\(R_1, R_2, F_V, F_A\))—fitting the discrete momentum data over the entire phase space might become more challenging~\cite{Gagliardi:2022szw}. Our method, which permits the computation of form factors at arbitrary momenta, offers an alternative approach to determine these form factors.

\section{Numerical Results\label{sec4}}
\subsection{The Lattice Setup}
We use two $N_f=2+1$ domain wall fermion ensembles (48I and 64I), both with physical pion masses generated by the RBC/UKQCD collaboration~\cite{RBC:2014ntl}. Table~\ref{table:ens} summarizes their parameters. These two ensembles have similar volumes but different lattice spacings. The kaon masses on the 48I and 64I ensembles differ, which may affect the continuum extrapolation based on these two ensembles, as the observed differences in the form factors could also stem from the mismatch in the kaon mass. To address this issue, we also analyze a partially quenched version of the 64I ensemble (``64Ipq''), in which the valence quark masses are chosen to be different 
from the sea quark masses.
For both the 64I and 64Ipq ensembles, we compute the kaon form factors $F_V(x_\gamma)$ and $F_A(x_\gamma)$ using the same 31 gauge configurations. 
The extracted kaon form factors from the 64I and 64Ipq ensembles at each $x_\gamma$ are used to determine the coefficients $c_{A,V}^{(\prime),K}(x_\gamma,a^2)$ at $a = a_{\text{64I}}$ with the chiral extrapolation form~\cite{Desiderio:2020oej},
\begin{equation}\label{chiral_extra}
    F_{A,V}(x_\gamma,a^2) = \frac{m_K}{f_K}\left[\,c^K_{A,V}(x_\gamma,a^2) 
    + c_{A,V}^{\prime\, K}(x_\gamma,a^2)\,\frac{m_\pi^2}{(4\pi f_\pi)^2}\right].
\end{equation}
Higher-order terms proportional to $O(m_\pi^2 m_K^2)$ or $O(m_\pi^4)$ with logarithmic corrections are neglected, as the pion and kaon masses in these ensembles lie close to the physical point. 

These coefficients are then applied to the 64I ensemble to obtain form factors extrapolated to the physical point, specified by the physical meson masses~\cite{ParticleDataGroup:2024cfk} and the $N_f=2+1$ FLAG averages of the decay constants~\cite{FlavourLatticeAveragingGroupFLAG:2024oxs}. We note that the kaon mass on the 48I ensemble also deviates slightly from its physical value. For the 48I ensemble, the corresponding extrapolation is performed by neglecting the lattice-spacing dependence of the coefficients. Taking $x_\gamma=0.5$ as an example, Table~\ref{table:chiral} lists the coefficients $c_{A,V}^{(\prime),K}(x_\gamma,a_{\text{64I}}^2)$ and illustrates the impact of the extrapolation. The shift in $F_V$ on the 64I ensemble is larger than its statistical uncertainty, while the shifts in the other cases remain consistent within errors. Since the induced shift in the 48I results is comparable to the statistical uncertainty, the approximation of neglecting the lattice-spacing dependence in the coefficients $c_{A,V}^{(\prime),K}(x_\gamma,a^2)$ is adequate at the present level of precision. The table also indicates that part of the difference between the kaon $F_V$ values obtained on the 48I and 64I ensembles can be attributed to the kaon-mass mismatch.

\begin{table}
	\centering
	\begin{tabular}{cccccc}
		\hline\hline
		Ensembles&$a^{-1}$[GeV]&$L^3\times T$&$m_\pi$[MeV]&$m_K$[MeV]&$N_{\text{conf}}$\\
		\hline
		$48 \mathrm{I}$&1.730(4)&$48^3\times 96$&139.55(19)&499.21(24)&112\\
		$64 \mathrm{I}$&2.359(7)&$64^3\times 128$&139.18(14)&507.98(35)&119\\
        $64 \mathrm{I}$-pq&2.359(7)&$64^3\times 128$&135.14(19)&496.50(81)&31\\
		\hline\hline
	\end{tabular}
	\caption{\label{table:ens}Parameters of the lattice ensembles used in this study. For each ensemble, we provide the inverse lattice spacing $a^{-1}$ (in GeV), the lattice volume $L^3\times T$, the pion mass $m_\pi$, the kaon mass $m_K$, and the number of configurations $N_{\text{conf}}$ used in this work.
}
\end{table}

\begin{table}
\centering
\begin{tabular}{c|cc|cc|cc}
\hline\hline
\multirow{2}{*}{Form factor} & \multirow{2}{*}{$c_{A,V}^K|_{x_\gamma=0.5}$} & \multirow{2}{*}{$c_{A,V}^{\prime\,K}|_{x_\gamma=0.5}$} & \multicolumn{2}{c|}{$F_{A,V}|_{x_\gamma=0.5}$ on 48I} & \multicolumn{2}{c}{$F_{A,V}|_{x_\gamma=0.5}$ on 64I} \\
\cline{4-7}
 & & & Raw & Extrapolated & Raw & Extrapolated \\
\hline
$F_A$ for $K$ & $0.012(7)$ & $-0.1(9)$ & $0.0385(20)$ & $0.0382(20)$ & $0.0393(23)$ & $0.0385(23)$ \\
$F_V$ for $K$ & $0.030(2)$ & $1.0(3)$ & $0.1168(8)$ & $0.1159(8)$ & $0.1190(10)$ & $0.1165(10)$\\
\hline\hline
\end{tabular}
\caption{Coefficients $c_{A,V}^{K}$ and $c_{A,V}^{\prime\,K}$ determined at $x_\gamma=0.5$, and the corresponding kaon form factors $F_{A,V}$ on the 48I and 64I ensembles before and after extrapolation to the meson masses and decay constants given in Ref.~\cite{ParticleDataGroup:2024cfk,FlavourLatticeAveragingGroupFLAG:2024oxs}. \label{table:chiral}}
\end{table}

\subsection{Infinite-Volume Reconstruction}
\begin{figure} 
	\centering
	\includegraphics[width=0.75\textwidth]{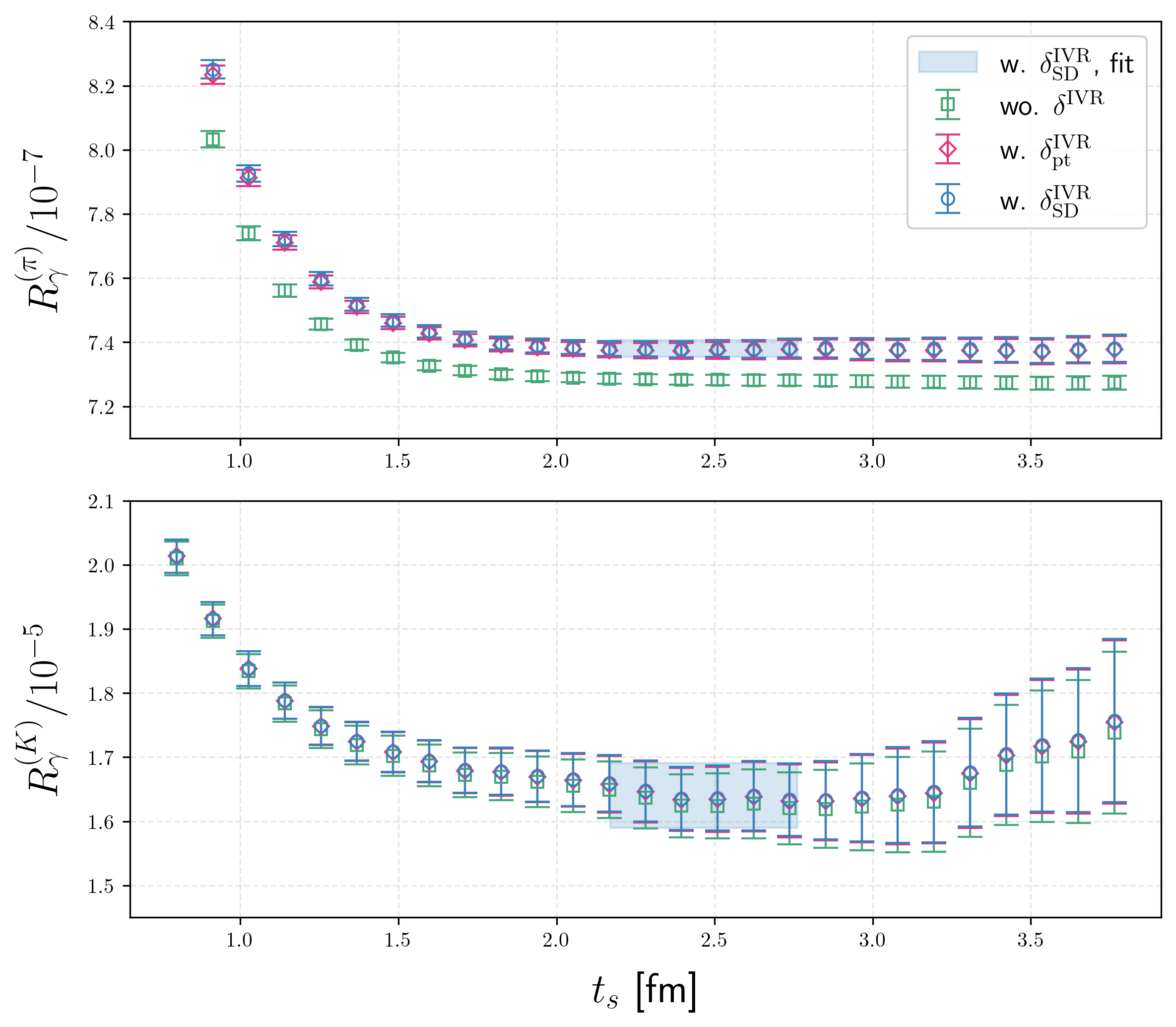}
	\caption{Results for $R_\gamma^{(\pi)}$ and $R_\gamma^{(K)}$ on the 48I ensemble. The time integral is calculated in the range $t\geq -t_s$. The green points correspond to the results without finite-volume correction, whereas the red and blue points represent results with correction $\delta^{\text{IVR}}_{\text{pt}}$ and $\delta^{\text{IVR}}_{\text{SD}}$, respectively. The blue bands indicate the fits to the plateau region $t_s\in[2.2~\text{fm},\,2.8~\text{fm}]$ of the results with $\delta^{\text{IVR}}_{\text{SD}}$ correction. 
		\label{fig:GIVR_Kpi}
	}
\end{figure}

We present results for $\pi \to e\nu_e\gamma$ in region O and $K \to e\nu_e\gamma$ in regions 1–5 on the 48I ensemble as examples to illustrate the impact of infinite-volume reconstruction. The definitions of these phase-space regions are given in Appendix~\ref{sec:phasespace}.
As discussed in Sec.~\ref{sec2.3}, when restricting to transverse photon polarizations, the temporal reconstruction (i.e., $\tilde{I}_i^{(l)}$) vanishes and only the spatial reconstruction (i.e., $\delta_i^{\mathrm{IVR}}$) is needed. 
Fig.~\ref{fig:GIVR_Kpi} shows the results before and after this correction for $R_\gamma^{(\pi)}$ and $R_\gamma^{(K)}$ on the 48I ensemble. 
We only present the results after applying the $O(\alpha^2 L_e)$ collinear radiative correction in Eq.~(\ref{R_RC}); the effect of this correction will be discussed in the following subsections. In these figures, the green points show results obtained with the integration range \( t \geq -t_s \), without applying finite-volume corrections. The red and blue points include the point-particle correction \(\delta^{\text{IVR}}_{\text{pt}}\) and the structure-dependent correction \(\delta^{\text{IVR}}_{\text{SD}}\), respectively. The blue bands indicate the fits to the plateau region of the results with $\delta^{\text{IVR}}_{\text{SD}}$ correction. For both $\pi$ and $K$ decays, a plateau exists for $t_s\in[2.2~\text{fm},\,2.8~\text{fm}]$, and we therefore adopt this interval to extract the results. For $R_\gamma^{(K)}$, the statistical error increases substantially for $t_s>3~\text{fm}$, because the statistical error of determining $f_K^{\text{3pt}}$ grows rapidly at larger $t_s$ in Eq.~\eqref{f3pt}.

\begin{table}
	\centering
	\begin{tabular}{c|c|ccc}
		\hline \hline  
                 &Region& wo. $\delta^{\mathrm{IVR}}$ & w. $\delta^{\text{IVR}}_{\text{pt}}$ & w. $\delta^{\text{IVR}}_{\text{SD}}$\\
		\hline 
		$R_\gamma^{(\pi)}/10^{-7}$ &O & $7.283(17)$ & $7.375(26)$ & $7.380(26)$ \\
		$R_\gamma^{(K)}/10^{-5}$ &1-5   & $1.628(50)$ & $1.639(51)$ & $1.640(51)$ \\
		\hline \hline
	\end{tabular}
	\caption{\label{table:FV} Results with or without finite-volume correction on the 48I ensemble. We show the results without finite-volume correction, with point-particle correction $\delta^{\text{IVR}}_{\text{pt}}$ and with structure-dependent correction $\delta^{\text{IVR}}_{\text{SD}}$. The results include only the statistical errors.}
\end{table}

Table \ref{table:FV} summarizes the fitting results of the plateau region for the different finite-volume corrections on the 48I ensemble. For our current lattice volume $L\simeq 5.4~\text{fm}$, the finite-volume correction of $R_\gamma^{(\pi)}$ is approximately $1\%$, but it is still sizable compared to the current statistical precision. The discrepancy between the point-particle correction $\delta^{\text{IVR}}_{\text{pt}}$ and the structure-dependent correction $\delta^{\text{IVR}}_{\text{SD}}$ is negligible relative to the statistical errors. In the case of $R_\gamma^{(K)}$, this finite-volume correction is negligible compared to the statistical uncertainty. 
Thus, the finite-volume effects from the ground-state \(\pi\) or \(K\) contribution are well controlled. Finite-volume effects from higher excited states, such as \(\rho\) or \(K^*\) intermediate states, may still be present, but are exponentially suppressed by their larger masses and are thus neglected in this work.

\subsection{Results for $\pi\to e\nu_e\gamma$}
\begin{figure}
	\centering
	\includegraphics[width=0.85\textwidth]{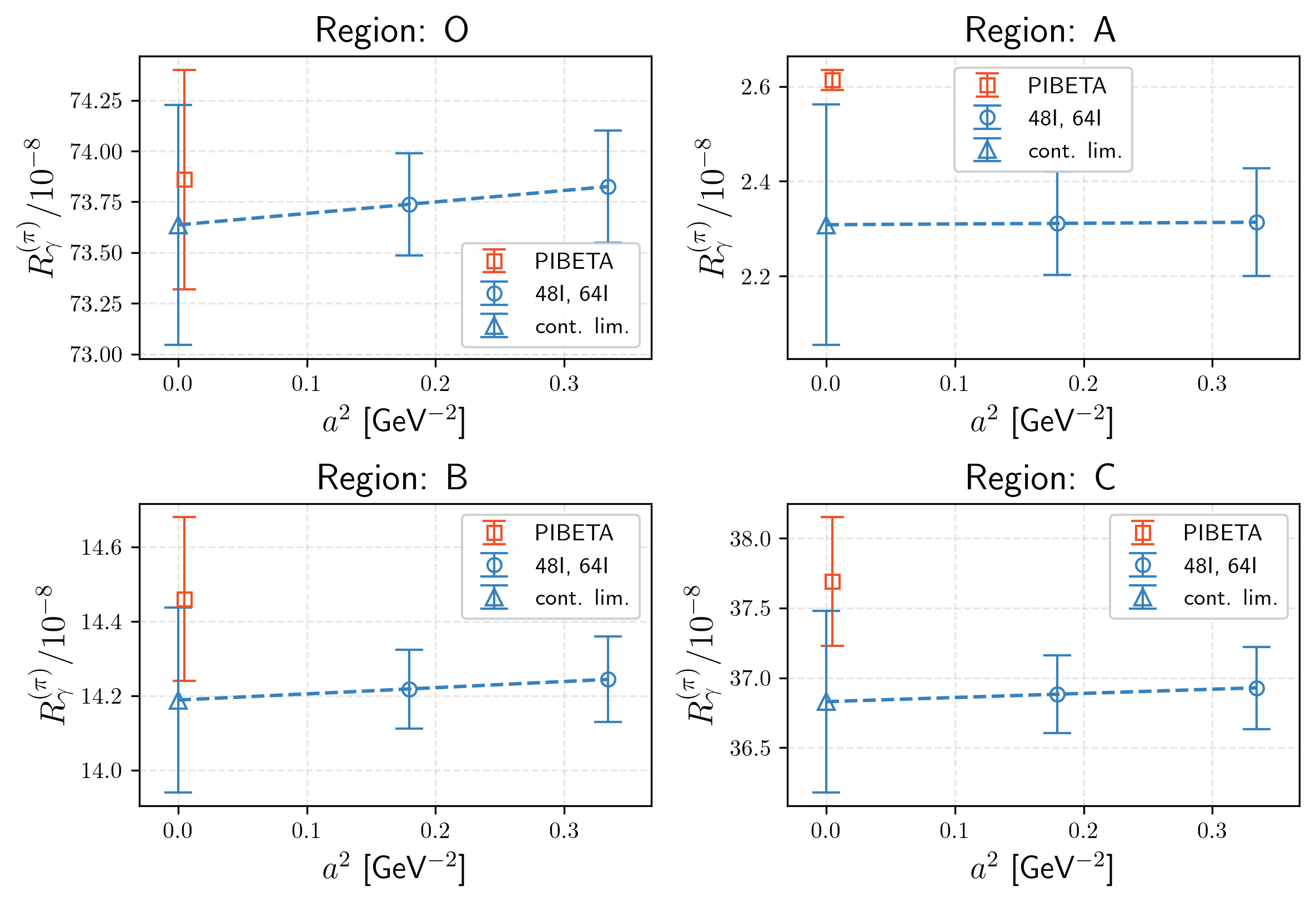}
	\caption{Continuum extrapolation of $R_\gamma^{(\pi)}$ in four phase space regions by a linear fit in \(a^2\)}. The lattice results are calculated by including the $O(\alpha^2 L_e)$ radiative correction, as defined in Eq.~\eqref{R_RC}. For comparison, the results from PIBETA experiment are also shown~\cite{Bychkov:2008ws}.
		\label{fig:cont_lim_pi}
	
\end{figure}
Fig.~\ref{fig:cont_lim_pi} shows the continuum extrapolation for $R_\gamma^{(\pi)}$ in the $\pi \to e\nu_e\gamma$ decay channel, obtained using a linear fit in $a^2$. The $O(\alpha^2 L_e)$ collinear corrections are included, and the second photon is treated inclusively, as defined in Eq.~\eqref{R_RC}. For comparison, the corresponding results from PIBETA experiment are also shown~\cite{Bychkov:2008ws}. Our lattice results in all four phase-space regions agree with experiment within statistical errors. 

\begin{figure}
	\centering
	\includegraphics[width=0.6\textwidth]{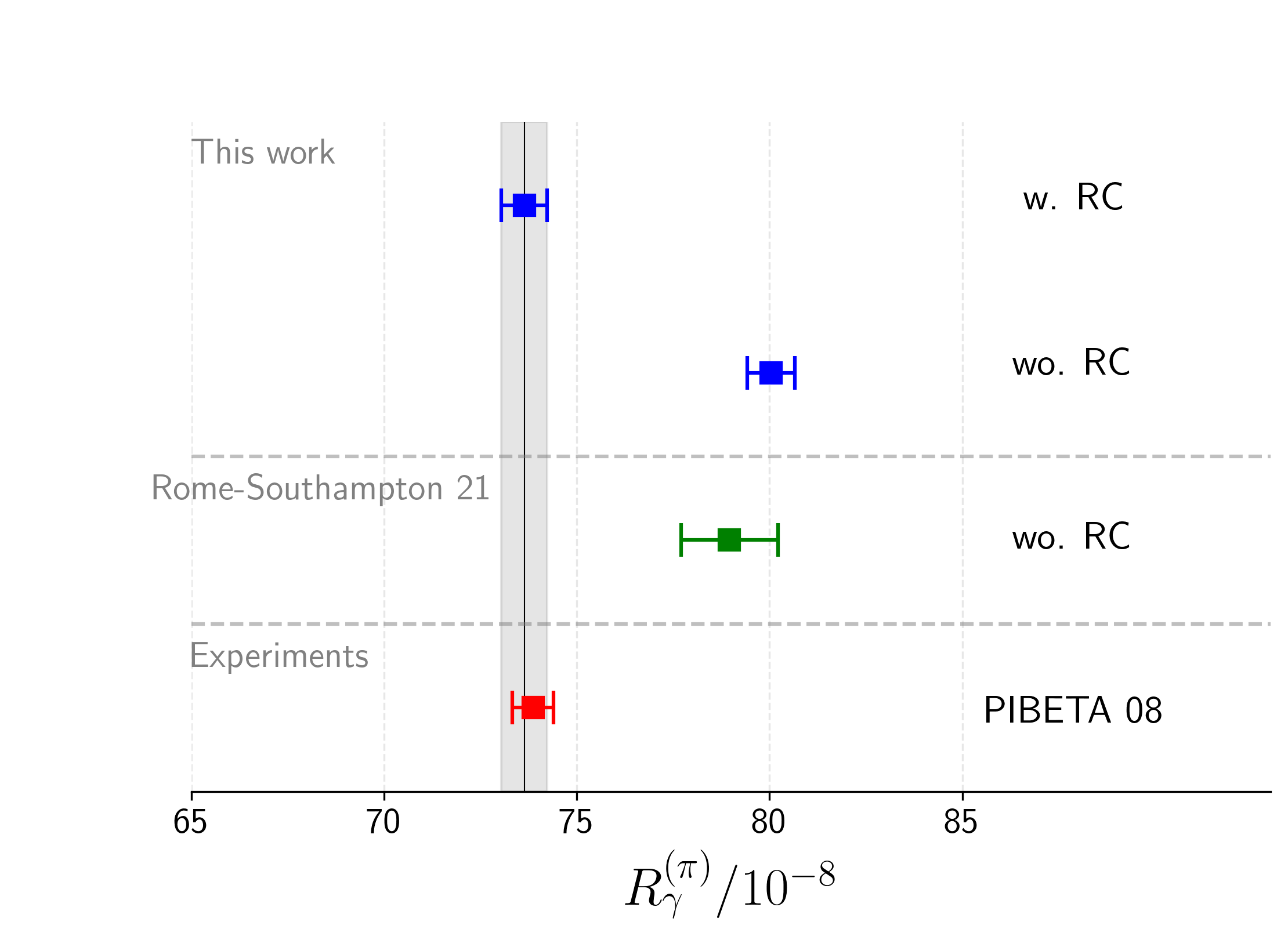}
	\caption{Comparison of the branching ratios \(R_\gamma^{(\pi)}\) for region O from lattice calculations and the PIBETA experiment~\cite{Bychkov:2008ws}. Lattice results are presented both without radiative corrections (``wo. RC'' as defined in Eq.~(\ref{R0})), using data from this work and Ref.~\cite{Frezzotti:2020bfa}, and with leading-order collinear radiative corrections (``w. \(O(\alpha^2 L_e)\) RC'' as defined in Eq.~(\ref{R_RC})) from this work.
		\label{fig:Rpi_compare}
	}
\end{figure}

Our analysis demonstrates that including collinear radiation corrections is essential to achieve agreement with the experimental branching ratios. Table~\ref{table: RC_pi} presents a comparison of results obtained without radiative corrections (Eq.~\eqref{R0}, labeled “wo. RC”) and with leading-order collinear radiative corrections (Eq.~\eqref{R_RC}, labeled “\(O(\alpha^2 L_e)\) RC”). The corresponding results for region~O are also displayed in Fig.~\ref{fig:Rpi_compare}. As shown in Table~\ref{table: RC_pi} and Fig.~\ref{fig:Rpi_compare}, our result without radiative corrections is consistent with the previous lattice result from Ref.~\cite{Frezzotti:2020bfa} but deviates from the experimental result. After including the radiative corrections, the lattice result agrees well with the experimental result for all phase-space regions within statistical errors, indicating that the discrepancy between lattice calculations of Ref.~\cite{Frezzotti:2020bfa} and PIBETA experimental measurements~\cite{Bychkov:2008ws} was primarily due to the absence of $O(\alpha^2 L_e)$ radiative corrections. 

\begin{table}[h]
	\centering
	\begin{tabular}{c|cc|c|c}
		\hline \hline 
		Region & \multicolumn{2}{c|}{This work} & Rome-Southampton 21~\cite{Frezzotti:2020bfa} 
        & PIBETA~\cite{Bychkov:2008ws}  
        \\
		& wo.~RC & \(O(\alpha^2 L_e)\)~RC & wo.~RC &  \\
		\hline
		A &  2.48(27)  & 2.31(26)  & 2.32(40) & 2.614(21) \\
		B & 15.02(26)  & 14.19(25)  & 14.59(54) & 14.46(22) \\
		C  & 40.70(70)  & 36.83(65)  & 40.15(1.04) & 37.69(46) \\
		O  & 80.04(62) & 73.64(59)  & 78.96(1.26) & 73.86(54) \\
		\hline \hline
	\end{tabular}
	\caption{Comparison of the branching ratios \(R_\gamma^{(\pi)}\) from this work and from the lattice calculation of Ref.~\cite{Frezzotti:2020bfa}, as well as the PIBETA measurement~\cite{Bychkov:2008ws}.
    Lattice results are shown for two cases: without radiative corrections (``wo. RC'' as defined in Eq.~(\ref{R0})), and with radiative corrections (``w. \(O(\alpha^2 L_e)\) RC'' as defined in Eq.~(\ref{R_RC})). All values are in units of \(10^{-8}\). The results include only statistical uncertainties.} \label{table: RC_pi}
\end{table}

\subsection{Results for $K\to e\nu_e\gamma$}
\begin{figure} 
	\centering
	\includegraphics[width=0.55\textwidth]{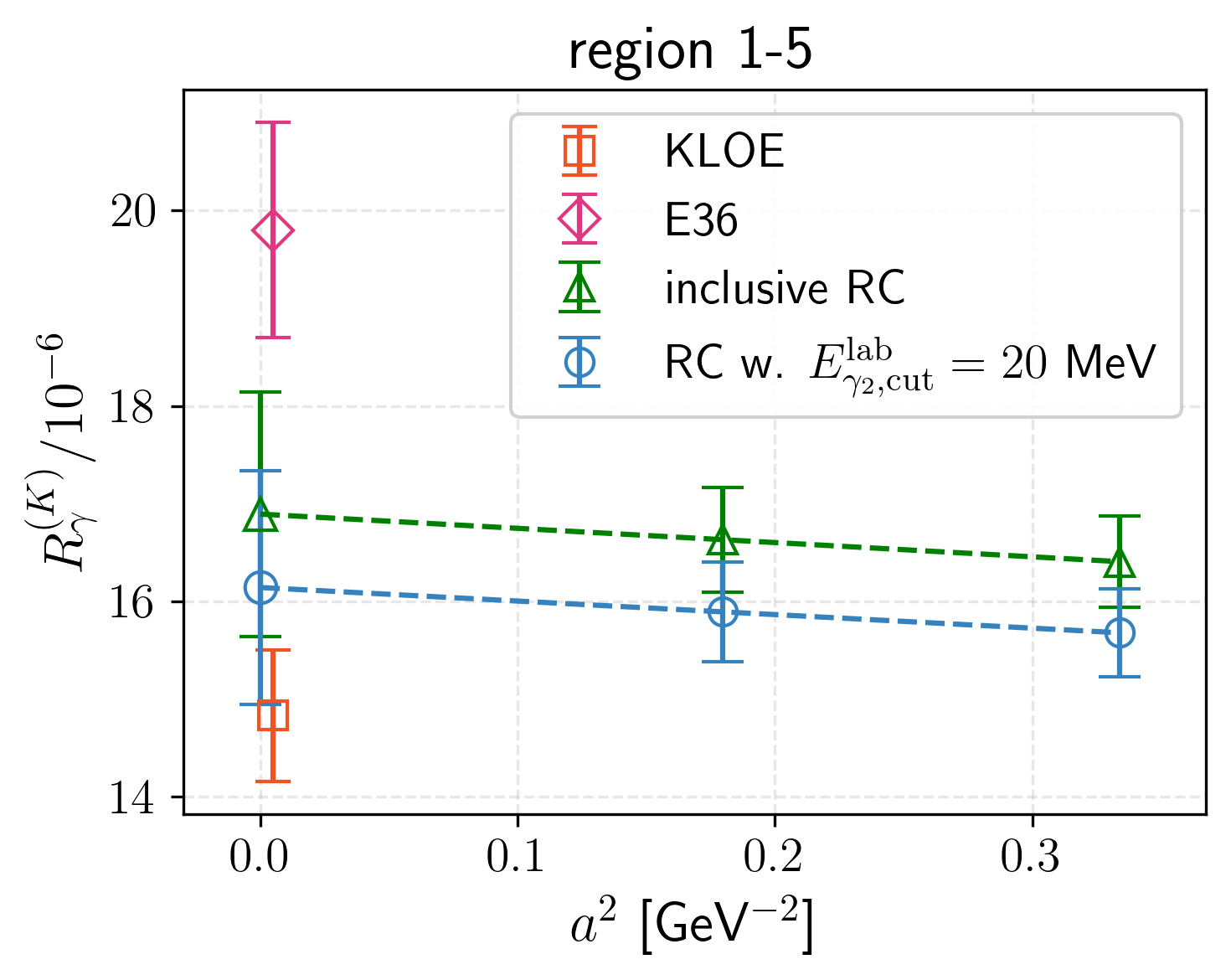}
	\caption{
    Continuum extrapolation of $R_\gamma^{(K)}$ in phase-space region 1–5 by a linear fit in \(a^2\). Results are shown for (i) inclusive with respect to the second photon (Eq.~\eqref{R_RC}, denoted as ``inclusive RC'') and (ii) with a laboratory-frame energy cut on the second photon (Eq.~\eqref{R_RC_cut} with $\vec{p}^{~\text{lab}} = 100$~MeV and $E_{\gamma_2,\text{cut}}^{\text{lab}} = 20$~MeV, denoted as ``RC w. $E_{\gamma_2,\text{cut}}^{\text{lab}}$''). For comparison, measurements from the KLOE~\cite{KLOE:2009urs} and E36~\cite{J-PARCE36:2022wfk} experiments are also shown.
\label{fig:cont_lim_K}}
\end{figure}
Fig.~\ref{fig:cont_lim_K} shows the continuum extrapolation of \(R_\gamma^{(K)}\) in the \(K\to e\nu_e\gamma\) channel, obtained from a linear fit in \(a^2\). The \(O(\alpha^2 L_e)\) radiative corrections are included. The phase-space region is chosen to be region 1-5 defined in Table~\ref{table:PSK}. Results are shown for both (i) inclusive with respect to the second photon (denoted as ``inclusive RC''), as defined in Eq.~\eqref{R_RC}, and (ii) with a laboratory‑frame energy cut on the second photon (denoted as ``RC w. $E_{\gamma_2,\text{cut}}^{\text{lab}}$''), as given in Eq.~\eqref{R_RC_cut} with $\vec{p}^{~\mathrm{lab}} = 100~\text{MeV}$ and $E_{\gamma_2,\text{cut}}^{\text{lab}} = 20~\text{MeV}$.
The measurements from the KLOE and E36 experiments are also shown~\cite{KLOE:2009urs,J-PARCE36:2022wfk}. 

\begin{figure} 
	\centering
	\includegraphics[width=0.65\textwidth]{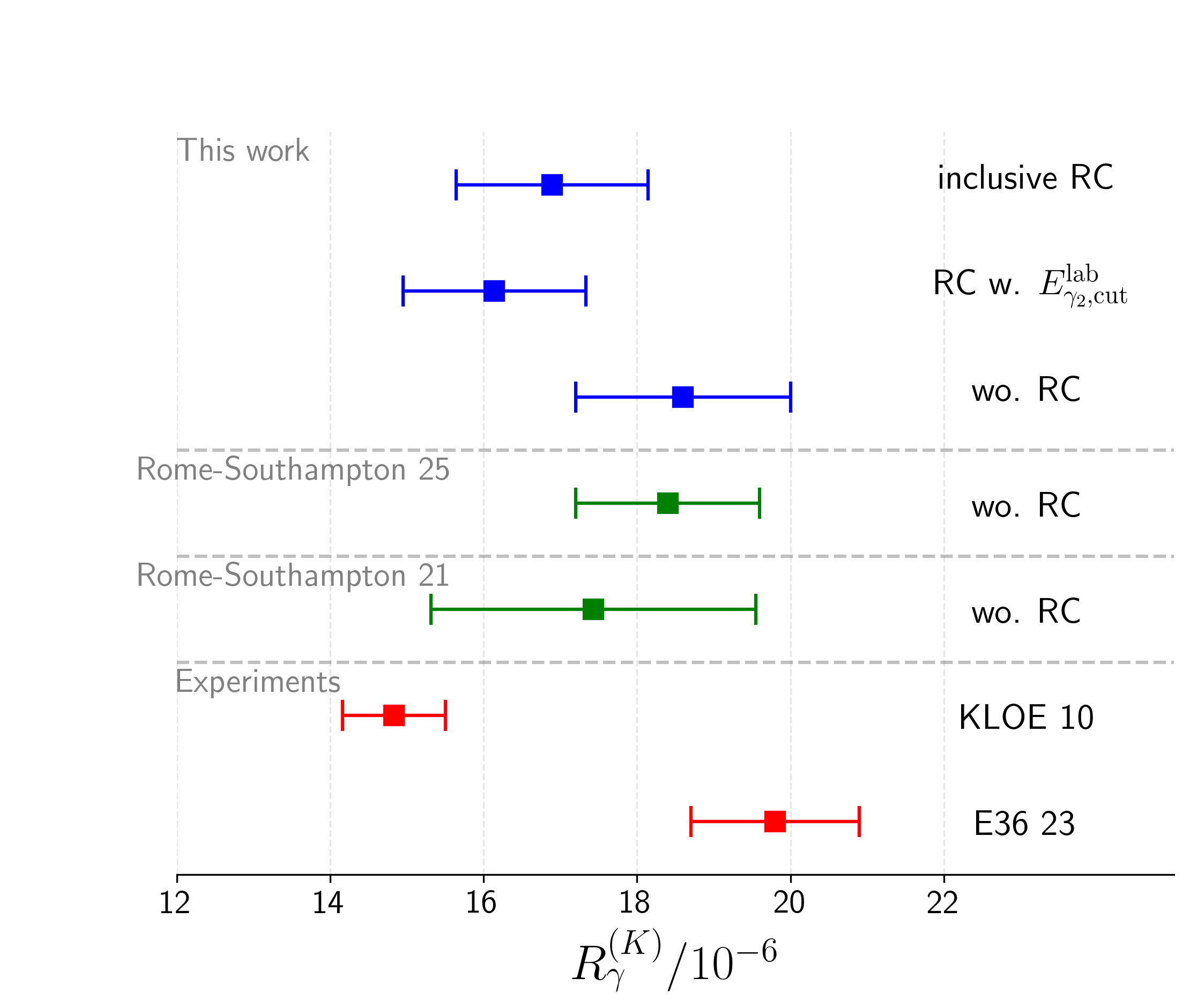}
	\caption{Comparison of the branching ratios \(R_\gamma^{(K)}\) for region 1-5 from this work and from the lattice calculations of Ref.~\cite{Frezzotti:2020bfa,DiPalma:2025iud}, as well as the KLOE and E36 experimental measurements~\cite{KLOE:2009urs,J-PARCE36:2022wfk}. Lattice results are presented both without radiative corrections (``wo. RC'', defined in Eq.~\ref{R0}) and with radiative corrections inclusive with respect to the second photon (``inclusive RC'', defined in Eq.~\ref{R_RC}) or with a laboratory-frame energy cut on the second photon (``RC w. $E_{\gamma_2,\text{cut}}^{\text{lab}}$'', defined in Eq.~\ref{R_RC_cut} with $\vec{p}^{\text{~lab}} = 100~$MeV and $E_{\gamma_2,\text{cut}}^{\text{lab}} = 20~$MeV). 
		\label{fig:RK_compare}
	}
\end{figure}

Table~\ref{table:RC_K} summarizes the continuum-extrapolated results with and without radiative corrections in all six phase-space regions. The results for region 1-5 are also shown in Fig.~\ref{fig:RK_compare}. For comparison, we also present the lattice results without radiative corrections from Refs.~\cite{Frezzotti:2020bfa,DiPalma:2025iud}, as well as the KLOE and E36 experimental measurements~\cite{KLOE:2009urs,J-PARCE36:2022wfk}. As shown in Table~\ref{table:RC_K} and Fig.~\ref{fig:RK_compare}, although different lattice actions and computational methods are adopted, our results without radiative corrections are in good agreement with those reported in Refs.~\cite{Frezzotti:2020bfa,DiPalma:2025iud}. Whether or not an energy cut is imposed, the $O(\alpha^2 L_e)$ radiative corrections are non-negligible compared to the statistical uncertainties. Although we adopt a simplified assumption of an angle-independent laboratory-frame energy cut, our lattice results with $E_{\gamma_2,\text{cut}}^{\text{lab}} = 20$~MeV are consistent with the KLOE measurements within $1\sigma$ across all phase-space regions. Our lattice result inclusive with respect to the second photon in region 1–5 shows a $1.7\sigma$ tension with the E36 measurement.

\begin{table}
	\centering
	\begin{tabular}{c|ccc|c|c|c}
		\hline \hline 
		Region & \multicolumn{3}{c|}{This work} 
        &Rome-Southampton 25~\cite{DiPalma:2025iud} 
        & KLOE~\cite{KLOE:2009urs} 
        & E36~\cite{J-PARCE36:2022wfk}  
        \\
		& wo. RC & inclusive RC & RC w. $E_{\gamma_2,\text{cut}}^{\text{lab}}$ & wo. RC & & \\
		\hline
		1   &  1.309(18)  & 1.057(17)  & 1.043(17)  &1.31(2)  &  0.94(30)  & -- \\
		2   &  2.68(20)   & 2.46(19)  & 2.34(19)    & 2.69(16)  &  2.03(22)  & -- \\
		3   &  5.59(45)   & 5.12(41) & 4.88(41)   & 5.60(36)   & 4.47(30)  & -- \\
		4   &  6.23(50)   & 5.69(46) & 5.43(46)   & 6.13(46)   & 4.81(37)  & -- \\
		5   &  2.84(22)   & 2.59(20) & 2.45(20)   & 2.69(24)   & 2.58(26)  & -- \\
		1-5 & 18.6(1.4)  & 16.9(1.3)   & 16.1(1.3)   & 18.4(1.2)  & 14.83(67) & 19.8(1.1) \\
		\hline \hline
	\end{tabular}
	\caption{Comparison of the branching ratios \(R_\gamma^{(K)}\) from this work and from the lattice calculation of Ref.~\cite{DiPalma:2025iud}, as well as the KLOE and E36 experimental measurement~\cite{KLOE:2009urs,J-PARCE36:2022wfk}. Lattice results are shown for three cases: without radiative corrections (``wo. RC'' in Eq.~(\ref{R0})), with radiative corrections inclusive with respect to the second photon (``inclusive RC'', defined in Eq.~(\ref{R_RC})), and with radiative corrections with a laboratory-frame energy cut on the second photon (``RC w. $E_{\gamma_2,\text{cut}}^{\text{lab}}$'', defined in Eq.~(\ref{R_RC_cut}) with $\vec{p}^{~\text{lab}} = 100$~MeV and $E_{\gamma_2,\text{cut}}^{\text{lab}} = 20$~MeV). All values are given in units of \(10^{-6}\). The results include only statistical uncertainties. \label{table:RC_K}}
\end{table}

\subsection{Results for $K\to \mu\nu_\mu\gamma$}
For the $K\to \mu\nu_\mu\gamma$ decay channel, radiative corrections are free of large logarithmic enhancements and can be neglected. Therefore, we only show the results without radiative corrections, as defined in Eq.~\eqref{R0}.

\begin{figure} 
	\centering
	\includegraphics[width=0.75\textwidth]{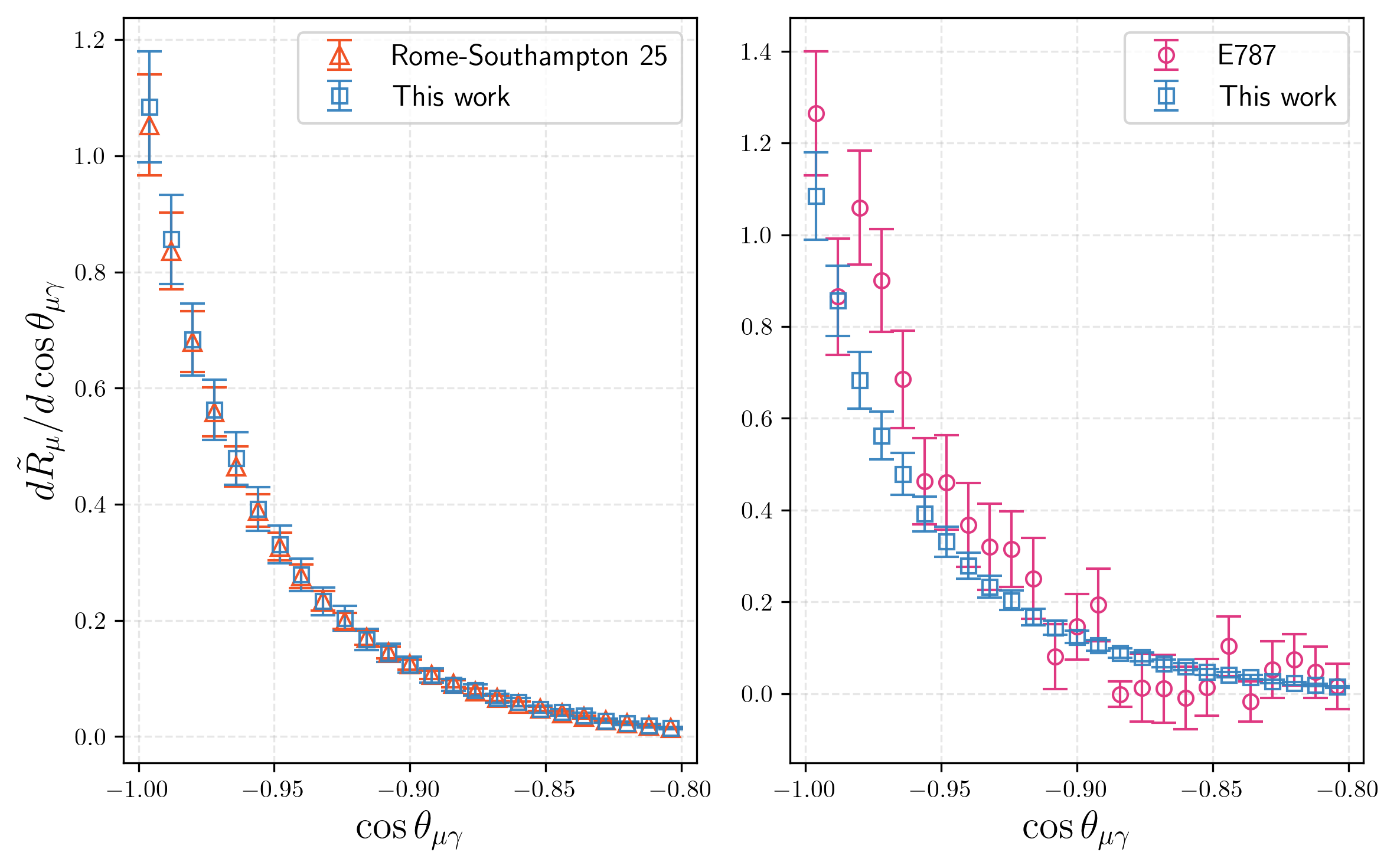}
	\caption{Continuum-extrapolated values of the differential branching ratio $\tfrac{d \tilde{R}_\mu}{d \cos \theta_{\mu \gamma}}$ in the E787 phase-space regions. The left panel compares our results with the lattice calculation of Ref.~\cite{DiPalma:2025iud}, while the right panel compares them with the E787 experimental measurements~\cite{E787:2000ehx}, whose values are given in Ref.~\cite{Frezzotti:2020bfa}.\label{fig:cont_lim_Kmunug_E787}}
\end{figure}
For the phase-space region of the E787 experiment, specified in Appendix~\ref{sec:phasespace}, 
we calculate the differential branching ratio with the inner-bremsstrahlung contribution subtracted, 
following Refs.~\cite{Frezzotti:2020bfa,DiPalma:2025iud}: 
\begin{equation}
    \frac{d \tilde{R}_\mu}{d \cos \theta_{\mu \gamma}}
    = \frac{1}{\Gamma^{(0)}\!\left[K\to\mu\nu_\mu\right]}
      \left(
        \frac{d \Gamma(K \rightarrow \mu \nu_\mu \gamma)}{d \cos \theta_{\mu \gamma}}
        - \frac{d \Gamma^{\text{IB}}(K \rightarrow \mu \nu_\mu \gamma)}{d \cos \theta_{\mu \gamma}}
      \right).
\end{equation}
This quantity is evaluated at the fixed photon–muon angles $\theta_{\mu\gamma}$ given in Table~\ref{table:PSK_E787}. The continuum-extrapolated results are shown in Fig.~\ref{fig:cont_lim_Kmunug_E787}. In the left panel of Fig.~\ref{fig:cont_lim_Kmunug_E787}, we compare our results with those from the lattice calculation in Ref.~\cite{DiPalma:2025iud}. Despite using different lattice actions and computational methods, the two lattice results are consistent with each other. In the right panel, we compare with the E787 experimental measurements~\cite{E787:2000ehx}, whose numerical values have been tabulated in Table V of Ref.~\cite{Frezzotti:2020bfa}. Our results confirm the previously observed tension between lattice predictions and the E787 data in regions with large muon–photon angles.

\begin{figure} 
	\centering
	\includegraphics[width=0.85\textwidth]{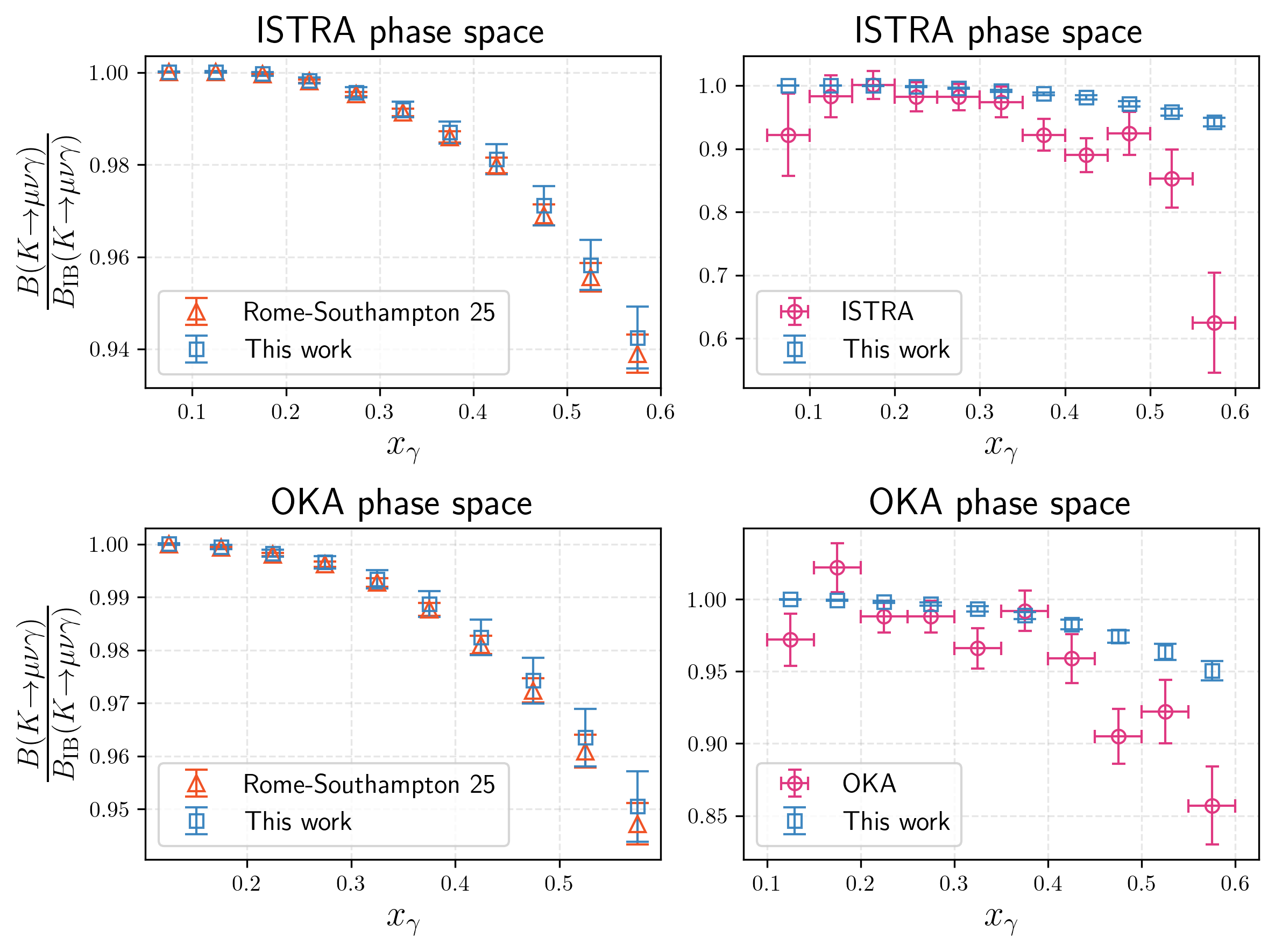}
	\caption{Continuum extrapolated values of the ratio $B(K\to\mu\nu_\mu\gamma)/B_{\text{IB}}(K\to\mu\nu_\mu\gamma)$ in the ISTRA and OKA phase-space regions. For comparison, the results from lattice calculation of Ref.~\cite{DiPalma:2025iud} and the ISTRA and OKA experimental measurements~\cite{OKA:2019gav,ISTRA:2010smy} are also shown. \label{fig:cont_lim_Kmunug}}
\end{figure}
For the ISTRA and OKA phase-space regions listed in Table~\ref{table:PSKmu}, 
the continuum-extrapolated results of the ratio 
\(
B(K\to\mu\nu_\mu\gamma)/B_{\mathrm{IB}}(K\to\mu\nu_\mu\gamma) 
\) are presented in Fig.~\ref{fig:cont_lim_Kmunug}. 
Fig.~\ref{fig:cont_lim_Kmunug} compares our results with the lattice calculation of Ref.~\cite{DiPalma:2025iud} 
and with the ISTRA and OKA experimental measurements~\cite{OKA:2019gav,ISTRA:2010smy}, 
shown in the left and right panels, respectively. 
The numerical values of the experimental data are given in Table~VIII of Ref.~\cite{Frezzotti:2020bfa}. 
The two lattice results are consistent with each other across all phase-space regions, 
but both deviate from the ISTRA and OKA measurements at large 
$x_\gamma = 2E_\gamma/m_K$.

\subsection{Results for Form Factors}
In this subsection, we present our lattice results for the form factors \(F_{V/A}(x_\gamma)\). For real photon emissions, these form factors are commonly parametrized by a linear expansion:
\begin{equation}
    F_{V/A}(x_\gamma)=F_{V/A}(1)\left[1+\lambda_{V/A}(1-x_\gamma)\right],
\end{equation}
where the \(\lambda_{V/A}\) are the slopes of form factors. In our work, Eq.~\eqref{FV} and Eq.~\eqref{FA} enable us to determine the form factors over the phase space for arbitrary values of \(x_\gamma\), thus we can directly check the linear behavior of form factors.

\begin{figure} 
	\centering
	\includegraphics[width=0.9\textwidth]{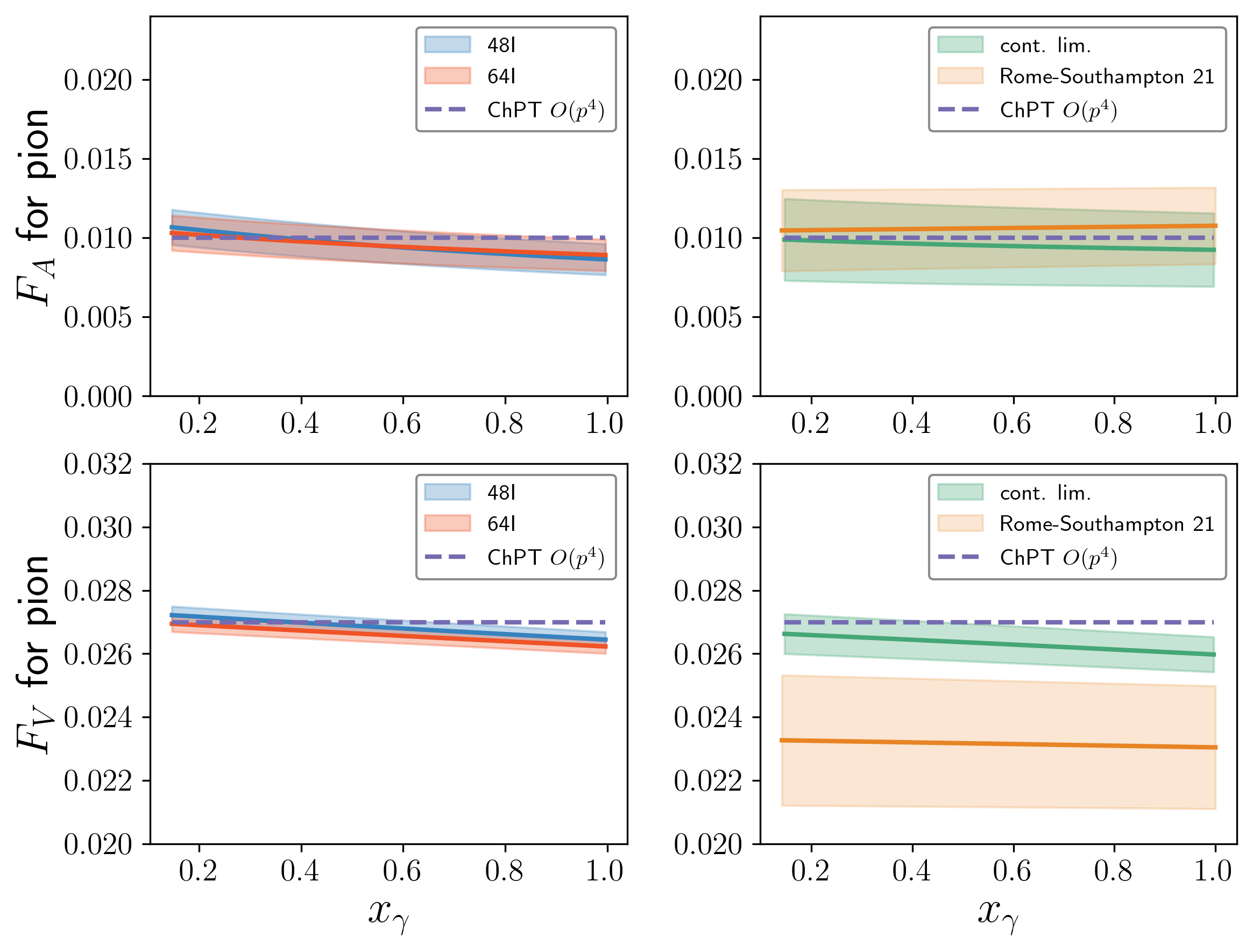}
	\caption{Form factors \(F_A\) and \(F_V\) for the pion. The left panels show the results in the 48I and 64I ensembles. The right panels show the continuum-extrapolated results. The dashed lines representing the \(O(p^4)\) $\chi$PT predictions~\cite{Unterdorfer:2008zz}. For comparison, we also show the results of previous lattice calculations in Ref.~\cite{Desiderio:2020oej} (denoted as ``Rome-Southampton 21''). We choose 128 uniformly spaced values of \(x_\gamma\) selected within the phase space region \(O\) to calculate form factors. \label{fig:fAVpi}}
\end{figure}

\begin{figure} 
	\centering
	\includegraphics[width=0.9\textwidth]{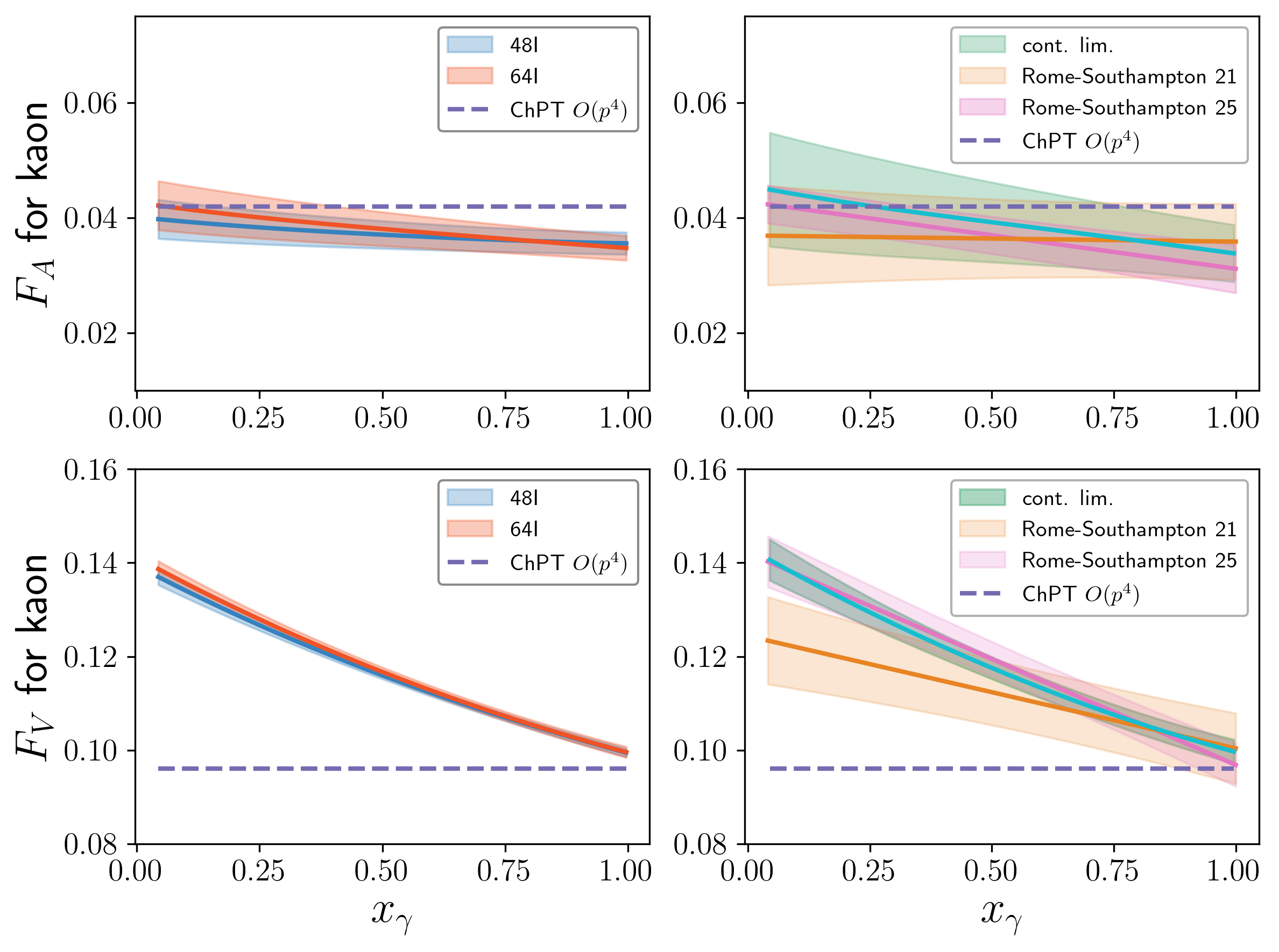}
	\caption{Form factors \(F_A\) and \(F_V\) for the kaon. The left panels show the results in the 48I and 64I ensembles. The right panels show the continuum-extrapolated results. The dashed lines representing the \(O(p^4)\) $\chi$PT predictions~\cite{Bijnens:1992en}. For comparison, we also show the results of previous lattice calculations in Ref.~\cite{Desiderio:2020oej} (denoted as ``Rome-Southampton 21'') and Ref.~\cite{DiPalma:2025iud} (denoted as ``Rome-Southampton 25''). We choose 128 uniformly spaced values of 
    \(x_\gamma\) selected within the phase space region 1-5 to calculate form factors. \label{fig:fAVK}}
\end{figure}

Fig.~\ref{fig:fAVpi} and Fig.~\ref{fig:fAVK} show our lattice results for the form factors for the pion and kaon, respectively. In the left panels, we show the lattice results obtained in the 48I (blue band) and 64I (red band) ensembles, with 128 uniformly spaced values of \(x_\gamma\) selected within the phase space region \(O\) for \(\pi\) decay and regions \(1\)–\(5\) for \(K\) decay. The dashed lines representing the \(O(p^4)\) $\chi$PT predictions. The right panels show the continuum-extrapolated results (green band). For comparison, we also show the results of previous lattice calculations in Ref.~\cite{Desiderio:2020oej,Frezzotti:2020bfa} (denoted as ``Rome-Southampton 21'') and Ref.~\cite{DiPalma:2025iud} (denoted as ``Rome-Southampton 25''). In their papers, the form factors were expanded around \(x_\gamma=0\) as a linear function \(F_{V/A}=C_{V/A}+D_{V/A}\,x_\gamma\). Using the fitted results of \(C_{V/A}\) and \(D_{V/A}\) along with their correlation matrices as given in Ref.~\cite{Frezzotti:2020bfa,DiPalma:2025iud}, we generate the statistical samples with the same distributions and then reconstruct their results of $F_{V/A}(x_\gamma)$.

From these figures, we observe that the form factors \(F_A\) and \(F_V\) for the pion and \(F_A\) for the kaon have an approximately linear dependence on \(x_\gamma\). The corresponding lattice results are also in agreement with the $O(p^4)~\chi$PT predictions. On the other hand, \(F_V\) for the kaon shows a mild deviation from linearity, suggesting the emergence of a pole-like behavior due to low-lying resonances. The lattice results for kaon \(F_V\) are larger than the \(O(p^4)\) \(\chi\)PT result for small $x_\gamma$.

Our lattice results are consistent with those reported by the Rome-Southampton collaboration within statistical uncertainties. Compared with the recent lattice calculation of the kaon form factors \(F_V\) and \(F_A\) in Ref.~\cite{DiPalma:2025iud} (denoted as ``Rome-Southampton 25''), our statistical uncertainty is larger for \(F_A\), while that for \(F_V\) is comparable. This difference may arise from a combination of factors, including the number of configurations, the ensembles used, the lattice actions, and the analysis strategies. Nevertheless, the agreement between the two calculations is highly nontrivial and provides an important cross-check of the lattice results.

To provide more lattice QCD inputs for phenomenology, we perform a linear fit of the form factors, as summarized in Table~\ref{table: ren1}:
\begin{equation}
	F_{i}(x_\gamma)=F^{\text{fit}}_{i}(1)\left(1+\lambda^{\text{fit}}_{i}(1-x_\gamma)\right),
\end{equation}
where $i \in \{V, A, +, -\}$, and $F_{\pm}(x_\gamma)$ are defined as $F_V(x_\gamma) \pm F_A(x_\gamma)$. The fit parameters are $F^{\text{fit}}_{i}(1)$ and $\lambda^{\text{fit}}_{i}$. We also list the results obtained directly from the lattice calculations at $x_\gamma=1$. Since the linear behavior holds well for the form factors, $F_{i}(1)$ and the fitted $F^{\text{fit}}_{i}(1)$ are consistent within the statistical errors. For the kaon $F_V(1)$ (and thus kaon $F_{\pm}(1)$), the central values obtained from the direct calculation and from the linear fit show a slight difference, reflecting a mild deviation from linearity as seen in Fig.~\ref{fig:fAVK}.

\begin{table}[htbp]
\centering
\label{table:ren1}

\begin{minipage}[t]{1\textwidth}
\centering
\textbf{Direct calculation at $x_\gamma=1$}\\[3pt]
\begin{tabular}{c|cccc}
\hline\hline
 meson& $F_V(1)$ & $F_A(1)$ & $F_{+}(1)$ & $F_{-}(1)$ \\
\hline
$\pi$ & 0.02598(55) & 0.0092(23) & 0.0352(23) & 0.0167(25) \\
$K$   & 0.0997(26)  & 0.0338(50) & 0.1335(55) & 0.0659(57) \\
\hline\hline
\end{tabular}
\end{minipage}
\vspace{1em}

\begin{minipage}[t]{1\textwidth}
\centering
\textbf{Linear-fit results}\\[3pt]
\begin{tabular}{c|cccccccc}
\hline\hline
 meson& $F^{\text{fit}}_V(1)$ & $F^{\text{fit}}_A(1)$ & $F^{\text{fit}}_+(1)$ & $F^{\text{fit}}_-(1)$ &  $\lambda_V^{\text{fit}}$ & $\lambda_A^{\text{fit}}$&  $\lambda_+^{\text{fit}}$ & $\lambda_-^{\text{fit}}$ \\
\hline
$\pi$ & 0.02598(55) & 0.0092(23) & 0.0352(23)& 0.0168(25)& 0.0295(32) & 0.080(95)& 0.043(20)& 0.002(44)\\
$K$   & 0.0975(25)  & 0.0338(48) & 0.1314(52)& 0.0640(55)& 0.430(49)  & 0.33(20)  &0.402(59)&0.48(12)\\
\hline\hline
\end{tabular}
\end{minipage}
\caption{Results from linear fits of the form factors \(F_{V/A}(x_\gamma)\) and the combinations $F_{\pm}(x_\gamma)=F_V(x_\gamma) \pm F_A(x_\gamma)$, obtained using the form \(F^{\text{fit}}_{i}(1)\left(1+\lambda^{\text{fit}}_{i}(1-x_\gamma)\right)\), together with the values calculated directly at $x_\gamma=1$, \(F_{V/A}(1)\) and \(F_{\pm}(1) = F_V(1) \pm F_A(1)\).
    \label{table: ren1}}
\end{table}

\section{Conclusions\label{sec5}}
In this work, we perform a lattice QCD calculation of the radiative leptonic decays $P \to \ell \nu_\ell \gamma$ ($P = \pi, K$) using domain-wall fermion ensembles generated by the RBC and UKQCD collaborations at the physical pion mass. We employ the infinite-volume reconstruction (IVR) method, which extends finite-volume lattice data to infinite volume and effectively controls the finite-volume effects. We compute the branching ratios for the decays \(\pi \to e\nu_e\gamma\), \(K \to e\nu_e\gamma\), and \(K \to \mu\nu_\mu\gamma\) with or without radiative corrections, and compare them with the previous lattice calculations~\cite{Desiderio:2020oej,Frezzotti:2020bfa,DiPalma:2025iud} and experimental measurements~\cite{Bychkov:2008ws,KLOE:2009urs,J-PARCE36:2022wfk,E787:2000ehx,ISTRA:2010smy,OKA:2019gav}. We also determine the momentum dependence of the vector and axial-vector form factors, \(F_V(x_\gamma)\) and \(F_A(x_\gamma)\), across the full phase space, providing insight into the hadronic structure of light mesons. The IVR method allows us to evaluate the form factors at all chosen values of $x_\gamma$. 

Despite adopting a different lattice action and computational strategy, our results for the form factors and branching ratios without radiative corrections are in good agreement with those reported in Ref.~\cite{Desiderio:2020oej,Frezzotti:2020bfa,DiPalma:2025iud} within statistical uncertainties, demonstrating the consistency of lattice QCD calculations. For \(K \to \mu\nu_\mu\gamma\) decays, for which collinear effects are negligible, our lattice results in the regions of phase space in which the E787, ISTRA and OKA experiments reported measurements are consistent with those of the Rome-Southampton collaboration and confirm the previously observed tension between lattice calculations and the ISTRA and OKA experimental measurements at large photon energies, as well as the deviation from the E787 results at large angles between the muon and the photon~\cite{Frezzotti:2020bfa,DiPalma:2025iud}. 

For decays involving a final-state electron, we emphasize the importance of collinear radiative corrections, which are significantly enhanced by the large logarithmic factors and therefore cannot be neglected at the current level of experimental precision~\cite{Kuraev:2003gq}.
In the case of \(\pi \to e\nu_e\gamma\) decays, the inclusion of radiative corrections resolves the previously observed discrepancy between lattice predictions without such corrections~\cite{Frezzotti:2020bfa} and the PIBETA measurement~\cite{Bychkov:2008ws}.
For $K\to e\nu_e\gamma$ decays, we note that the different treatments of additional inner-bremsstrahlung photons in the event selection procedures of the KLOE and E36 experiments lead to different radiative corrections, which may account for part of the observed $4\sigma$ discrepancy between their measurements. Regardless of whether an energy cut is imposed on the second photon, the radiative correction remains larger than \(O(10\%)\).  Although we adopt a simplified assumption of an angle-independent laboratory-frame energy cutoff, \(E_{\gamma_2,\text{cut}}^{\text{lab}} = 20~\text{MeV}\), with $\vec{p}^{~\text{lab}}=100~\mathrm{MeV}$, our lattice predictions including \(O(\alpha^2 L_e)\) radiative corrections are consistent with the KLOE measurements. The lattice prediction with the fully inclusive radiative correction with respect to the second photon has $1.7\sigma$ tension from the E36 result.

Further improvements of the present calculation will require additional ensembles at different volumes and lattice spacings. For finite-volume effects, the difference between the results obtained with \(\delta^{\mathrm{IVR}}_{\mathrm{pt}}\) and \(\delta^{\mathrm{IVR}}_{\mathrm{SD}}\), which probes the structure-dependent part of the correction, is much smaller than the current statistical uncertainty. However, finite-volume effects from heavier excited states have not been estimated and will require future calculations on additional ensembles at multiple volumes. In this work we use two lattice spacings for the continuum extrapolation; a reliable assessment of the residual discretization errors will require calculations at a third lattice spacing. In addition, disconnected-diagram contributions, which were included in the calculation of Ref.~\cite{DiPalma:2025iud}, are not included here and are left for future work.

The application of the infinite-volume reconstruction (IVR) method to the full radiative corrections in leptonic decays, including contributions from virtual-photon loop diagrams, is currently in progress. Although restricting to transverse photon polarizations significantly suppresses ground-state finite-volume effects, leading to small IVR corrections for the real-photon emissions studied in this work, the IVR method becomes essential when computing the complete radiative corrections to leptonic decays. In such cases, inclusion of all photon polarization states in real-photon emissions is required to achieve infrared cancellation with virtual-photon loop contributions in the Feynman gauge. 
The IVR method reduces the power-law suppressed finite-volume effects associated with the photon propagator to exponentially suppressed ones~\cite{Christ:2023lcc}, thereby improving very significantly the precision of existing lattice QCD calculations of radiative corrections \cite{Giusti:2017dwk, DiCarlo:2019thl, Boyle:2022lsi}. 
Such improvements are crucial for more precise determinations of the CKM matrix elements \(V_{us}\) and \(V_{ud}\), and for tests of the first-row unitarity of the CKM matrix. 

Furthermore, the methods developed in this work can also be applied to processes involving virtual photon emission, such as \(P \to \ell\nu_\ell\gamma^*\to \ell\nu_\ell\ell'\ell'\), offering additional insights into the hadronic structure of mesons and enabling precision tests of the Standard Model through comparisons with experimental measurements.
In these decays, the additional challenge is to properly account for the power-law suppressed finite-volume effects arising from the intermediate states \(P \to \ell\nu_\ell\pi\pi \to \ell\nu_\ell\gamma^*\). To address this issue, the power-law finite-volume corrections and exponentially growing terms associated with low-lying $\pi\pi$ states has been investigated in Ref.~\cite{Tuo:2024bhm}.

\acknowledgements
We would like to thank our RBC and UKQCD Collaboration colleagues for helpful discussions and support. 
We also thank Suguru Shimizu from the E36 collaboration and Angela Romano, Evgueni Goudzovski and Tomas Husak from the NA62 collaboration for clarifying remarks concerning their treatment of possible additional photons in the final state.
P.B. and T.I. were supported in part by US DOE Contract DESC0012704(BNL) and the Scientific Discovery
through Advanced Computing (SciDAC) program LAB
22-2580. X.F. has been supported in part by NSFC of China under Grant No. 12125501. 
L.J. acknowledge the support of DOE Office of Science Early Career Award DE-SC0021147 and DOE
grant DE-SC0010339, DE-SC0026314. C.T.S. is partially supported by STFC consolidated grant ST/X000583/1.
X.Y.T has been supported by US DOE Contract
DESC0012704(BNL).

\newpage
\appendix
\section{Definition of the regions of phase space considered in the experiments\label{sec:phasespace}}
For the $\pi \to e\nu_e\gamma$ decay channel, the regions of phase space for which the results are presented by the PIBETA experiment~\cite{Bychkov:2008ws} are defined by cuts on the electron and photon energies, $E_e = m_\pi y/2$ and $E_\gamma = m_\pi x/2$ respectively as shown in 
Table~\ref{table:PSpi}. Here the energies are defined in the rest frame of the pion. This experiment also requires that the angle between the final-state electron and photon satisfies $\theta_{e\gamma} > 40^\circ$.

\begin{table}
	\centering
	\begin{tabular}{c|cccc}
		\hline \hline  
		Regions & A & B& C& O \\
		\hline 
		$E_{e,\text{cut}}~[\text{MeV}]$ & $50$ &  $10$ & $50$ & $m_e$ \\
		$E_{\gamma,\text{cut}}~[\text{MeV}]$ & $50$ &  $50$ &$10$ &$10$ \\
		\hline \hline
	\end{tabular}
	\caption{\label{table:PSpi}Phase space regions for the $\pi\to e\nu_e\gamma$ process from the PIBETA experiment~\cite{Bychkov:2008ws}. The angle between the final-state electron and photon satisfies $\theta_{e\gamma} > 40^\circ$. Four phase-space regions are defined by $E_e = m_\pi y_e/2>E_{e,\text{cut}}$ and $E_\gamma = m_\pi x_\gamma/2>E_{\gamma,\text{cut}}$. Here the energies are defined in the rest frame of the pion.}
\end{table}

For the $K \to e\nu_e\gamma$ decay channel, we adopt the phase space definitions in the KLOE and E36 experiments~\cite{KLOE:2009urs,J-PARCE36:2022wfk}. As summarized in Table~\ref{table:PSK}, the experiments require the electron momentum to satisfy $p_e>200~\text{MeV}$ and the photon energy to be greater than 10\,MeV, $E_\gamma>10~\text{MeV}$. Here the photon energy is defined in the rest frame of the kaon. The phase space is divided into five regions corresponding to different ranges of $E_\gamma$. The KLOE experiment reports branching ratios in all five phase space regions, while the E36 experiment provides the total branching ratio in the ``1-5'' region.
\begin{table}
	\centering
	\begin{tabular}{c|cccccc}
		\hline \hline  
		Regions & 1 &2 & 3& 4& 5 &1-5 \\
		\hline 
		$[E^{\text{min}}_\gamma,E^{\text{max}}_\gamma]~[\text{MeV}]$ & $[10,50]$ & $[50,100]$ & $[100,150]$ & $[150,200]$& $[200,250]$& $[10,250]$ \\
		\hline \hline
	\end{tabular}
	\caption{\label{table:PSK}Phase space regions for the $K\to e\nu_e\gamma$ decay from KLOE and E36 experiments~\cite{KLOE:2009urs,J-PARCE36:2022wfk}, with the electron momentum $p_e>200~\text{MeV}$. The phase space is divided into five regions corresponding to different photon energy ranges of $E_\gamma^{\text{min}}<E_\gamma<E_\gamma^{\text{max}}$. Here the photon energy is defined in the rest frame of the kaon. The KLOE experiment reports branching ratios in all five phase space regions, while the E36 experiment provides the total branching ratio in the ``1-5'' region.}
\end{table}

For the $K \to \mu\nu_\mu\gamma$ decay channel, we focus on the ISTRA, OKA and E787 experiments~\cite{ISTRA:2010smy,OKA:2019gav,E787:2000ehx}. These experiments present measurements in multiple phase-space regions, whose definitions are provided in Ref.~\cite{Frezzotti:2020bfa,DiPalma:2025iud}. In this work, we adopt the same phase-space regions defined in Tables V, VI and VII of Ref.~\cite{Frezzotti:2020bfa}. For the ISTRA and OKA experiments, the phase-space regions are defined by the constraints 
$x_\gamma^{\text{min}} < x_\gamma < x_\gamma^{\text{max}}$, 
$y_\mu^{\text{min}} < y_\mu < y_\mu^{\text{max}}$, 
and on the photon–muon angle, $\cos\theta_{\mu\gamma} > \cos\theta_{\mu\gamma}^{\text{cut}}$, 
as summarized in Table~\ref{table:PSKmu}. 
For the E787 experiment, the differential decay rate
\begin{equation}
    \frac{d \Gamma(K \rightarrow \mu \nu_\mu \gamma)}{d \cos \theta_{\mu \gamma}}
    = \int_{\tfrac{2 E_\gamma^{\text{cut}}}{m_K}}^{1-r_\mu} d x_\gamma 
      \int_{\max \!\left(y_\mu^{\min}, \tfrac{2E_\mu^{\text{cut}}}{m_K}\right)}^{1+r_\mu} 
      d y_\mu \,
      \left[\frac{d^2 \Gamma}{d x_\gamma d y_\mu}\right]
      \delta\!\left(\cos \theta_{\mu \gamma}-\cos \theta_{\mu \gamma}(x_\gamma,y_\mu)\right)
\end{equation}
is measured at fixed photon–muon angles $\theta_{\mu\gamma}$ listed in Table~\ref{table:PSK_E787}, together with the constraints $E_\gamma^{\text{cut}}>90~\text{MeV}$ and $E_\mu^{\text{cut}}>m_\mu+137~\text{MeV}$. Here, $y_\mu^{\text{min}}=1-x_\gamma+r_\mu/(1-x_\gamma)$ is given in Eq.~\eqref{phasespace} and $\cos \theta_{\mu \gamma}(x_\gamma,y_\mu)$ is defined as
\begin{equation}
    \cos \theta_{\mu \gamma}(x_\gamma,y_\mu)=\frac{x_\gamma y_\mu-2(x_\gamma+y_\mu-r_\mu-1)}{x_\gamma\sqrt{y_\mu^2-4r_\mu}}.
\end{equation}

\begin{table}
\centering
\begin{minipage}{0.48\textwidth}
\centering
\begin{tabular}{c|ccc}
\hline\hline
Regions & $[x_\gamma^{\text{min}},x_\gamma^{\text{max}}]$ & $[y_\mu^{\text{min}},y_\mu^{\text{max}}]$ & $\cos(\theta^{\text{cut}}_{\mu\gamma})$ \\
\hline
01 & $[0.05,0.10]$ & $[0.90,1.10]$ & -0.8 \\
02 & $[0.10,0.15]$ & $[0.90,1.10]$ & -0.8 \\
03 & $[0.15,0.20]$ & $[0.85,1.00]$ & -0.8 \\
04 & $[0.20,0.25]$ & $[0.80,0.95]$ & -0.2 \\
05 & $[0.25,0.30]$ & $[0.75,0.90]$ & -0.3 \\
06 & $[0.30,0.35]$ & $[0.72,0.87]$ & -0.4 \\
07 & $[0.35,0.40]$ & $[0.65,0.85]$ & -0.3 \\
08 & $[0.40,0.45]$ & $[0.62,0.85]$ & -0.5 \\
09 & $[0.45,0.50]$ & $[0.57,0.80]$ & -0.7 \\
10 & $[0.50,0.55]$ & $[0.52,0.75]$ & -1.0 \\
11 & $[0.55,0.60]$ & $[0.48,0.70]$ & -1.0 \\
\hline \hline
\end{tabular}
\end{minipage}
\hfill
\begin{minipage}{0.48\textwidth}
\centering
\begin{tabular}{c|ccc}
\hline\hline
Regions & $[x_\gamma^{\text{min}},x_\gamma^{\text{max}}]$ & $[y_\mu^{\text{min}},y_\mu^{\text{max}}]$ & $\cos(\theta^{\text{cut}}_{\mu\gamma})$\\
\hline
01 & $[0.10,0.15]$ & $[0.89,1.01]$ & -0.8 \\
02 & $[0.15,0.20]$ & $[0.85,1.01]$ & -0.2 \\
03 & $[0.20,0.25]$ & $[0.80,1.00]$ & -0.2 \\
04 & $[0.25,0.30]$ & $[0.75,0.97]$ & -0.4 \\
05 & $[0.30,0.35]$ & $[0.70,0.93]$ & -0.4 \\
06 & $[0.35,0.40]$ & $[0.66,0.90]$ & -0.5 \\
07 & $[0.40,0.45]$ & $[0.62,0.88]$ & -0.5 \\
08 & $[0.45,0.50]$ & $[0.58,0.86]$ & -0.6 \\
09 & $[0.50,0.55]$ & $[0.54,0.83]$ & -0.6 \\
10 & $[0.55,0.60]$ & $[0.50,0.80]$ & -0.6 \\
\\
\hline\hline
\end{tabular}
\end{minipage}
\caption{\label{table:PSKmu}Phase space regions for the $K\to \mu\nu_\mu\gamma$ decay channel from ISTRA experiment (left table) and OKA experiment (right table)~\cite{ISTRA:2010smy,OKA:2019gav}. The definition of these regions are given in Table VI and Table VII in Ref.~\cite{Frezzotti:2020bfa}. Each phase-space region is defined by $x_\gamma^{\text{min}}<x_\gamma<x_\gamma^{\text{max}}$, $y_\mu^{\text{min}}<y_\mu<y_\mu^{\text{max}}$, and $\cos(\theta_{\mu\gamma})>\cos(\theta_{\mu\gamma}^{\text{cut}})$.}
\end{table}

\begin{table}
    \centering
    \begin{tabular}{c|ccccccccc}
        \hline\hline
        $\cos\theta_{\mu\gamma}$ 
        & $-0.996$ & $-0.988$ & $-0.980$ & $-0.972$ & $-0.964$ 
        & $-0.956$ & $-0.948$ & $-0.940$ & $-0.932$ \\
        & $-0.924$ & $-0.916$ & $-0.908$ & $-0.900$ & $-0.892$ 
        & $-0.884$ & $-0.876$ & $-0.868$ & $-0.860$ \\
        & $-0.852$ & $-0.844$ & $-0.836$ & $-0.828$ & $-0.820$ 
        & $-0.812$ & $-0.804$ \\
        \hline\hline
    \end{tabular}
    \caption{\label{table:PSK_E787} 
     The photon–muon angles $\theta_{\mu\gamma}$ used in the E787 experiment~\cite{E787:2000ehx}. These angles are given in Table~V of Ref.~\cite{Frezzotti:2020bfa}. 
    The phase space in the E787 experiment is also constrained by $E_\gamma^{\text{cut}}>90~\text{MeV}$ and $E_\mu^{\text{cut}}-m_\mu>137~\text{MeV}$.
    }
\end{table}

\section{Radiative Corrections in $P\to e\nu_e \gamma$ Decays\label{Appendix:RC}}
In this section we review the derivation of the expressions for the collinear radiative corrections at $O(\alpha^2)$ in $P\to\ell\nu_\ell\gamma(\gamma)$ decays which are enhanced by large logarithms. We start in Sec.\,\ref{subsec:inclusive} by considering the case in which all events with a possible second photon are included, corresponding, for example, to the experimental results of the PIBETA and E36 experiments. In Sec.\,\ref{subsec:laboratory} we consider an idealized case in which there is a cut-off on the energy of a possible second photon in the laboratory frame which is independent of the direction of its momentum. In more realistic situations, in which such a cut-off depends on the angle of emission and on details of detector acceptances, the experimental collaborations would have to implement the corrections themselves. Our results in Sec.\,\ref{subsec:laboratory} are indicative of the likely magnitudes of such corrections.

\subsection{Case (i): Inclusive with Respect to the Second Photon}\label{subsec:inclusive}
We first review the radiative corrections to the decay $P\to e\nu_e \gamma(\gamma)$, which is inclusive with respect to the second photon~\cite{Kuraev:2003gq}:
\begin{equation}\label{B_RC_full}
    \frac{d B^{\mathrm{RC}}[P\to e\nu_e \gamma]}{d x_\gamma\, d y_e}(x_\gamma,y_e)=\int_{y_e}^1 \frac{d t}{t} \frac{d B[P\to e\nu_e \gamma]}{d x_\gamma\, d y_e}(x_\gamma, t) D\left(\frac{y_e}{t}\right)\left(1+\frac{\alpha}{2 \pi} K(x_\gamma, y_e)\right),
\end{equation}
where $\tfrac{d B^{\mathrm{RC}}[P\to e\nu_e \gamma]}{d x_\gamma\,d y_e}(x_\gamma,y_e)$ denotes the differential branching ratio including radiative corrections at the phase-space point $(x_\gamma,y_e)$. The right-hand side is written as a convolution integral in which $\tfrac{d B[P\to e\nu_e \gamma]}{d x_\gamma\,d y_e}(x_\gamma,t)$ represents the $O(\alpha)$ 
differential branching ratio in the absence of radiative corrections at the phase space point $(x_\gamma,t)$. The convolution involves the so-called ``electron structure function'',
\begin{equation}
	D(z)=\delta(1-z)+\frac{\alpha}{2 \pi}\left(L_e-1\right) P^{(1)}(z)+\frac{1}{2!}\left(\frac{\alpha}{2 \pi}\right)^2\left(L_e-1\right)^2 P^{(2)}(z)+\cdots.
\end{equation}
Physically, \(D(z)\) is interpreted as the probability density for the electron to retain a fraction \(z\) of its original momentum after radiating one or more collinear photons. The leading term $\delta(1-z)$ corresponds to no collinear radiation, whereas the first-order term $P^{(1)}(z)$, proportional to $\frac{\alpha}{2\pi}(L_e-1)$, describes the leading collinear radiative correction:
\begin{equation}\label{P1}
P^{(1)}(z)=\lim _{\Delta \rightarrow 0}\left[\frac{1+z^2}{1-z} \theta(1-z-\Delta)+\delta(1-z)\left(2 \ln \Delta+\frac{3}{2}\right)\right]
\end{equation}
Here, $\Delta=\frac{2\Delta\epsilon}{y_e m_P}$, introduced in Ref.~\cite{Kuraev:2003gq}, separates the hard and soft regions of the second emitted photon. In the rest frame of $P$, a second emitted photon with energy $E_{\gamma_{2}}<\Delta\epsilon$ ($E_{\gamma_{2}}>\Delta\epsilon$) is classified as soft (hard). The first term in the square brackets of Eq.~\eqref{P1} represents the $O(\alpha^2 L_e)$ contribution from the hard region, whereas the second term accounts for the $O(\alpha^2 L_e)$ contribution from the soft region and the virtual diagrams. Because the total correction is independent of the hard–soft separation, we can take the limit $\Delta\to 0$ when evaluating $P^{(1)}(z)$.

The function \(K(x_\gamma,y_e)\) in Eq.~\eqref{B_RC_full} arises from \(O(\alpha^2 L_e^0)\) corrections that are not enhanced by the large logarithm \(L_e\), and is therefore neglected in our analysis. Its explicit form can be found in Ref.~\cite{Kuraev:2003gq}. On the other hand, although the constant term \(-1\) in the combination \((L_e - 1)\) is similarly not logarithmically enhanced, it is conventionally included in the definition of the electron structure function $D(z)$, and we retain it to maintain consistency with the standard formalism in Ref.~\cite{Kuraev:2003gq}. The physical origin of the $-1$ term is clarified in Ref.~\cite{Arbuzov:2019hcg}: large collinear logarithm arise from the angular integration over terms like \((p_{e,1}\cdot p_{e,2})/[(p_{e,1}\cdot k)(p_{e,2}\cdot k)]\), where \(p_{e,1}\) and \(p_{e,2}\) denote the electron momenta before and after emitting the bremsstrahlung photon, and \(k\) is the photon momentum. The contributions of the form \(m_{e}^2/[(p_{e,1}\cdot k)(p_{e,2}\cdot k)]\) always accompany these terms, effectively shifting the logarithm by unity.

By keeping only the collinear radiative correction at $O(\alpha^2 L_e)$, the convolution integral can be evaluated analytically, yielding the simplified expression:
\begin{equation}
	\frac{dB^{\mathrm{RC}}[P\to e\nu_e \gamma]}{dx_\gamma\,dy_e} = \frac{\alpha}{2\pi(1-r_e)^2}\,B[P\to e\nu_e(\gamma)] \left( A(x_\gamma,y_e) + \frac{\alpha}{2\pi}(L_e-1) A^{\mathrm{RC}}(x_\gamma,y_e) \right),
\end{equation}
where $B[P\to e\nu_e(\gamma)]$ denotes the branching ratio of $P\to e\nu_e(\gamma)$. Using $B[P\to e\nu_e(\gamma)]$ as the prefactor effectively includes the electroweak corrections to the decay constant $f_P$. This choice of prefactor is equivalent to the $A_{\text {exp.}}=e \frac{G_F}{\sqrt{2}} V_{\text {ud }} f_\pi^{\text {exp }} m_e$ used in Ref.~\cite{Kuraev:2003gq} and is adopted in the theoretical prediction of the PIBETA collaboration~\cite{Bychkov:2007pnf,Bychkov:2008ws}. Appendix B of Ref.~\cite{Kuraev:2003gq} provides the expression for $A^{\mathrm{RC}}(x_\gamma,y_e)$:
\begin{equation}\label{ARC}
	\begin{aligned}
		A^{\mathrm{RC}}(x_\gamma,y_e)=\,& f^{\mathrm{RC}}_{\mathrm{IB}}(x_\gamma,y_e)
		+\frac{1}{r_e}\left(\frac{m_P}{2f_P}\right)^2 \Bigl[\left(F_V+F_A\right)^2 f^{\mathrm{RC}}_{\mathrm{SD}^{+}}(x_\gamma,y_e) + \left(F_V-F_A\right)^2 f^{\mathrm{RC}}_{\mathrm{SD}^{-}}(x_\gamma,y_e) \Bigr] \\
		& - \left(\frac{m_P}{f_P}\right) \Bigl[\left(F_V+F_A\right) f^{\mathrm{RC}}_{\mathrm{INT}^{+}}(x_\gamma,y_e) + \left(F_V-F_A\right) f^{\mathrm{RC}}_{\mathrm{INT}^{-}}(x_\gamma,y_e) \Bigr].
	\end{aligned}
\end{equation}
Note that the form factors $(f_V,f_A)$ used in Ref.~\cite{Kuraev:2003gq} are related to $(F_V,F_A)$ defined here by $f_{V/A}=(m_P^2/2m_e f_P)F_{V/A}$. The functions $f^{\mathrm{RC}}_{i}(x_\gamma,y_e)$, with $i\in\{\mathrm{IB},\mathrm{SD}^{+},\mathrm{SD}^{-},\mathrm{INT}^{+},\mathrm{INT}^{-}\}$, are defined by
\begin{equation}
    f^{\mathrm{RC}}_{i}(x_\gamma,y_e)=\int_{y_e}^{1}\frac{dt}{t} P^{(1)}(y_e/t) f_{i}(x_\gamma,t),
\end{equation}
where the $f_i(x_\gamma,y_e)$ are defined in Eq.~\eqref{f0}. The convolution integral can be evaluated analytically in the $m_e\to 0$ limit~\cite{Kuraev:2003gq}, yielding
\begin{equation}\label{fRC}
	\begin{aligned}
		f^{\mathrm{RC}}_{\mathrm{IB}}(x_\gamma,y_e)=\,& \frac{1+\bar{x}_\gamma^2}{x_\gamma^2}\Biggl[\frac{3}{2}\frac{\bar{y}_e}{\bar{z}}+\frac{\bar{y}_e}{\bar{x}_\gamma}-\frac{\bar{x}_\gamma+x_\gamma y_e}{\bar{x}_\gamma^2}\ln y_e
		+2\,\frac{\bar{y}_e}{\bar{z}}\ln\frac{\bar{y}_e}{y_e}-\frac{x_\gamma\left(\bar{x}_\gamma^2+y_e^2\right)}{\bar{x}_\gamma^2\,\bar{z}}\ln\frac{x_\gamma}{\bar{z}}\Biggr], \\
		f^{\mathrm{RC}}_{\mathrm{SD}^+}(x_\gamma,y_e)=\,& \bar{x}_\gamma\Biggl[\frac{3}{2} \bar{z}^2+\frac{1-y_e^2}{2}+\bar{y}_e(y_e-2\bar{x}_\gamma)
		+\bar{x}_\gamma(\bar{x}_\gamma-2y_e)\ln y_e-\bar{x}_\gamma^2\bar{y}_e+2\bar{z}^2\ln\frac{\bar{y}_e}{y_e}\Biggr], \\
		f^{\mathrm{RC}}_{\mathrm{SD}^-}(x_\gamma,y_e)=\,& \bar{x}_\gamma\Biggl[\frac{3}{2}\bar{y}_e^2+\frac{1-y_e^2}{2}+\bar{y}_e(y_e-3)
		+(1-2y_e)\ln y_e+2\bar{y}_e^2\ln\frac{\bar{y}_e}{y_e}\Biggr], \\
		f^{\mathrm{RC}}_{\mathrm{INT}^+}(x_\gamma,y_e)=\,& \frac{\bar{x}_\gamma}{x_\gamma}\Biggl[\frac{\bar{y}_e}{2}-\bar{y}_e\ln y_e-2\bar{y}_e\ln\frac{\bar{y}_e}{y_e}\Biggr], \\
		f^{\mathrm{RC}}_{\mathrm{INT}^-}(x_\gamma,y_e)=\,& \frac{1}{x_\gamma}\Biggl[-\frac{1}{2}\bar{x_\gamma}\bar{y}_e+\frac{3}{2}\frac{x_\gamma^2\bar{y}_e}{\bar{z}}+\bar{x_\gamma}\Bigl(\bar{y}_e\ln y_e+2\bar{y}_e\ln\frac{\bar{y}_e}{y_e}\Bigr) \\
		&\quad + x_\gamma^2\Biggl(\frac{\bar{y}_e}{\bar{x}_\gamma}-\frac{\bar{x}_\gamma+x_\gamma y_e}{\bar{x}_\gamma^2}\ln y_e+2\,\frac{\bar{y}_e}{\bar{z}}\ln\frac{\bar{y}_e}{y_e}-\frac{x_\gamma\left(\bar{x}_\gamma^2+y_e^2\right)}{\bar{x}_\gamma^2\,\bar{z}}\ln\frac{x_\gamma}{\bar{z}}\Biggr)\Biggr].
	\end{aligned}
\end{equation}
Here the variables are defined as $\bar{z}=x_\gamma+y_e-1$, $\bar{x}_\gamma=1-x_\gamma$, and $\bar{y}_e=1-y_e$. 

Finally, we point out that terms proportional to $\ln(2E_{\gamma_2,\text{max}}^*/m_P)=\ln(1-y_e)$ always appear in the ratio $f_i^{\mathrm{RC}}(x_\gamma, y_e)/f_i(x_\gamma, y_e)$, leading to an enhancement of the radiative corrections when $y_e$ is close to~1. This logarithmic term originates from the convolution structure
\begin{equation}
\begin{aligned}
f_i^{\mathrm{RC}}\!\left(x_\gamma, y_e\right)
 &= \int_{y_e}^1 \frac{dz}{z}\, P^{(1)}(z)\, f_i\!\left(x_\gamma, \frac{y_e}{z}\right) \\[4pt]
 &= f_i\!\left(x_\gamma, y_e\right)\! \int_{y_e}^1 dz\, P^{(1)}(z)
    + \int_{y_e}^1 dz\, \frac{1+z^2}{1-z}
      \!\left[\frac{1}{z} f_i\!\left(x_\gamma, \frac{y_e}{z}\right)
      - f_i\!\left(x_\gamma, y_e\right)\right] \\[4pt]
 &= f_i\!\left(x_\gamma, y_e\right)
    \!\left[2\ln(1-y_e) + \frac{y_e(2+y_e)}{2}\right]
    + \int_{y_e}^1 dz\, \frac{1+z^2}{1-z}
      \!\left[\frac{1}{z} f_i\!\left(x_\gamma, \frac{y_e}{z}\right)
      - f_i\!\left(x_\gamma, y_e\right)\right].
\end{aligned}
\end{equation}
Here, in the first line, we make a change of variable from \(t\) to \(z = y_e / t\). 
The most singular region of \(P^{(1)}(z)\) is located around \(z \sim 1\), and its dominant behavior is well captured by the first integral in the second line, 
which gives rise to the logarithmic term \(\ln(1-y_e)\) in the third line. 
In contrast, the second integral is regular around \(z \sim 1\) and does not produce any logarithmic enhancement as \(y_e \to 1\).

\subsection{Case (ii): a laboratory-frame energy cutoff on the second photon}\label{subsec:laboratory}
To incorporate experimental conditions more realistically, such as those in the KLOE experiment~\cite{KLOE:2009urs}, we extend the $O(\alpha^2 L_e)$ radiative correction to the case in which the second photon is subject to a laboratory-frame energy cutoff. We first define the kinematics. Let the meson $P$ carry momentum $\vec{p}^{~\text{lab}}$ in the laboratory frame, and impose an angle-independent, laboratory-frame energy cutoff on the second photon, $E_{\gamma_2}^{\text{lab}} < E^{\text{lab}}_{\gamma_2,\text{cut}}$. We define $\theta_{P e}$ ($\theta_{P\gamma_2}$) as the angle between $\vec{p}^{~\text{lab}}$ and the momentum of the electron (momentum of the second photon) in the rest frame of $P$.

As discussed in Ref.~\cite{Kuraev:2003gq}, the $O(\alpha^2 L_e)$ radiative corrections originate solely from the phase-space region where the second photon is emitted collinearly with the electron. Contributions from other angular regions are absorbed into the $O(\alpha^2 L_e^0)$ corrections contained in the function $K(x_\gamma, y_e)$. Accordingly, we focus on those contributions where the second photon satisfies the collinearity condition $\theta_{P\gamma_2} \approx \theta_{P e}$. In the rest frame of the meson, the energy $E_{\gamma_2}^*$ of such a collinear photon has an angular-dependent cutoff $E_{\gamma_2}^* < E^*_{\gamma_2,\text{cut}}(\theta_{P e})$, given by
\begin{equation}
	E^*_{\gamma_2,\text{cut}}(\theta_{P e}) = \frac{E^{\text{lab}}_{\gamma_2,\text{cut}}}{\gamma(1 + \beta \cos \theta_{P e})},
\end{equation}
where the standard Lorentz factors are defined by $\gamma = \sqrt{(\vec{p}^{~\text{lab}})^2 + m_P^2}/m_P$ and $\beta = \sqrt{1 - \gamma^{-2}}$.

In the derivation of Eq.~\eqref{B_RC_full} in the previous section, it was assumed that the differential decay rate is independent of the emission angle $\theta_{P e}$ of the electron in the rest frame of the meson. Although this assumption holds at the level of the decay amplitude, the angular dependence introduced by the cutoff $E^*_{\gamma_2,\text{cut}}(\theta_{P e})$ in the phase-space integration leads to a nontrivial $\theta_{P e}$ dependence in the radiative correction. Consequently, an average over the electron’s angular distribution is required. Taking this into account, the $O(\alpha^2 L_e)$ correction in Eq.~\eqref{B_RC_full} generalizes to
\begin{equation}\label{B_RC_cut}
    \frac{d B_{\text{cut}}^{\mathrm{RC}}[P\to e\nu_e \gamma]}{d x_\gamma\, d y_e}=\frac 12\int_{-1}^1 d\cos\theta_{P e}\int_{y_e}^{y_e^\text{max}(\theta_{P e})} \frac{d t}{t} \frac{d B[P\to e\nu_e \gamma]}{d x_\gamma\, d y_e}(x_\gamma, t) D\left(\frac{y_e}{t}\right).
\end{equation}
Here, we denote the differential branching ratio with the energy cutoff on the second photon as $\frac{dB^{\mathrm{RC}}_{\text{cut}}[P\to e\nu_e \gamma]}{dx_\gamma\,dy_e}$. The upper limit of the convolution is given by the kinematic constraint $y_e^{\text{max}}(\theta_{P e})=\min\{1,y_e+2 E^*_{\gamma_2,\text{cut}}(\theta_{P e})/ m_P\}$. 

As in the previous subsection, Eq.~\eqref{B_RC_cut} can be simplified to
\begin{equation}
\begin{aligned}
    \frac{dB^{\mathrm{RC}}_{\text{cut}}[P\to e\nu_e \gamma]}{dx_\gamma\,dy_e} &= \frac{\alpha}{2\pi(1-r_e)^2}\,B[P\to e\nu_e(\gamma)]\\&\times \left( A(x_\gamma,y_e) + \frac{\alpha}{2\pi}(L_e-1) A^{\mathrm{RC}}_{\text{cut}}(x_\gamma,y_e;\vec{p}^{~\text{lab}},E^{\text{lab}}_{\gamma_2,\text{cut}}) \right),
\end{aligned}
\end{equation}
where the cutoff-dependent correction function $A^{\mathrm{RC}}_{\text{cut}}(x_\gamma,y_e;\vec{p}^{~\text{lab}},E^{\text{lab}}_{\gamma_2,\text{cut}})$ is given by
\begin{equation}
	\begin{aligned}
		&A^{\mathrm{RC}}_{\text{cut}}(x_\gamma,y_e;\vec{p}^{~\text{lab}},E^{\text{lab}}_{\gamma_2,\text{cut}})=\, f^{\mathrm{RC}}_{\text{cut},\mathrm{IB}}(x_\gamma,y_e;\vec{p}^{~\text{lab}},E^{\text{lab}}_{\gamma_2,\text{cut}}) \\
		&
		+\frac{1}{r_e}\left(\frac{m_P}{2f_P}\right)^2 \Bigl[\left(F_V+F_A\right)^2 f^{\mathrm{RC}}_{\text{cut},\mathrm{SD}^{+}}(x_\gamma,y_e;\vec{p}^{~\text{lab}},E^{\text{lab}}_{\gamma_2,\text{cut}}) + \left(F_V-F_A\right)^2 f^{\mathrm{RC}}_{\text{cut},\mathrm{SD}^{-}}(x_\gamma,y_e;\vec{p}^{~\text{lab}},E^{\text{lab}}_{\gamma_2,\text{cut}}) \Bigr]\\
		& - \left(\frac{m_P}{f_P}\right) \Bigl[\left(F_V+F_A\right) f^{\mathrm{RC}}_{\text{cut},\mathrm{INT}^{+}}(x_\gamma,y_e;\vec{p}^{~\text{lab}},E^{\text{lab}}_{\gamma_2,\text{cut}}) + \left(F_V-F_A\right) f^{\mathrm{RC}}_{\text{cut},\mathrm{INT}^{-}}(x_\gamma,y_e;\vec{p}^{~\text{lab}},E^{\text{lab}}_{\gamma_2,\text{cut}}) \Bigr],
	\end{aligned}
\end{equation}
with
\begin{equation}
    f^{\mathrm{RC}}_{\text{cut},i}(x_\gamma,y_e;\vec{p}^{~\text{lab}},E^{\text{lab}}_{\gamma_2,\text{cut}})=\frac{1}{2}\int_{-1}^1 d\cos\theta_{P e}\int_{y_e}^{y_e^{\text{max}}(\theta_{P e})}\frac{dt}{t} P^{(1)}(y_e/t) f_{i}(x_\gamma,y_e),
\end{equation}
for $i \in \{\mathrm{IB},\mathrm{SD}^{+},\mathrm{SD}^{-},\mathrm{INT}^{+},\mathrm{INT}^{-}\}$. The integration in $f^{\mathrm{RC}}_{\text{cut},i}(x_\gamma,y_e;\vec{p}^{~\text{lab}},E^{\text{lab}}_{\gamma_2,\text{cut}})$ can also be evaluated analytically. However, due to the complexity of their explicit forms, we do not present them here.

\section{Scalar Function Method\label{Appendix:scalar}}
In this section, we briefly review the method of constructing hadronic matrix elements using the ``scalar function method'' as proposed in Ref.~\cite{Tuo:2021ewr}. This derivation is in the infinite-volume and continuum limit. The finite-volume corrections are discussed in Sec.~\ref{sec:IVR} and Appendix.~\ref{Appendix:FV}.

We first define the Euclidean-space counterparts of the scalar functions $\tilde{I}_i(x_\gamma)$ defined in Eq.~\eqref{IM}:
\begin{equation}\label{IE_m}
	\begin{aligned}
		& \tilde{I}_{E,1}(x_\gamma)=\delta^{\mu\nu}\,m_P^2\,H_E^{\mu \nu}(k_E, p_E), 
		\quad \tilde{I}_{E,2}(x_\gamma)=-p_{E}^\mu\,p_{E}^\nu\,H_E^{\mu \nu}(k_E, p_E), \\
		& \tilde{I}_{E,3}(x_\gamma)=-k_{E}^\mu\,p_{E}^\nu\,H_E^{\mu \nu}(k_E, p_E), 
		\quad \tilde{I}_{E,4}(x_\gamma)=-p_{E}^\mu\,k_{E}^\nu\,H_E^{\mu \nu}(k_E, p_E), \\
		& \tilde{I}_{E,5}(x_\gamma)=-k_{E}^\mu\,k_{E}^\nu\,H_E^{\mu \nu}(k_E, p_E), 
		\quad \tilde{I}_{E,6}(x_\gamma)=\varepsilon_E^{\mu\nu \alpha \beta}\,k_E^\alpha p_E^\beta\,H_E^{\mu \nu}(k_E, p_E).
	\end{aligned}
\end{equation}
with $p_E=(im_P,\vec{0})$ and $k_E=(ik^0,\vec{k})$ the Euclidean momenta of the initial-state meson and the photon. $H_E^{\mu \nu}(k_E, p_E)$ is defined in Eq.~\eqref{Direct}.
If, as is the case for $P\to\ell\nu_\ell\gamma$ decays, there are no intermediate hadronic states lighter than the initial state, 
the relation 
$H_M^{\mu \nu}(k,p) = c_{M\to E}^{\mu\nu} H_E^{\mu \nu}(k_E, p_E)$ holds, 
where $c_{M\to E}^{\mu\nu}$ encodes the difference between Euclidean and Minkowski gamma matrices. Using the values of $c_{M\to E}^{\mu\nu}$ given below Eq.\,\eqref{Direct2} and the definitions of momenta and the Levi-Civita symbol in Euclidean and Minkowski spaces, one can verify that the scalar functions coincide, namely $\tilde{I}_{i}(x_\gamma)=\tilde{I}_{E,i}(x_\gamma)$.

We multiply both sides of Eq.~\eqref{Direct} by the appropriate Lorentz structures in Eq.~\eqref{IE_m} to obtain expressions for the scalar functions. The scalar functions defined in Eq.~\eqref{IE_m} are independent of the direction of the momentum \(\vec{k}\), which allows us to perform angular averages over \(\vec{k}\) as follows:
\begin{equation}\label{k_ave}
	\begin{aligned}
	\int d^3 x\, e^{-i \vec{k} \cdot \vec{x}} & \rightarrow \int \frac{d \Omega_{\hat{k}}}{4 \pi} \int d^3 x\, e^{-i \vec{k} \cdot \vec{x}} = \int d^3 x\, j_0(\varphi)\,, \quad \varphi = |\vec{k}|\, |\vec{x}|\,, \\
	k^i \int d^3 x\, e^{-i \vec{k} \cdot \vec{x}} & \rightarrow \int \frac{d \Omega_{\hat{k}}}{4 \pi} \int d^3 x\, \left(i \partial_i\right) e^{-i \vec{k} \cdot \vec{x}} = \int d^3 x\, \left(i \partial_i\right) j_0(\varphi) \\
	& = -i\, |\vec{k}| \int d^3 x\, j_1(\varphi) \frac{x_i}{|\vec{x}|}\,, \\
	k^i k^j \int d^3 x\, e^{-i \vec{k} \cdot \vec{x}} & \rightarrow \int \frac{d \Omega_{\hat{k}}}{4 \pi} \int d^3 x\, \left(-\partial_i \partial_j\right) e^{-i \vec{k} \cdot \vec{x}} = \int d^3 x\, \left(-\partial_i \partial_j\right) j_0(\varphi) \\
	& = |\vec{k}| \int d^3 x\, j_1(\varphi) \frac{\delta_{ij}}{|\vec{x}|} - |\vec{k}|^2 \int d^3 x\, j_2(\varphi) \frac{x_i x_j}{|\vec{x}|^2}\,,
	\end{aligned}
\end{equation}
Here \(\int d\Omega_{\hat{k}}\cdots~\) denotes integrals over the direction of \(\vec{k}\). The definitions of the spherical Bessel functions $j_i(\varphi)$ can be found in Eq.\,\eqref{eq:j123def}. 

To illustrate the application of the angular averaging in Eq.\,(\ref{k_ave})  we express the first scalar function as
\begin{equation}
\begin{aligned}
	\tilde{I}_1(x_\gamma) &= \tilde{I}_{E,1}(x_\gamma)\\
    &=\delta^{\mu\nu}\, m_P^2\, H_E^{\mu\nu}(k_E,p_E)\\[1mm]
	& = -i\, m_P^2\, \int_{-\infty}^{\infty} dt \int d^3\vec{x}\, e^{k^0 t - i \vec{k} \cdot \vec{x}}\, \delta^{\mu\nu}\, H_E^{\mu\nu}(x)\\[1mm]
	& = -i\, m_P^2\, \int_{-\infty}^{\infty} dt \int d^3\vec{x}\, e^{k^0 t}\, j_0(\varphi)\, I_1(|\vec{x}|,t)\,.
\end{aligned}
\end{equation}
Similarly, for the other scalar functions, we can use Eq.~\eqref{k_ave} to simplify the expressions and obtain Eqs.\,(\ref{scalar_calc}) and~(\ref{jphi}) in the main text.

Next, we construct the hadronic matrix element by
\begin{equation}
	H_M^{\mu\nu}(k,p)=\sum_{i=1}^6 \omega_i^{\mu\nu}(k,p)\,\tilde{I}_i(x_\gamma),
\end{equation}
To derive $\omega_i^{\mu\nu}(k,p)$, we first express the hadronic matrix element as a general Lorentz decomposition:
\begin{equation}
	\begin{aligned}
	H_M^{\mu \nu}(k, p) &= a(x_\gamma)\, k^\mu p^\nu + b(x_\gamma)\, k^\nu p^\mu + c(x_\gamma)\, k^\mu k^\nu \\
	&+ d(x_\gamma)\, p^\mu p^\nu + e(x_\gamma)\, g^{\mu \nu}\, m_P^2 + f(x_\gamma)\, \varepsilon^{\mu \nu \alpha \beta} k_\alpha p_\beta.
	\end{aligned}
\end{equation}
Multiplying both sides of the above equation by the Lorentz structures in Eq.~\eqref{IM} yields linear equations that relates the scalar functions \(\tilde{I}_i(x_\gamma)\) (\(i=1,\ldots,6\)) to the coefficients \(a(x_\gamma),\ldots, f(x_\gamma)\). By solving these equations, \(a(x_\gamma),\ldots, f(x_\gamma)\) can be expressed in terms of the scalar functions \(\tilde{I}_i(x_\gamma)\). Substituting them back into the Lorentz decomposition and rearranging the terms then gives the explicit expressions for \(\omega_i^{\mu\nu}(k,p)\) as
\begin{equation}
\begin{aligned}
    \omega_1^{\mu\nu}(k,p)&=\frac{g^{\mu\nu}}{2 m_P^2}-\frac{k^\mu p^\nu+k^\nu p^\mu}{m_P^4 x_\gamma}+\frac{2 k^\mu k^\nu}{m_P^4 x_\gamma^2},\\
    \omega_2^{\mu\nu}(k,p)&=\frac{4 k^\mu k^\nu}{m_P^4 x_\gamma^2},\\
    \omega_3^{\mu\nu}(k,p)&=-\frac{g^{\mu\nu}}{m_P^2 x_\gamma}+\frac{2 k^\mu p^\nu+6 k^\nu p^\mu}{m_P^4 x_\gamma^2}-\frac{12 k^\mu k^\nu}{m_P^4 x_\gamma^3},\\
    \omega_4^{\mu\nu}(k,p)&=-\frac{g^{\mu\nu}}{m_P^2 x_\gamma}+\frac{6 k^\mu p^\nu+2 k^\nu p^\mu}{m_P^4 x_\gamma^2}-\frac{12 k^\mu k^\nu}{m_P^4 x_\gamma^3},\\
    \omega_5^{\mu\nu}(k,p)&=\frac{2 g^{\mu\nu}}{m_P^2 x_\gamma^2}-\frac{12 (k^\mu p^\nu+k^\nu p^\mu)}{m_P^4 x_\gamma^3}+\frac{24 k^\mu k^\nu}{m_P^4 x_\gamma^4}+\frac{4 p^\mu p^\nu}{m_P^4 x_\gamma^2},\\
    \omega_6^{\mu\nu}(k,p)&=\frac{2i \epsilon^{\mu\nu\alpha\beta} k_\alpha p_\beta}{m_P^4 x_\gamma^2}.
\end{aligned}
\end{equation}

\section{Finite-Volume Corrections\label{Appendix:FV}}
In this section, we use the pion decay as an example to illustrate how to correct for finite-volume effects. The finite-volume effect $\delta_i^{\text{IVR}}(L)$ in Eq.~\eqref{scalar_calc2} is defined by
\begin{equation}\label{deltaIVR}
\begin{aligned}
    \delta_i^{\text{IVR}}(L)&=\tilde{I}_i(x_\gamma)-\tilde{I}^{\text{IVR}}_i(x_\gamma;L)\\
    &=\lim_{L_\infty\to\infty}\left(\tilde{I}^{\text{IVR}}_i(x_\gamma;L_\infty)-\tilde{I}^{\text{IVR}}_i(x_\gamma;L)\right)
\end{aligned}
\end{equation}
In our numerical calculations, the finite-volume correction $\delta_i^{\mathrm{IVR}}(L)$ is obtained by comparing ground-state models computed in a finite volume $V=L^3$ with those in a sufficiently large reference volume $V_\infty = L_\infty^3$, where finite-volume effects are negligible.
We take $L_\infty = 22~\mathrm{fm}$ and have verified that the residual finite-volume effects beyond $V_\infty$ are negligible. 

The ground-state models used to estimate finite-volume effects are motivated by the point-particle contribution to the infinite-volume Minkowski hadronic function:
\begin{equation}\label{Hpt_F}
    H_{M,\text{pt}}^{\mu\nu}(k,p) = f_\pi \left[ g^{\mu\nu} - \frac{(2p - k)^\mu (p - k)^\nu}{(p - k)^2 - m_\pi^2} \right],
\end{equation}
where the second term in the bracket corresponds to the scalar QED vertex and propagator structure. The first term, proportional to \( g^{\mu\nu} \), ensures the Ward identity \( k_\mu H_{M,\text{pt}}^{\mu\nu}(k,p) = f_\pi p^\nu \) is satisfied.

To implement the point-particle contribution in a finite volume, we first express it as a sum over time-ordered contributions. Specifically, the contributions from the time-ordered diagrams in Fig.~\ref{fig:diagrams} can be written as 
\begin{equation}\label{Hpt_TO}
\begin{aligned}
    H_{M,A}^{\mu\nu}(k,p) &= f_\pi\frac{\langle 0\vert J_{W,M}^\nu \vert \pi(-\vec{k})\rangle
		\langle \pi(-\vec{k})\vert J_{\text{em},M}^\mu\vert \pi(\vec{0})\rangle}{2E_\pi(\vec{k})(k^0 + E_\pi(\vec{k}) - m_\pi)}\\
    &= f_\pi F^{(\pi)}\bigl((p_\pi - p)^2\bigr)\frac{(p_\pi + p)^\mu p_\pi^\nu}{2E_\pi(\vec{k})(k^0 + E_\pi(\vec{k}) - m_\pi)}, \quad p_\pi = (E_\pi(\vec{k}), -\vec{k}), \\
    H_{M,B}^{\mu\nu}(k,p) &= -f_\pi\frac{\langle 0\vert J_{\text{em},M}^\mu\vert\pi(\vec{k})\pi(\vec{0})\rangle
		\langle \pi(\vec{k})\pi(\vec{0})\vert J_{W,M}^\nu\vert\pi(\vec{0})\rangle}{2E_\pi(\vec{k})2m_\pi(k^0 - E_\pi(\vec{k}) - m_\pi)}\\
    &\simeq -f_\pi F^{(\pi)}\bigl((p'_\pi + p)^2\bigr)\frac{(p_\pi' - p)^\mu p_\pi'^\nu}{2E_\pi(\vec{k})(k^0 - E_\pi(\vec{k}) - m_\pi)}, \quad p_\pi' = (E_\pi(\vec{k}), \vec{k}),
\end{aligned}
\end{equation}
where \( E_\pi(\vec{k}) = \sqrt{\vec{k}^2 + m_\pi^2} \). The subscripts ``A'' and ``B'' denote the diagram A and B in Fig.~\ref{fig:diagrams}. In deriving the contribution from diagram~B, we restrict our consideration to the momentum configuration \(\pi(\vec{k})\pi(\vec{0})\) and approximate
\begin{equation}\label{pipi_nonint}
    \langle \pi(\vec{k})\,\pi(\vec{0}) \vert J_{W}^\nu(0) \vert \pi(\vec{0}) \rangle
\simeq 
\langle \pi(\vec{k}) \vert J_{W}^\nu(0) \vert 0 \rangle \,
\langle \pi(\vec{0}) \vert \pi(\vec{0}) \rangle,
\end{equation}
under the assumption that the intermediate \(\pi(\vec{k})\pi(\vec{0})\) states are nearly non-interacting, since these are the only parts corresponding to the point-particle contribution.
Then $H_{M,\text{pt}}^{\mu\nu}(k,p)$ in Eq.~\eqref{Hpt_F} can be decomposed into the sum of the time-ordered diagrams~A and~B when the electromagnetic form factor is set to $F^{(\pi)}(q^2) = 1$:  
\begin{equation}\label{HM_pt}
\begin{aligned}
    H_{M,\text{pt}}^{\mu\nu}(k,p)&=f_\pi(g^{\mu\nu}-\delta^{\mu 0}\delta^{\nu 0})+H_{M,A_{\text{pt}}}^{\mu\nu}(k,p)+H_{M,B_{\text{pt}}}^{\mu\nu}(k,p)\\
    H_{M,A_{\text{pt}}}^{\mu\nu}(k,p) &= f_\pi \frac{(p_\pi + p)^\mu p_\pi^\nu}{2E_\pi(\vec{k})(k^0 + E_\pi(\vec{k}) - m_\pi)} \\
    H_{M,B_{\text{pt}}}^{\mu\nu}(k,p) &= -f_\pi \frac{(p_\pi' - p)^\mu p_\pi'^\nu}{2E_\pi(\vec{k})(k^0 - E_\pi(\vec{k}) - m_\pi)}.
\end{aligned}
\end{equation}
An additional contact term, \( f_\pi \bigl(g^{\mu\nu} - \delta^{\mu 0} \delta^{\nu 0}\bigr) \), accounts for the contribution from heavier vector-like meson states required to restore the Ward identity and is expected to induce negligible finite-volume effects; in fact, the determination of \( f^{\text{3pt}} \) in Eq.~\eqref{f3pt} relies on this term. 
In contrast, the dominant finite-volume effects originate from the contribution of diagram~A in the $t<0$ time ordering and from diagram~B in the $t>0$ time ordering.

Next, we express the contributions from diagrams A and B in Euclidean coordinate space within a finite volume. For the \( t<0 \) time ordering, the contribution from diagram A is
\begin{equation}\label{HLA}
	\begin{aligned}
		&H_{E,A}^{(L),\mu\nu}(x)
		= \frac{1}{L^3}\sum_{\vec{k}\in\Gamma}\frac{1}{2E_\pi(\vec{k})}
		\langle 0\vert J_{W,E}^\nu(0)\vert \pi(-\vec{k})\rangle
		\langle \pi(-\vec{k})\vert J_{\text{em},E}^\mu(\vec{x},t)\vert \pi(\vec{0})\rangle\\
		=& \frac{1}{L^3}\sum_{\vec{k}\in\Gamma}\frac{1}{2E_\pi(\vec{k})}\,\left(f_\pi\,p_{E,\pi}^\nu\right)\,\left(-iF^{(\pi)}\bigl(-(p_{E,\pi} - p_E)^2\bigr)\,
		\bigl(p_{E,\pi} + p_E\bigr)^\mu\right)\,
		e^{(E_\pi(\vec{k}) - m_\pi)\,t}\,
		e^{i\vec{k}\cdot\vec{x}},
	\end{aligned}
\end{equation}
where $p_{E}=(im_\pi,\vec{0})$ and $p_{E,\pi}=(iE_\pi(\vec{k}),-\vec{k})$ 
denote the Euclidean momenta of the initial and intermediate pions, respectively. The momentum sum runs over all modes $\Gamma = \left\{\vec{k} \,\big|\, \vec{k} = \frac{2 \pi}{L} \vec{n} \right\}$ in the finite-volume, 
with $\vec{n}$ denoting three-dimensional integer vectors. It can be shown that inserting $H_{E,A}^{(L),\mu\nu}(x)$ into Eq.~\eqref{Direct} and Eq.~\eqref{Direct2} 
reproduces the Minkowski-space hadronic function $H_{M,A}^{\mu\nu}(k,p)$ given in Eq.~\eqref{Hpt_TO} in the infinite-volume limit. 

\begin{figure} 
	\centering
	\includegraphics[width=0.47\textwidth]{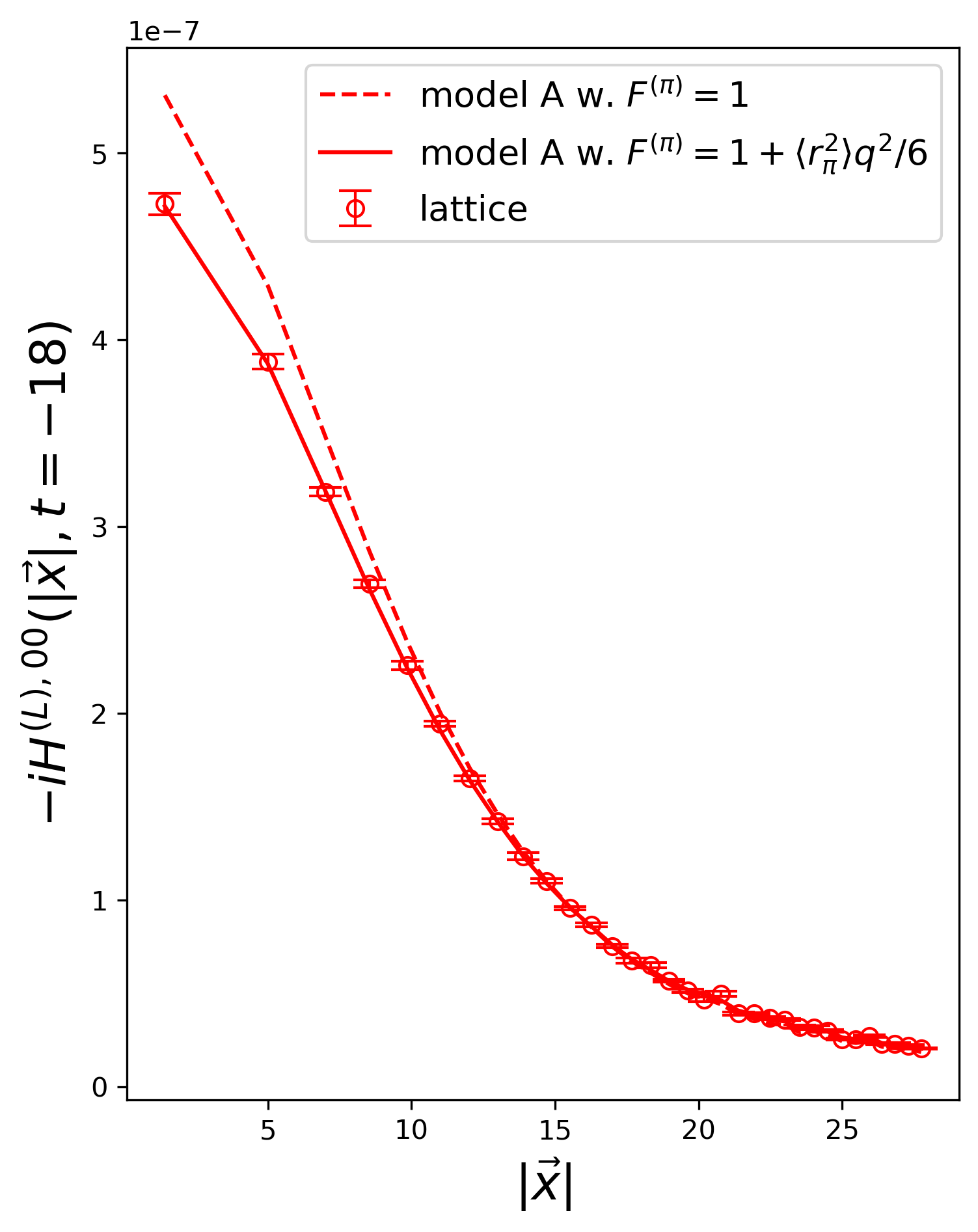}
	\includegraphics[width=0.49\textwidth]{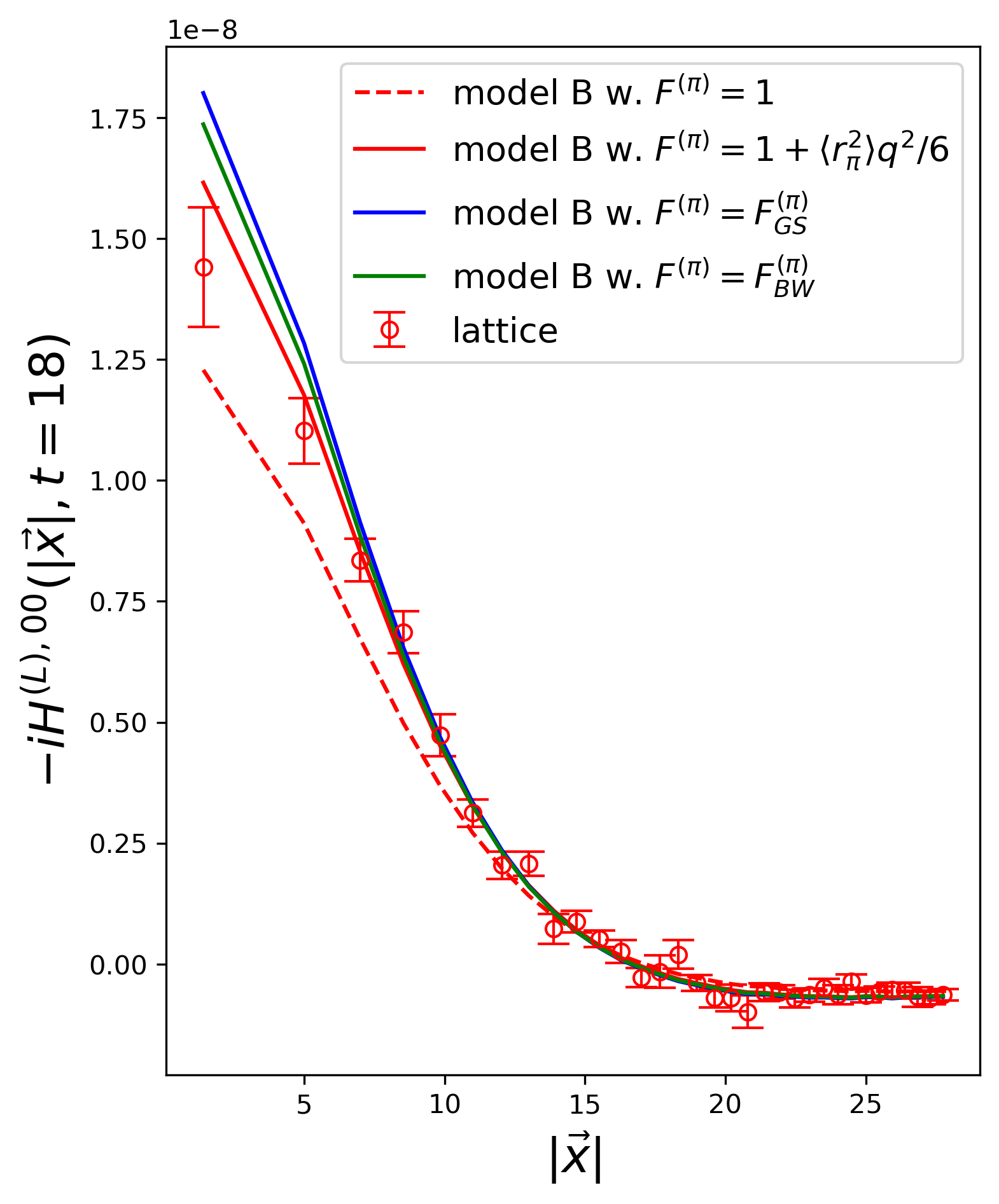}
	\caption{Comparison of lattice data with the point-particle approximation (dashed lines) and the structure-dependent model (solid lines) in the long-distance region.
		\label{fig:HAB}
	}
\end{figure}

In the calculation of $H_{E,A}^{(L),\mu\nu}(x)$, setting \(F^{(\pi)}(q^2) = 1\) corresponds to the point-particle contribution. Alternatively, structure-dependent contribution can be incorporated by expanding \(F^{(\pi)}(q^2)\) linearly in \(q^2\) using the pion charge radius, $F^{(\pi)}(q^2) = 1 + \frac{\langle r_\pi^2 \rangle}{6}\,q^2$. For illustration, these two models are compared with lattice data at the time slice \(t=-18\) in the 48I ensemble in the left panel of Fig.~\ref{fig:HAB}. The dashed and solid lines correspond to \(F^{(\pi)}(q^2) = 1\) and \(F^{(\pi)}(q^2) = 1 + \frac{\langle r_\pi^2 \rangle}{6}\,q^2\), respectively. The pion charge radius $\sqrt{\langle r_\pi^2\rangle}=0.659(4)~\text{fm}$ is taken from the PDG review~\cite{ParticleDataGroup:2024cfk}. The results indicate that both models agree well with the lattice data in the long-distance region where \(\vert \vec{x}\vert\) is large. Since finite-volume effects arise primarily from the long-distance region near the lattice boundaries, both models are suitable for predicting finite-volume corrections, with the dominant contribution described by the point-particle approximation. Eq.~\eqref{HLA} can be straightforwardly generalized to the kaon channel.

Next, we consider the contribution from B diagram in \(t>0\) time ordering:
\begin{equation}
	\begin{aligned}
		&H_{E,B}^{(L),\mu\nu}(x)
		=\frac{1}{L^3}\sum_{\vec{k}\in\Gamma}\frac{1}{2E_\pi(\vec{k})2m_\pi}
		\langle 0\vert J_{\text{em}}^\mu(\vec{x},t)\vert\pi(\vec{k})\pi(\vec{0})\rangle
		\langle \pi(\vec{k})\pi(\vec{0})\vert J_{W}^\nu(0)\vert\pi(\vec{0})\rangle\\
		&\simeq\frac{1}{L^3}\sum_{\vec{k}\in\Gamma}\frac{1}{2E_\pi(\vec{k})}\,\left(-iF^{(\pi)}\bigl(-(p^\prime_{\pi,E}+p_E)^2\bigr)\,
		(p^\prime_{\pi,E} - p_E)^\mu\right)\,\left(f_\pi\,p^{\prime \nu}_{\pi,E}\right)
		e^{-(E_\pi(\vec{k})+m_\pi)t}\,
		e^{i\vec{k}\cdot\vec{x}}.
	\end{aligned}
\end{equation}
Here, $p^\prime_{\pi,E}=(iE_\pi(\vec{k}),\vec{k})$ denotes the Euclidean momentum of the intermediate pion. In the second line, we also employ the approximation that the $\pi(\vec{k})\pi(\vec{0})$ states are non-interacting, as given in Eq.~\eqref{pipi_nonint}. Inserting $H_{E,B}^{(L),\mu\nu}(x)$ into Eq.~\eqref{Direct} and Eq.~\eqref{Direct2}
gives the Minkowski-space hadronic function $H_{M,B}^{\mu\nu}(k,p)$ defined in Eq.~\eqref{Hpt_TO} in the infinite-volume limit. 

In $H_{E,B}^{(L),\mu\nu}(x)$, the energy-momentum transfer for the pion form factor is in the timelike region. In the right panel of Fig.~\ref{fig:HAB}, we take $t=18$ as an example to compare the lattice data in the long-range region with various parameterizations of \(F^{(\pi)}(q^2)\), including the point-particle contribution \(\bigl(F^{(\pi)}(q^2)=1\bigr)\), the linear expansion using the charge radius \(\bigl(F^{(\pi)}(q^2)=1+\frac{\langle r_\pi^2\rangle}{6}q^2\bigr)\), 
as well as the Gounaris--Sakurai (GS) model~\cite{Gounaris:1968mw} and the Breit--Wigner (BW) model~\cite{Breit:1936zzb,Guo:2008nc}. The figure shows that the long-distance behavior with large $|\vec{x}|$ is dominated by the point-particle contribution \(F^{(\pi)}(q^2)=1\). The other three models yield consistent results in the long-distance region and also agree well with the lattice data. Although modeling the \(\pi\pi\) as non-interacting particles is just a approximation, it can still be used to estimate the finite-volume effect from Fig.~\ref{fig:intstate}(B), since the magnitude of this finite-volume effect is comparable to the statistical errors and the model accurately describes the lattice data in the long-distance region. For the kaon channel, because $2m_K$ is significantly
larger, the finite-volume effects from B diagram can be neglected.

We estimate the finite-volume corrections $\delta_{i,\mathrm{pt}}^{\mathrm{IVR}}$ and $\delta_{i,\mathrm{SD}}^{\mathrm{IVR}}$ in Eq.~(\ref{deltaIVR}) using the point-particle approximation (dashed lines in Fig.~\ref{fig:HAB}) and the structure-dependent model (solid lines in Fig.~\ref{fig:HAB}), respectively.
As shown in Sec.~\ref{sec4}, the difference between the two choices is much smaller than the statistical errors.

\section{Determination of $\Delta T$\label{Appendix:deltaT}}

\begin{figure} 
	\centering
	\includegraphics[width=0.9\textwidth]{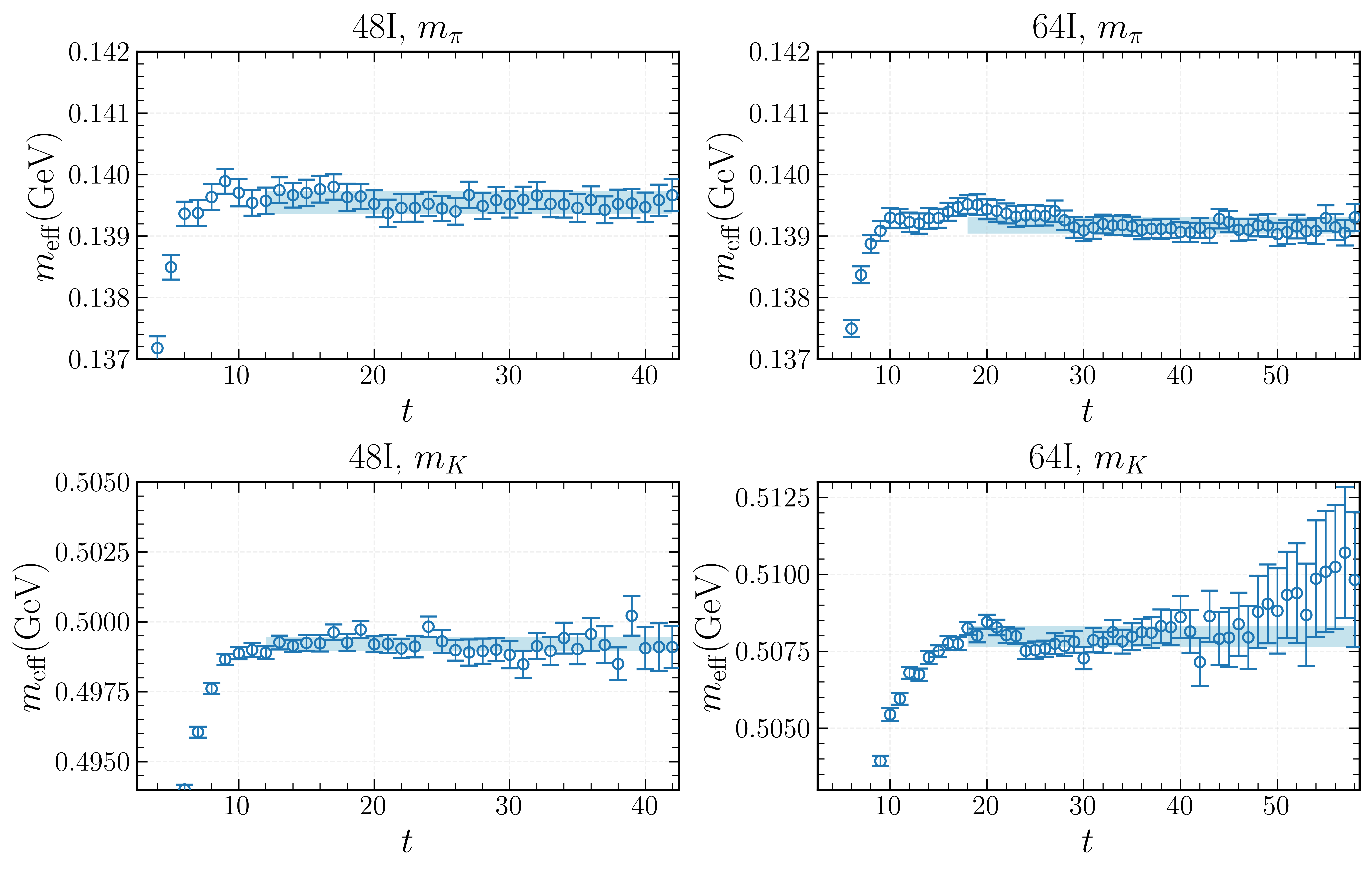}
	\caption{Effective-mass plots for the $\pi$ and $K$ mesons obtained from the two-point function $\langle \phi_P(t)\phi_P^\dagger(0)\rangle$ with the wall-source--point-sink setup. The blue bands indicate the meson masses extracted from fits to the two-point function over the range $t\in [\Delta T, T-\Delta T]$.} \label{fig:meff}
\end{figure}

In practice, $\Delta T$ is determined from the effective-mass plateau of the two-point function $\langle\phi_P(t)\phi^\dagger_P(0)\rangle$, computed with a Coulomb-gauge-fixed wall source and a point sink. As shown in Fig.~\ref{fig:meff}, we plot the effective masses of the $\pi$ and $K$ mesons extracted from this two-point function. They are obtained by solving the following equation for $m_P$ using adjacent data points at $t$ and $t+1$ as input:
\begin{equation}
    \frac{\langle\phi_P(t+1)\phi^\dagger_P(0)\rangle}{\langle\phi_P(t)\phi^\dagger_P(0)\rangle}=\frac{\cosh(m_P(t+1-T/2))}{\cosh(m_P(t-T/2))}.
\end{equation}

Based on the effective-mass plateau, we choose $\Delta T=12$ and $\Delta T=18$ for the 48I and 64I ensembles, respectively. These values correspond to a physical separation of approximately $1.5~\text{fm}$, which is sufficient to isolate the ground state in the two-point functions. In Fig.~\ref{fig:meff}, the blue bands indicate the meson masses obtained from fits to the two-point functions over the time range $t\in [\Delta T, T-\Delta T]$.

In this work, the three-point and two-point functions use the same Coulomb-gauge-fixed wall-source operator $\phi^\dagger_P$ and therefore couple to states with the same quantum numbers. Although the overlap coefficients of these states can differ between them, the excited-state energies and the associated exponential time dependence are the same. We therefore use the value of $\Delta T$ determined from the two-point function as the time separation between $\phi^\dagger_P$ and the nearest current operators in the three-point function, and assume that this separation is also sufficient to suppress excited-state contamination there.

\newpage
\bibliography{ref}
	
\end{document}